\documentclass{nature}

\usepackage{caption}
\usepackage{graphicx}
\usepackage[scaled]{helvet}
\usepackage{amsmath}
\usepackage{amssymb}
\usepackage{bm}
\usepackage{xfrac}

\newcommand*{\myfont}{\fontfamily{phv}\selectfont}
\usepackage{hyperref}

\usepackage[dvipsnames]{xcolor}

\usepackage{lineno}

 \usepackage{color}

\definecolor{myGray}{rgb}{.9,.9,.9}

\title{The physics of brain network structure, function, and control}

\author{Christopher W. Lynn$^1$ \& Danielle S. Bassett$^{1,2,3,4,5,*}$}

\begin{document}

%\linenumbers

\maketitle

\begin{affiliations}
\item Department of Physics \& Astronomy, College of Arts \& Sciences, University of Pennsylvania, Philadelphia, PA 19104, USA
\item Department of Bioengineering, School of Engineering \& Applied Science, University of Pennsylvania, Philadelphia, PA 19104, USA
\item Department of Electrical \& Systems Engineering, School of Engineering \& Applied Science, University of Pennsylvania, Philadelphia, PA 19104, USA
\item Department of Neurology, Perelman School of Medicine, University of Pennsylvania, Philadelphia, PA 19104, USA
\item Department of Psychiatry, Perelman School of Medicine, University of Pennsylvania, Philadelphia, PA 19104, USA
\end{affiliations}

\newpage

\begin{abstract}

The brain is a complex organ characterized by heterogeneous patterns of structural connections supporting unparalleled feats of cognition and a wide range of behaviors. New noninvasive imaging techniques now allow these patterns to be carefully and comprehensively mapped in individual humans and animals. Yet, it remains a fundamental challenge to understand how the brain's structural wiring supports cognitive processes, with major implications for the personalized treatment of mental health disorders. Here, we review recent efforts to meet this challenge that draw on intuitions, models, and theories from physics, spanning the domains of statistical mechanics, information theory, and dynamical systems and control. We begin by considering the organizing principles of brain network architecture instantiated in structural wiring under constraints of symmetry, spatial embedding, and energy minimization. We next consider models of brain network function that stipulate how neural activity propagates along these structural connections, producing the long-range interactions and collective dynamics that support a rich repertoire of system functions. Finally, we consider perturbative experiments and models for brain network control, which leverage the physics of signal transmission along structural wires to infer intrinsic control processes that support goal-directed behavior and to inform stimulation-based therapies for neurological disease and psychiatric disorders. Throughout, we highlight several open questions in the physics of brain network structure, function, and control that will require creative efforts from physicists willing to brave the complexities of living matter.

\end{abstract}

\newpage

It is our good fortune as physicists to seek to understand the nature of the observable world around us. In this inquiry, we need not reach to contemporary science to appreciate the fact that our perception of the world around us is inextricably linked to the world within us: the mind. Indeed, even Aristotle c. 350 B.C. noted that it is by mapping the structure of the world that the human comes to understand their own mind \cite{lear1988desire}. ``Mind thinks itself because it shares the nature of the object of thought; for it becomes an object of thought in coming into contact with and thinking its objects, so that mind and object of thought are the same'' \cite{metaphysics}. Over the ensuing 2000-plus years, it has not completely escaped notice that the mappers of the world have unique contributions to offer the mapping of the mind (from Thales of Miletus, c. 624--546 B.C., to Leonardo Da Vinci, 1452--1519). More recently, it is notable that nearly all famous physicists of the early $\text{20}^{\text{th}}$ century -- Albert Einstein, Niels Bohr, Erwin Schroedinger, Werner Heisenberg, Max Born -- considered the philosophical implications of their observations and theories \cite{stenger2015physicists}. In the post-war era, philosophical musings turned to particularly conspicuous empirical contributions at the intersection of neuroscience and artificial intelligence, spanning polymath John von Neumann's work enhancing our understanding of computational architectures \cite{neumann1958computer} and physicist John Hopfield's invention of the associative neural network, which revolutionized our understanding of collective computation \cite{hopfield1982neural}.

In the contemporary study of the mind and its fundamental organ -- the brain -- nearly all of the domains of physics, perhaps with the exception of relativity, are not only relevant but truly essential, motivating the early coinage of the term \emph{neurophysics} some four decades ago \cite{scott1977neurophysics}. The fundamentals of electricity and magnetism prove critical for building theoretical models of neurons and the transmission of action potentials \cite{koch1983theoretical}. These theories are being increasingly informed by mechanics to understand how force-generating and load-bearing proteins bend, curl, kink, buckle, constrict, and stretch to mediate neuronal signaling and plasticity \cite{tyler2012mechanobiology}. Principles from thermodynamics come into play when predicting how the brain samples the environment (action) or shifts the distribution of information that it encodes (perception) \cite{friston2006free}. Collectively, theories of brain function are either buttressed or dismantled by imaging, with common tools including magnetic resonance imaging \cite{plewes2012physics} and magnetoencephalography \cite{hari2012magnetoencephalography}, the latter being built on superconducting quantum interference devices and next-generation quantum sensors that can be embedded into a system that can be worn like a helmet, revolutionizing our ability to measure brain function while allowing free and natural movement \cite{boto2018moving}. Moreover, recent developments in nanoscale analysis tools and in the design and synthesis of nanomaterials have generated optical, electrical, and chemical methods to explore brain function by enabling simultaneous measurement and manipulation of the activity of thousands or even millions of neurons \cite{alivisatos2013nanotools}. Beyond its relevance for continued imaging advancements \cite{piazza2018enhanced}, optics has come to the fore of neuroscience over the last decade with the development of optogenetics, an approach that uses light to alter neural processing at the level of single spikes and synaptic events, offering reliable, millisecond-timescale control of excitatory and inhibitory synaptic transmission \cite{boyden2005millisecond}. 

Such astounding advances, enabled by the intersection of physics and neuroscience, have motivated the construction of a National Brain Observatory at the Argonne National Laboratory (Director: Peter Littlewood, previously of Cavendish Laboratories) funded by the National Science Foundation, as well as frequent media coverage including titles in the APS News such as ``Physicists, the Brain is Calling You.''\cite{popkin2016physicists} And as physicists answer the call, our understanding of the brain deepens and our ability to mark and measure its component parts expands. Yet alongside this growing systematization and archivation, we have begun to face an increasing realization that it is the interactions between hundreds or thousands of neurons that generate the mind's functional states \cite{alivisatos2013nanotools}. Indeed, from interactions among neural components emerge computation \cite{mcculloch1943logical}, communication \cite{fries2015rhythms}, and information propagation \cite{betzel2018specificity}. We can confidently state of neuroscience what Henri Poincare, the French mathematician, theoretical physicist, and philosopher of science, states of science generally: ``The aim of science is not things themselves, as the dogmatists in their simplicity imagine, but the relations among things; outside these relations there is no reality knowable.''\cite{poincare1905science} The overarching goal of mapping these interactions in neural systems has motivated multibillion-dollar investments across the United States (the Brain Initiative generally, and the Human Connectome Project specifically \cite{essen2013wuminn}), the European Union (the Blue Brain Project \cite{markram2015reconstruction}), China (the China Brain Project \cite{poo2016china}), and Japan (Japan's Brain/MINDS project \cite{okano2015brain}).

While it is clear that interactions are paramount, exactly how the functions of the mind arise from these interactions remains one of the fundamental open questions of brain science \cite{bassett2011understanding}. To the physicist, such a question appears to exist naturally within the purview of statistical mechanics \cite{sethna2006statistical}, with one major caveat: the interaction patterns observed in the brain are far from regular, such as those observed in crystalline structures, and are also far from random, such as those observed in fully disordered systems \cite{bassett2016small}. Indeed, the observed heterogeneity of interaction patterns in neural systems -- across a range of spatial and temporal scales -- generally limits the utility of basic continuum models or mean-field theories, which would otherwise comprise our natural first approaches. Fortunately, similar observations of interaction heterogeneity have been made in other technological, social, and biological systems, leading to concerted efforts to develop a statistical mechanics of complex networks \cite{albert2002statistical}. The resultant area of inquiry includes criteria for building a network model of a complex system \cite{butts2009revisiting}, statistics to quantify the architecture of that network \cite{costa2006characterization}, models to stipulate the dynamics that can occur both in and on a network \cite{gross2008adaptive, zhang2017random, hackett2011cascades}, and theories of network function and control \cite{newman2003structure, motter2015networkcontrology}.

Here, we provide a brief review for the curious physicist, spanning the network-based approaches, statistics, models, and theories that have recently been used to understand the brain. Importantly, the interpretations that can be rationally drawn from all such efforts depend upon the nature of the network representation \cite{butts2009revisiting}, including its descriptive, explanatory, and predictive validity -- topics that are treated with some philosophical rigor elsewhere \cite{bassett2018nature}. Following a simple primer on the nature of network models, we discuss the physics of brain network structure, beginning with an exposition regarding measurement before turning to an exposition regarding modeling. In a parallel line of discourse, we then discuss the physics of brain network function, followed by a description of perturbation experiments and brain network control. In each section we separate our remarks into the known and the unknown, the past and the future, the fact and the speculation. Our goal is to provide an accessible introduction to the field, and to inspire the younger generation of physicists to courageously tackle some of the most pressing open questions surrounding the inner workings of the mind.

\section*{The physics of brain network structure}

We begin with a discussion of the architecture, or structural wiring, of networks in the brain, focusing on the measurement and modeling of their key organizational features (see Box 1 for a simple primer on networks). Each edge in a structural brain network represents a physical connection between two elements. For example, synapses support the propagation of information between neurons \cite{pereda2014electrical} and white matter tracts define physical pathways of communication between brain regions \cite{avena2017communication}. In physics, it has long been recognized that the organization of such structural connections can determine the qualitative large-scale features of a system \cite{albert2002statistical}. In the Ising model, for instance, a one-dimensional lattice remains paramagnetic across all temperatures \cite{ising1925beitrag}, while in two dimensions or more, the system spontaneously breaks symmetry, yielding the type of bulk magnetization exhibited by magnets on a refrigerator \cite{onsager1944crystal, brush1967history}. Similarly, the organization of structural wiring in the brain largely determines the types of mental processes and cognitive functions that can be supported \cite{sporns2004organization, medaglia2015cognitive, sporns2014contributions, petersen2015brain, misic2016from}, from memory \cite{wallace2013randomly, rajan2016recurrent, chaudhuri2016computational} to learning \cite{hermundstad2011learning,tesileanu2017rules}, and from vision \cite{takemura2013visual} to motion \cite{zhen2015locomotion}. However, unlike many physics applications, which assume simple lattice or random network architectures, the wiring of the brain is highly heterogeneous, often making symmetry arguments and mean-field descriptions far from applicable \cite{bassett2016small}. While this heterogeneity presents a unique set of challenges, in what follows we review some powerful experimental and theoretical tools that allow us to distill the brain's structural complexity to a number of fundamental organizing principles.

\begin{center}
[Box 1 here]
\end{center}

\noindent \textbf{Measuring brain network structure.} Some of the earliest empirical measurements of the brain's structural connectivity can be traced to Camillo Golgi, who in 1873 soaked blocks of brain tissue in silver-nitrate solution to provide among the first glimpses of the intricate branching of nerve cells \cite{golgi1885sulla}. Soon after, Santiago Ram\'{o}n y Cajal combined Golgi's method with light microscopy to achieve stunning pictures establishing that neurons do not exist in solitude; they instead combine to form intricate networks of physical connections \cite{y1888estructura}. This notion that the brain comprises a complex web of distinct components, known as the neuron doctrine \cite{shepherd2015foundations}, established the foundation upon which modern network neuroscience has flourished. The introduction of the electron microscope in the 1930s provided even more detailed measurements of the physical connections between neurons. Perhaps the most impressive application remains the complete mapping of interconnections between the 302 neurons in the nematode \textit{C. elegans} \cite{white1986structure}. Since this achievement, reconstructions of the synaptic connectivity in other animals have evolved rapidly, from a mapping of the optic medulla in the visual system of the fruit fly \textit{Drosophila} to the enumeration of connections between 950 distinct neurons in the mouse retina \cite{takemura2013visual,helmstaedter2013connectomic}. Efforts continue to press forward toward the ultimate goal of reconstructing the neuronal wiring diagram of an entire human brain \cite{sporns2005human}.

\begin{figure}
\centering
\includegraphics[width = \textwidth]{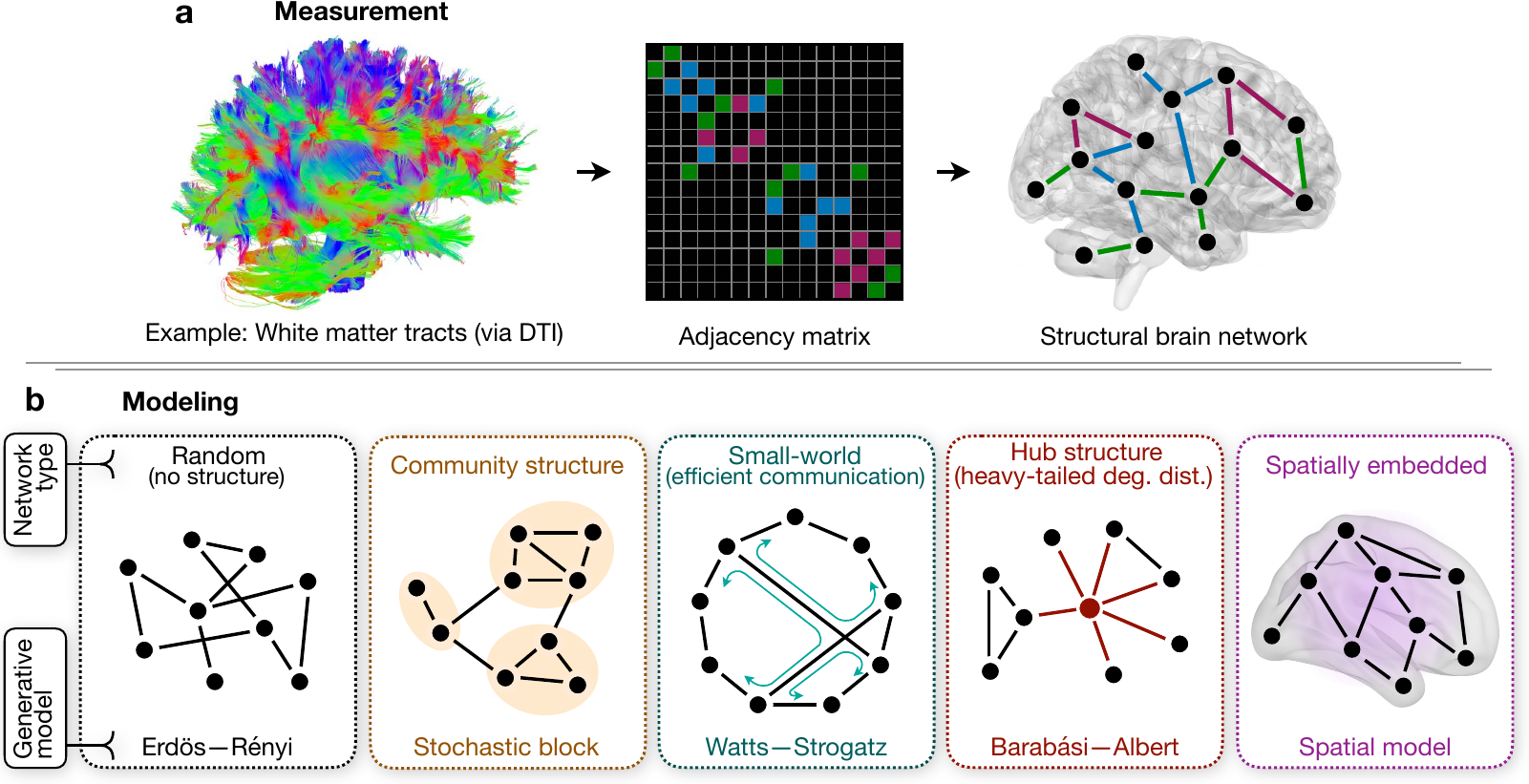} \\
\captionsetup{labelformat=empty}
{\spacing{1.25} \caption{\label{structure} \myfont Figure 1 $\vert$ \textbf{Measuring and modeling brain network structure.} \textbf{a} $\vert$ The measurement of brain network structure begins with experimental data specifying the physical interconnections between neurons or brain regions. As an example, we consider a dataset of white matter tracts measured via DTI. First, the data is discretized into non-overlapping gray matter volumes representing distinct nodes. Then, one constructs an adjacency matrix $\mathbf{A}$, where $A_{ij}$ represents the connection strength between nodes $i$ and $j$. This adjacency matrix, in turn, defines a structural brain network constructed from our original measurements of physical connectivity. \textbf{b} $\vert$ To capture an architectural feature of structural brain networks, we utilize generative network models. The simplest generative network model is the Erd\"{o}s--R\'{e}nyi model, which has no discernible non-random structure. Networks with modular structure, divided into communities with dense connectivity, are constructed using the stochastic block model. Small-world networks, which balance efficient communication and high clustering, are generated using the Watts--Strogatz model. Networks with hub structure, characterized by a heavy-tailed degree distribution, are typically constructed using a preferential attachment model such as the Barab\'{a}si--Albert model. Spatially embedded networks, whose connectivity is constrained to exist within a physical volume, are generated through the use of spatial network models. $\vert$ }}
\end{figure}

Concurrently with these achievements using electron microscopy, complimentary efforts in tract tracing have revealed the mesoscale structure of the macaque \cite{stephan2001advanced, markov2014weighted}, cat \cite{young1994analysis}, mouse \cite{oh2014mesoscale}, and fly \cite{shih2015connectomics}. Particularly important for our understanding of human cognition are recent advances in noninvasive imaging that have allowed unprecedented views of the mesoscale structure of the brain \textit{in vivo}. Introduced in the 1970s, computerized axial tomography (CAT) provided among the most detailed anatomic images of the human brain to date \cite{hsieh2009computed}. Soon after, the development of magnetic resonance imagining (MRI) sparked an explosion of refinements, a notable example being diffusion tensor imaging (DTI) \cite{peirpaoli1996diffusion}. While standard CAT and MRI techniques capture cross-sectional images of the brain, DTI traces the diffusion of water molecules through white matter tracts to reconstruct the large-scale neural pathways connecting distinct brain regions \cite{basser2000invivo, behrens2005relating}. Given measurements of the anatomical wiring connecting a set of neural elements, such as synapses linking neurons or white matter tracts connecting brain regions, researchers can build a structural brain network by forming edges between elements that share a physical connection (Fig. \ref{structure}a). Ongoing experimental efforts to acquire these measurements continue to provide rich network datasets detailing the brain's structural organization.

\noindent \textbf{Modeling brain network structure.} A first glance at the brain's wiring reveals that it is far from homogeneous -- a fact that is not surprising considering the array of physical, energetic, and cognitive constraints that it is required to balance \cite{bullmore2012economy}. To handle this heterogeneity, researchers have increasingly turned to the field of network science for mathematical tools and intuitions \cite{betzel2017generative, bassett2017network}. The primary goal of this interdisciplinary effort has been to distill the explosion of experimental data, spanning structural brain networks in \emph{C. elegans} \cite{nicosia2013phase}, the mouse \cite{henriksen2016simple}, cat \cite{beul2015predictive}, macaque \cite{ercsey2013predictive, beul2017predictive}, and human \cite{betzel2016generative}, down to a number of cogent organizing principles. Here we review some important properties that are thought to characterize structural brain networks and introduce several generative network models that help to explain how these properties arise from underlying biological mechanisms (Fig. \ref{structure}b).

\noindent \textit{Random structure.} While healthy members of a species exhibit anatomical similarities in brain structure, the specific instantiation of physical connections in each individual is far from deterministic. Indeed, \textit{in vivo} imaging techniques in humans, such as DTI described above, have revealed not only stark differences in brain structure between individuals \cite{thompson2001genetic}, but also within the same individual over time \cite{raz2005regional, gong2009age}. Importantly, these structural differences have been linked to variability in a wide range of behaviors \cite{kanai2011structural}, including empathy \cite{banissy2012inter}, introspection \cite{fleming2010relating}, fear acquisition \cite{hartley2011brain}, and even political orientation \cite{kanai2011political}. To study the mathematical properties of random networks, and to understand the types of biological mechanisms that can give rise to qualitative structural properties, it is useful to consider generative network models \cite{betzel2017generative}. The simplest and most common model for generating random networks is the Erd\"{o}s--R\'{e}nyi (ER) model \cite{erdos1960evolution}, wherein each pair of nodes is connected independently with a fixed probability $P$. While the ER model has a number of interesting mathematical properties, such as a binomial degree distribution, it has no discernible structure and does not reflect the mechanisms by which most networks grow in the brain. Accordingly, if we wish to understand some of the principles underlying naturally occurring brain networks, we must consider generative models that yield networks with realistic properties.

\noindent \textit{Community structure.} Perhaps the brain's most well-studied structural property is its division into distinct anatomical regions, which are widely thought to be responsible for specialized cognitive functions \cite{sherrington1906integrative}. Interestingly, by studying the large-scale structure of brain networks in several mammalian species, researchers have shown that the organization of connections tends to partition the networks into densely-connected communities separated by sparse inter-community connectivity \cite{sporns2000theoretical, hilgetag2000anatomical, sporns2004small, sporns2016modular}. Moreover, these clusters of high connectivity closely resemble postulated anatomical subdivisions \cite{hilgetag2000anatomical}. It has therefore been argued that the so-called community structure of brain networks segregates the brain into subnetworks with specific cognitive functions \cite{bassett2010efficient, taylor2017within, lesicko2016connectional, sohn2011topological, azulay2016elegans}. Practically speaking, in order to extract the community structure of a real-world network, one must employ algorithms for community detection -- a vibrant branch of research that is now applied throughout network neuroscience \cite{betzel2017multi, khambhati2017modeling}. From a complimentary perspective, to generate networks with a defined community structure, researchers predominantly use the stochastic block (SB) model, wherein nodes are assigned to distinct communities and an edge is placed between each pair of nodes with a probability that depends on the nodes' community assignments \cite{aicher2015learning, betzel2018diversity}. Such SB networks are often used as null models to distinguish between properties of brain networks that are implied simply by their community structure and those that require additional biological mechanisms \cite{betzel2017generative, betzel2018diversity}.

\noindent \textit{Small-world structure.} Seemingly in contradiction to their striking community structure, large-scale brain networks also exhibit average path lengths between all nodes that are much shorter than a typical random network \cite{bullmore2012economy, van2013network, liao2017small}. This competition between high clustering and short average paths is thought to facilitate the simultaneous segregation and integration of information in the brain \cite{deco2015rethinking}, possibly minimizing the total number of computational steps needed to process external stimuli \cite{latora2001efficient, kaiser2006nonoptimal}. Seeking an explanation for similar ``small-world" topologies exhibited by other real-world systems (most notably social networks \cite{travers1967small}), Duncan Watts and Steven Strogatz developed a model for generating random networks with both high clustering and short average path lengths \cite{watts1998collective}. Generally, the Watts-Strogatz (WS) model supposes that small-world networks are an interpolation between two extreme configurations: a ring lattice, wherein nodes are arranged along a circle and connected to their $k$ nearest neighbors on either side, and an ER random network. Notably, the presence of small-world structure in the brain suggests that efficient communication emerges from a finely-tuned balance of lattice-like organization and structural disorder.

\noindent \textit{Hub structure.} In addition to their modular and small-world structure, many large-scale brain networks also feature high-degree ``hubs", which form a densely interconnected structural core \cite{gong2008mapping}. Acting as bridges between structurally distinct communities, these specialized hub regions are thought to help minimize overall path lengths across the network \cite{sporns2004small} and facilitate the integration of information \cite{deco2015rethinking}. Supporting the notion of a centralized core, many studies have identified hubs within the parietal and prefrontal regions, areas that are often active during a wide range of cognitive functions \cite{gong2008mapping, wedeen2005mapping}. Such core-periphery architecture is characterized by a heavy-tailed degree distribution, such as that observed in scale-free networks, in some cases arising through preferential attachment mechanisms \cite{price1965networks}. In the Barab\`{a}si--Albert (BA) model \cite{barabasi1999emergence}, for instance, nodes are added to a network in sequential order, and each new node $i$ forms an edge with each existing node $j$ with a probability proportional to the degree of node $j$. In this way, new nodes preferentially attach to existing nodes of high degree, creating a ``rich club" of centralized hubs that link otherwise distant regions of the network.

\noindent \textit{Spatial structure.} Thus far, we have focused exclusively on the topological properties of brain networks, which are thought to be driven primarily by the simultaneous functional pressures of information segregation and integration \cite{deco2015rethinking}. However, brain networks are also physically constrained to exist within a tight three-dimensional volume and their structural connections are metabolically driven to minimize total wiring distance \cite{kaiser2006nonoptimal, bassett2010efficient, bullmore2012economy}. Such physical and metabolic constraints are captured by spatial (or geometric) network models, which embed networks into three-dimensional Euclidean space and penalize the formation of long-distance connections \cite{betzel2017generative}. The simplest such model assumes that the probability of two nodes $i$ and $j$ forming an edge is proportional to $d_{ij}^{-\alpha}$, where $d_{ij}$ is the physical distance between $i$ and $j$, and $\alpha\ge 0$ tunes the metabolic cost associated with constructing connections of a given length \cite{dall2002random}. If we keep the number of nodes and edges fixed, one can see that, much like the WS model, this spatial model interpolates between a lattice-like structure, in which nodes only connect to their nearest neighbors ($\alpha \rightarrow \infty$), and an ER random network ($\alpha = 0$).

\noindent \textit{Competition between structural properties.} As the brain grows and adapts to changing cognitive demands, it is widely thought that the underlying network evolves to balance the trade-off between topological value and metabolic wiring cost \cite{bullmore2012economy}. Thus, while the modular, small-world, heavy-tailed, and inherently physical properties of brain networks provide simple organizing principles, in reality the brain is constantly and dynamically weighing these pressures against one another. Accordingly, an accurate generative model should aim to explain multiple real-world properties at once \cite{betzel2017generative}. With this goal in mind, recent work has shown that an impressive range of topological properties can be understood as arising from a competition between two competing factors: a metabolic penalty for the formation of long-distance connections and a topological incentive to connect regions with similar inputs \cite{vertes2012simple}. Notably, investigations of the human, \textit{C. elegans}, and mouse connectomes have revealed that the total wiring distance is consistently greater than minimal, supporting the notion that brain networks weigh the costs of long-distance connections against the functional benefits of an integrated network topology \cite{bassett2010efficient, rubinov2015wiring}. Together, these efforts toward a comprehensive generative model are vital for our understanding of healthy brain network structure, with important clinical implications for the diagnosis, prognosis, prevention, and treatment of disorders of mental health \cite{kaiser2017mechanisms, stam2014modern}.

\noindent \textbf{The future of brain network structure.} Current advances in neuroimaging techniques and network science continue to expand our ability to measure and model the architecture of structural connections in the brain. As experimental measurements become increasingly detailed, an important direction is the bridging of brain network structure at different spatiotemporal scales \cite{scholtens2014linking, chaudhuri2015large, breakspear2017dynamic}. Such cross-scale approaches could link protein interaction networks within neurons to the wiring of synaptic connectivity between neurons to mesoscale networks connecting brain regions and all the way to social networks linking distinct organisms (Box 2). The goal of such cross-scale integration is to understand how the architecture of connectivity at each of these scales emerges from the scale below. Practically, researchers have begun to address this goal by employing hierarchical network models \cite{bentley2016multilayer}, which treat each node at the macroscale as an entire subnetwork at the microscale \cite{mejias2016feedforward}.

\begin{center}
[Box 2 here]
\end{center}

Perhaps the most ambitious future goal is the reconstruction of the entire human connectome at the scale of individual neurons, pressing the current boundaries of 3D electron microscopy and statistical image reconstruction \cite{sporns2005human}. Extensive mapping efforts in other species have revealed notable and quantifiable neuronal diversity \cite{seung2014neuronal, arnatkeviciute2018hub}, suggesting the importance of extending network models to include non-identical units. At the mesoscale, advances in noninvasive imaging have allowed researchers to begin tracking changes in structural connectivity over time \cite{nicosia2013phase, scholz2009training, baum2017modular, zuo2017human}. To analyze these temporally ordered measurements, network scientists have extended standard static graph theoretic tools to study networks with dynamically evolving connections \cite{khambhati2017modeling}. Notably, these so-called temporal networks \cite{holme2012temporal} were recently shown to be easier to control, requiring less energy to attain a desired pattern of neural activity, than their static counterparts \cite{li2017fundamental}.

Properly modeling the dynamics \textit{of} brain networks requires also understanding the functional dynamics occurring \textit{on} brain networks. For instance, dating to Donald Hebb's 1949 book \textit{The Organization of Behavior}, it has been posited that the strength of a synaptic connection increases with the persistent synchronized firing of its pre- and postsynaptic neurons \cite{hebb1949organization}. Such Hebbian plasticity has been observed \textit{in vitro} \cite{magee1997synaptically} and is thought to explain many aspects of brain network structure \cite{montague1996framework, song2000competitive}. More generally, Hebb's postulate highlights the fact that a complete understanding of the brain cannot simply include a description of its structural wiring; it must also stipulate the types of dynamics supported by this physical circuitry.

\section*{The physics of brain network function}

While structural brain networks represent the physical wiring between neural elements (e.g., between individual neurons or brain regions), knowledge of this circuitry alone is not sufficient to understand how the brain \textit{works}. For this reason, we turn our attention to models of brain network function that stipulate how neural activity propagates along structural connections. Just as the neuron doctrine postulates that the brain's structure is divided into a network of distinct nerve cells, it is also widely expected that the brain's array of cognitive functions emerges from the collective activity of individual neurons \cite{chialvo2010emergent, alivisatos2013nanotools, bassett2011understanding, fries2015rhythms, tononi2016integrated}. To understand how the firing of simple nerve cells can give rise to the brain's rich repertoire of cognitive functions \cite{abbott2001theoretical}, analogies are often drawn with notions of emergence in statistical mechanics \cite{chialvo2010emergent, dechery2017emergent, bassett2011understanding}. Developed concurrently with the neuron doctrine in the late $\text{19}^{\text{th}}$ century, statistical mechanics established (among other achievements) that the thermodynamic laws governing the macroscopic behavior of gas molecules can be derived from the microscopic dynamics of the molecules themselves \cite{reif2009fundamentals}. Similarly, growing evidence suggests that the dynamics of individual neurons and brain regions, when embedded in networks of structural connections, can produce the types of long-range correlations and collective patterns of activity that we observe in the brain \cite{brody1999correlations, brody1999disambiguating, sporns2000connectivity, schneidman2006weak, levina2007dynamical, chialvo2010emergent, hoevel2014functional}. Here we traverse what is known about brain network function in relatively broad strokes, from the dynamics of distinct neurons to the networked activity of the entire brain.

\noindent \textbf{Measuring brain network function.} The first measurements of the brain's functional organization date to 1815, when Marie-Jean-Pierre Flourens pioneered the use of localized lesions in the brains of living animals to observe their effects on behavior. Through his experiments, Flourens discovered that the cerebellum regulates motor control, the cerebral cortex supports higher cognition, and the brain stem controls vital functions \cite{flourens1842recherches}. The remainder of the $\text{19}^{\text{th}}$ century brought increasingly detailed measurements of the brain's functional organization, from the demonstration that the occipital lobe regulates vision \cite{panizza1855osservazioni} to the discovery that the left frontal lobe is essential for speech \cite{broca1861remarques}. These discoveries, combined with the early images of neural circuits captured by Ram\'{o}n y Cajal \cite{y1888estructura}, culminated in Thomas Scott Sherrington's book \textit{The Integrative Action of the Nervous System}, which proposed the idea that neurons behave in functional groups \cite{sherrington1906integrative}.

Meanwhile, in 1849 the physicist Hermann von Helmholtz achieved the first electrical measurements of a nerve impulse \cite{helmholtz1850Vorlaufiger}, sparking a wave of experiments investigating the electrical properties of the nervous system. Through invasive measurements in animals using newly-developed electroencephalography (EEG) techniques \cite{haas2003hans}, it quickly became clear that individual neurons communicate with one another via electrical signals \cite{caton1875electrical, beck1890strome, lorente1934studies}, thus providing a clear mechanism explaining how information is propagated and manipulated in the brain. Today, scientists possess a rich menu of experimental techniques for measuring brain dynamics across a range of scales. At the neuronal level, the development of invasive methods in animals, such as electrophysiological recordings of brain slice preparations \textit{in vitro} \cite{green1982circadian, edwards1989thin} and calcium imaging of neuronal activity \textit{in vivo} \cite{stosiek2003vivo, grewe2010high}, have vastly expanded our understanding of synaptic communication. At the regional level, complimentary minimally-invasive imaging techniques have identified fundamental properties of information processing in humans \cite{penny2011statistical}. Interestingly, these advances in mesoscale functional imaging can largely be traced to the efforts of physicists. MEG methods, for instance, use superconducting quantum interference devices (SQUIDS) to directly measure the magnetic fields generated by electrical currents in the brain \cite{hamalainen1993magnetoencephalography, boto2018moving}; and PET techniques measure the positron emission of radioisotopes produced in cyclotrons to reconstruct the metabolic activity of neural tissue \cite{bailey2005positron}. Over the last twenty years, measurements of brain dynamics have been increasingly dominated by functional MRI (fMRI) \cite{raichle1998behind}, which estimates neural activity by calculating contrasts in blood oxygen levels, without relying on the invasive injections and radiation that limit the applicability of other imaging techniques \cite{zarahn1997empirical}. This modern progress in functional brain imaging has galvanized the field of network neuroscience by making detailed datasets of large-scale neural activity widely accessible.

\begin{figure}
\centering
\includegraphics[width = \textwidth]{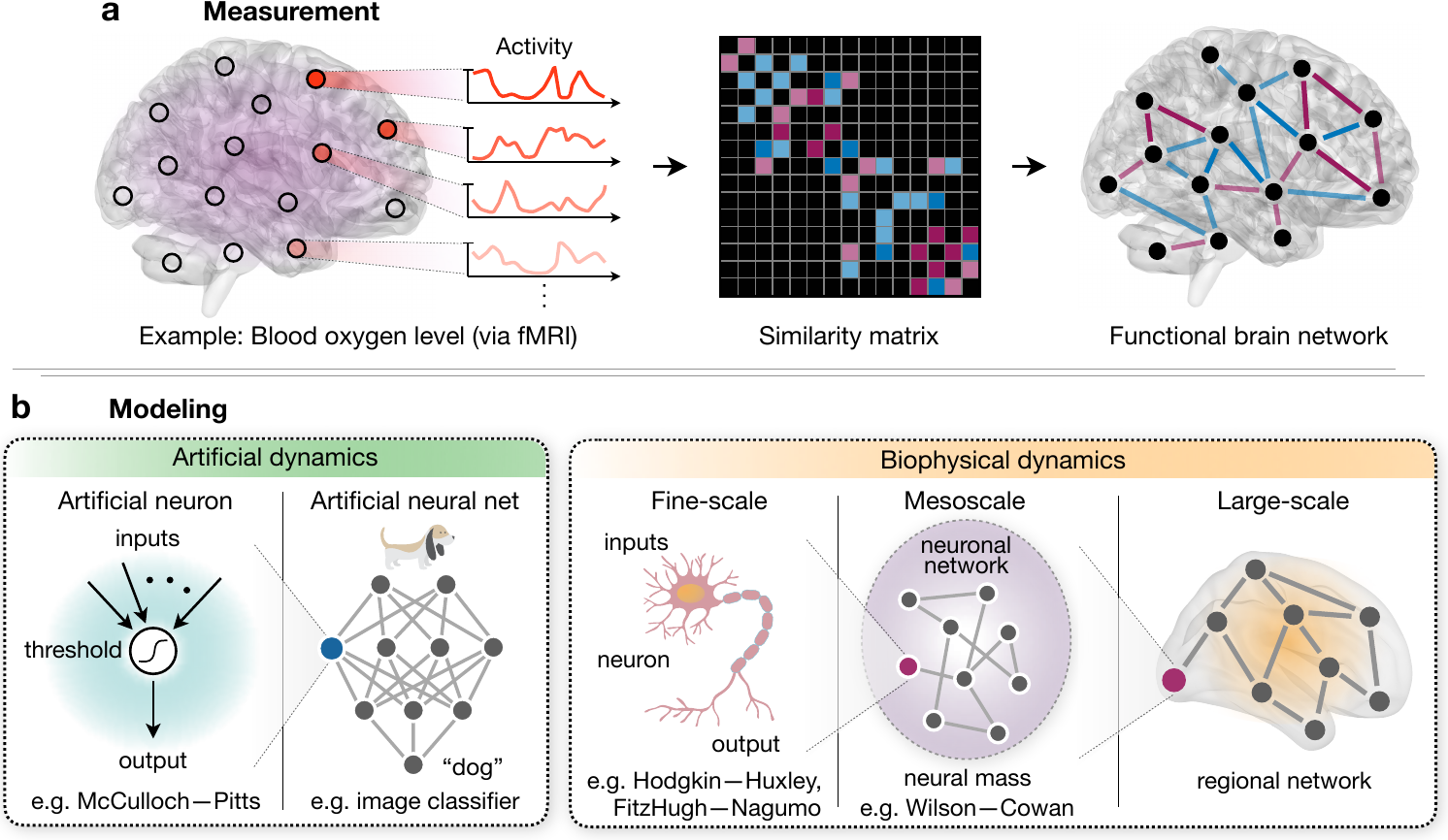} \\
\captionsetup{labelformat=empty}
{\spacing{1.25} \caption{\label{function} \myfont Figure 2 $\vert$ \textbf{Measuring and modeling brain network function.} \textbf{a} $\vert$ The measurement of brain network function begins with experimental data specifying the activity of neurons or brain regions. As an example, we consider variations in blood oxygen level in different parts of the brain measured via fMRI. Calculating the similarity (e.g., correlation or synchronization) between pairs of activity time series, one arrives at a similarity matrix. This matrix, in turn, defines a functional brain network constructed from our original measurements of neural activity. \textbf{b} $\vert$ We divide models of neural activity into two classes: abstract models with artificial dynamics (\emph{left}) and biophysical models with realistic dynamics (\emph{right}). Models of artificial neurons, such as the MP neuron, typically take in a weighted combination of inputs and pass the inputs through a nonlinear threshold function to generate an output. Networks of artificial neurons, from deep neural networks to Hopfield networks, have been shown to reproduce key aspects of human information processing, such as learning from examples and storing memories. By contrast, biophysical models of individual neurons, such as the Hodgkin--Huxley or FitHugh--Nagumo models, capture realistic functional features such as the propagation of the nerve impulse. When interconnected with artificial synapses, researchers are able to simulate entire neuronal networks. Complimentary mesoscale approaches, including neural mass models such as the Wilson--Cowan model, average over all neurons in a population to derive a mean firing rate. To simulate the large-scale activity of an entire brain, researchers use neural mass models to represent brain regions and embed them into a network with connectivity derived from measurements of neural tracts (e.g., as measured via DTI). $\vert$ }}
\end{figure}

One particularly important application of functional brain imaging has been the study of so-called functional brain networks \cite{van2010exploring}, which have allowed researchers to investigate the organization of neural activity using tools from network science. In functional brain networks, as in their structural counterparts, nodes represent physical neural elements, ranging in size from individual neurons to distinct brain regions \cite{bullmore2009complex}. However, whereas structural brain networks define the connectivity between elements based on physical measures of neural wiring (e.g., synapses between neurons or white matter tracts between brain regions), functional brain networks define connectivity based on the similarity between two elements' dynamics \cite{bullmore2009complex}. To see how this works, we briefly consider the common example of a large-scale functional brain network calculated from fMRI measurements of regional activity \cite{van2010exploring} (Fig. \ref{function}a). First, blood oxygen levels indirectly reflecting neural activity are measured within three-dimensional non-overlapping voxels, spatially contiguous collections of which each represent a distinct brain region. After preprocessing the signal to correct for sources of systematic noise such as fluctuations in heart rate, the activity of each brain region is discretized in time, yielding a vector (or time series) of neural activity. Finally, to quantify functional connectivity, one computes the similarity between each pair of brain regions, for example using the quite simple Pearson correlation between the two regions' activity time series \cite{brody1999correlations, zalesky2012correlation}. The end result, even for different types of functional data and different choices for the preprocessing steps and similarity metric, is a functional brain network representing the organization of neural activity.

After constructing a functional brain network, researchers can utilize techniques from network science to study its key organizing features. Such efforts have demonstrated that large-scale functional brain networks, much like structural networks, exhibit signs of modular, small-world, heavy-tailed, and metabolically constrained organization \cite{he2009uncovering, salvador2005neurophysiological, achard2006resilient, bettencourt2007functional, van2010exploring}. The existence of strong functional community structure, for instance, further supports the hypothesis that brain networks segregate into subnetworks with specialized cognitive functions \cite{sadovsky2013scaling, yue2017brain}. Moreover, the presence of high clustering and short average path lengths, combined with the existence of high-degree hub regions, highlights the competing functional pressures of information segregation and integration in the brain \cite{achard2006resilient, bassett2006small}. Metabolic constraints on the brain's structural wiring are also evident in its functional connectivity \cite{rosenbaum2017spatial}, with spatially localized brain regions generally supporting more strongly correlated activity than distant regions \cite{bullmore2012economy}. In light of the similarities between the brain's functional and structural organization, it is tempting to suspect that functional brain networks closely resemble the physical wiring upon which they exist \cite{goni2014resting, honey2009predicting}. However, the relationship between brain function and structure is highly nonlinear \cite{medaglia2018functional}, and understanding how a functional brain network arises from its underlying structural connectivity remains a subject of intense academic focus \cite{park2013structural, breakspear2017dynamic}.

\noindent \textbf{Modeling brain network function.} To understand how the web of physical connections in the brain gives rise to its functional properties, statistical mechanical intuition dictates that we should begin by studying the dynamics of individual elements. Once we have settled on accurate models of the interactions between individual neurons and brain regions, we can link these elements together in a network to predict macroscopic features of the brain's function from its underlying structure \cite{bassett2017network, bassett2018nature}. Interestingly, the history of modeling in neuroscience has followed precisely this path, beginning with models of neuronal dynamics \cite{mcculloch1943logical, hodgkin1952quantitative, fitzhugh1961impulses}, then increasing in scale to mean-field neural mass models of distinct brain regions \cite{beurle1956properties, wilson1972excitatory}, and eventually achieving models of entire networks of neurons and brain regions \cite{hopfield1982neural, schneidman2006weak, kuramoto2012chemical}. Here we review important developments in the modeling of neural dynamics, dividing the modeling techniques into two complimentary classes: those with artificial dynamics and those with biophysically realistic dynamics (Fig. \ref{function}b). As we will see, models from each of these two classes are able to reproduce important aspects of neural activity and system function that have been observed in a range of physiological and behavioral experiments.

\noindent \textit{Artificial models.} One of the earliest mathematical models of neural activity whatsoever was proposed in the mid-1940s by Warren McCulloch and Walter Pitts to describe the logical functioning of an individual neuron \cite{mcculloch1943logical}. Known as the MP neuron, their model accepted binary inputs, combined these inputs using linear weights, and produced a binary output reflecting whether or not the weighted sum of inputs exceeded a given threshold (Fig. \ref{function}b). Albeit a simple caricature of neuronal dynamics, this model has been shown to reproduce some important qualitative features of neuronal activity, including the linear summation of excitatory inputs \cite{cash1999linear} and the ``all-or-none" response to the resulting integrated signal \cite{ferrell1998biochemical}. Moreover, by connecting the inputs and outputs of multiple MP neurons, researchers have achieved deep insights about how brain networks perform basic cognitive functions. For example, soon after the introduction of the MP model, researchers demonstrated that networks of artificial neurons could be used to represent any Boolean function (i.e., any function mapping a list of binary variables to a binary output), thereby establishing the basic capability of neural networks to perform logical computations \cite{scott1977neurophysics}.

While their ability to perform basic computations was quickly realized, it was not clear at the outset whether artificial neural networks could reproduce other cognitive functions, such as the ability to learn or store memories. The former was established by Frank Rosenblatt in 1957, when he showed that the weights on the inputs to an MP neuron could be tuned such that the output defines a binary classifier. Known as the perceptron, this algorithm enabled a single MP neuron to segregate incoming data into one of two classes by learning from past examples. This remarkable result directly inspired more advanced learning algorithms, including support vector machines \cite{hearst1998support} and artificial neural networks \cite{kleene1951representation}, effectively setting in motion the study of machine learning. Today, deep neural networks, consisting of multiple layers of artificial neurons feeding in one direction from the input layer to the output layer (Fig. \ref{function}b), are able to learn a wide range of impressive cognitive functions that we have come to expect from the brain \cite{schmidhuber2015deep}. While the list of applications is ever-expanding, deep neural networks have been used to process and identify images of objects, scenes, and people \cite{egmont2002image}; recognize, interpret, and respond to spoken language \cite{hinton2012deep}; and formulate strategies and make decisions in adversarial settings \cite{silver2016mastering}.

In addition to performing computations and learning from examples, the physicist John Hopfield showed in 1982 that neural networks can also store and recall memories. Specifically, Hopfield demonstrated that the synaptic weights connecting a set of MP neurons could be adjusted in a Hebbian fashion such that the network is able to ``memorize" a number of desired activity states \cite{hopfield1982neural} (i.e., configurations of the network in which each neuron is either active or inactive). Notably, the number of memorized states grows linearly with the number of neurons in the network \cite{newman1988memory}, and errors in recall often yield states that are semantically similar to the target state, a phenomenon commonly observed in humans \cite{hertz1991introduction}. Interestingly, the memorized activity states can be interpreted as local minima of an associated energy function, making each Hopfield network equivalent to an Ising model at zero temperature \cite{brush1967history}. More recently, Ising-like models have also been used to explain the critical or avalanche-like behavior of activity in neural ensembles \cite{moosavi2015structural}, which is thought to support adaptation to environmental changes \cite{karimipanah2015adaptation}, information storage \cite{haldeman2005}, optimal information transmission \cite{beggs2003}, maximal dynamic range \cite{kinouchi2006, shew2009}, and computational power \cite{bertschinger2004}. Further building upon this connection to statistical mechanics, scientists have recently used maximum entropy techniques to construct data-based models of neuronal dynamics. These maximum entropy models, which are equivalent to networks of Ising spins with specially-chosen external fields and interaction strengths, have been shown to predict the observed long-range correlations within naturally occurring networks of neurons and brain regions \cite{schneidman2006weak, ganmor2011sparse}. Together, artificial models of neural dynamics, from simple MP neurons to artificial neural networks and data-driven maximum entropy models, continue to inform our understanding of brain networks as information processing systems.

\noindent \textit{Biophysical models.} While artificial models continue to generate insights about the nature of neural computation, they only vaguely resemble the complex biophysical mechanisms that guide observable neural activity. Among the first biophysically realistic models of the electrical behavior of an individual neuron was achieved nearly a decade after the introduction of the MP neuron by physiologists Alan Lloyd Hodgkin and Andrew Fielding Huxley \cite{hodgkin1952quantitative}. Beginning from a principled description of the initiation and propagation of action potentials in living neurons, the Hodgkin--Huxley (HH) model explains important qualitative aspects of neuronal behavior \cite{scott1977neurophysics}, including the spontaneous emergence of limit cycles or oscillations in activity \cite{lee1998coherence} and the presence of a Hopf bifurcation in the neuronal firing rate, which is thought to underlie the all-or-none principle \cite{hodgkin1952quantitative} (Fig. \ref{function}c). Subsequent extensions of the HH model expand biophysical realism by incorporating multiple ion channel populations \cite{hille2001ion}, the complex geometries of dendrites and axons \cite{plant1976mathematical}, and more realistic stochastic dynamics yielding thermodynamic and hybrid HH models \cite{andersen2009towards, pakdaman2010fluid}. Concurrent with these descriptive improvements, several simplified neuronal models were also developed, including the notable FitzHugh--Nagumo model \cite{fitzhugh1961impulses, nagumo1962active}, facilitating efficient large-scale simulations of groups of neurons.

Simplifications in neuronal modeling, paired with fine-scale measurements of the synaptic wiring in several animals, have spurred large-scale simulations of real neuronal circuits (Fig. \ref{function}b). For example, on the heels of mapping the entire \textit{C. elegans} connectome \cite{white1986structure}, researchers began simulating the 302-neuron network at the cellular level \cite{niebur1993theory}, eventually even including the nematode's entire muscular system and representations of its physical environment \cite{bryden2004simulation}. Despite these and other efforts simulating the \textit{Drosophila} brain \cite{arena2010insect} and the rat's neocortical column \cite{markram2006blue}, it remains unclear how networks of neurons combine to generate the complex range of behaviors observed even in these relatively simple organisms. This contrast between the simplicity of neuronal dynamics and the apparent complexity of large-scale neural behavior hints at the crucial role of emergence. To understand how macroscopic behaviors emerge within groups of neurons, researchers began developing mean-field descriptions of large neuronal populations. Known as neural mass models, these efforts culminated in the foundational Wilson--Cowan (WC) model of population dynamics \cite{wilson1972excitatory}. Whereas previous neural mass models only considered excitatory interactions between neurons, Wilson and Cowan also included inhibitory interactions, thereby enabling the WS model to predict the collective neural oscillations observed in experiments as well as the emergence of other key properties of neural behavior, including the existence of multiple stable states and hysteresis in the neural response to stimuli \cite{wilson1972excitatory}. This progress was further extended to include spatial fluctuations in activity, yielding neural field models that exhibit other behaviors typically observed in the brain, including regions of localized activity \cite{kishimoto1979existence} and traveling waves \cite{pinto2001spatially}.

In much the same way that neuronal circuits have been modeled using observable synaptic wiring in animals, one could imagine simulating a network of neural mass models whose connections are drawn based on non-invasive measures of regional connectivity in humans. By doing so, researchers are now able to simulate whole sections of the human brain (Fig. \ref{function}c), opening the door for comparisons with experimental measurements of regional activity. Precisely this approach has driven a deeper understanding of the structure-function relationship, including the demonstration that the broad spectrum of MEG/EEG recordings of electrical activity can be reproduced by networked models of neural masses \cite{david2003neural} and that the functional connectivity within such recordings depends critically on the coupling strength between neural masses \cite{david2004evaluation}. To facilitate large-scale simulations of the entire human brain, researchers have frequently turned to the Kuramoto model of oscillatory dynamics as a simplified neural mass model \cite{kuramoto1975lecture, kuramoto2012chemical}. These efforts have provided insights about the spontaneous synchronization of neural oscillations \cite{ward2003synchronous}, a phenomenon which is thought to play a critical role in neural communication \cite{fries2005mechanism}, information processing \cite{palmigiano2017flexible}, and motor coordination \cite{schnitzler2005normal}. Moreover, by embedding Kuramoto oscillators into a realistic map of the human connectome, researchers have shown that even this simple model is able to reproduce the patterned fluctuations in activity and long-range correlations observed in fMRI data \cite{cabral2011role}. Detailed biophysical models of neural dynamics, from descriptions of the electrical activity of individual neurons to networked neural mass models simulating the entire brain, continue to inform our understanding of how collective neural behavior and high-level cognitive functions arise from the brain's underlying physical circuitry.

\noindent \textbf{The future of brain network function.} Over the last two centuries, our understanding of the brain's functional organization and information processing capabilities has progressed immensely. Despite this progress, the modern neuroscientist remains fundamentally limited by the experimental and theoretical tools at their disposal \cite{petersson1999statistical1, petersson1999statistical2}. Invasive techniques such as intracranial electrocorticography, and even minimally invasive techniques such as stereotactic electroencephalography (sEEG) \cite{bancaud1973, chauvel1966stereo, todaro2018mapping}, provide immense precision in mapping human brain dynamics, but remain constrained to patients with medically refractory epilepsy. Other noninvasive imaging techniques all suffer from trade-offs between spatial and temporal resolution \cite{menon1999spatial}; methods that directly measure electromagnetic signals (e.g., EEG and MEG) have high temporal resolution but low spatial resolution, while measurements of blood flow and metabolic activity (e.g., via fMRI or PET) have relatively high spatial accuracy but poor resolution in time. Even fMRI -- widely considered the standard for high spatial resolution in humans -- integrates signals over hundreds of thousands of neurons and several seconds \cite{aguirre2014functional}. Consequently, any changes in neural activity that occur over tens of thousands of neurons or even over the span of a second are imperceivable on a standard fMRI scan.

To improve the precision of functional neuroimaging (fMRI in particular), recent efforts have leveraged modern advances in image processing to strengthen the signal and reduce background noise. For example, to minimize the inevitable effects of head movements and fluctuations in blood flow during scanning, fMRI signals are increasingly corrected using techniques similar to image stabilization in video cameras \cite{ciric2017benchmarking}. Additionally, in order to draw general conclusions from neuroimaging results across a group of subjects, impressive strides have been made to correct for inter-subject heterogeneities in brain structure \cite{avants2011reproducible}. Together, advances in image processing have begun to push neuroimaging from a tool exclusively used for academic research to one that can aid in the diagnosis and treatment of psychiatric disorders such as schizophrenia and Alzheimer's disease.

Beyond data collection, data analysis and models in network neuroscience have historically been limited to dyadic relationships between neural elements, such as synapses connecting pairs of neurons or Pearson correlations between pairs of brain regions \cite{bassett2017network, bassett2018nature}. While these dyadic notions of connectivity have provided important insights about the brain's circuitry, mounting evidence suggests that higher-order interactions between three or more elements are also crucial for understanding the large-scale behavior of entire brain networks \cite{amari2003synchronous, ganmor2011sparse, sizemore2017cliques}. In order to study these higher-order connections, recent efforts have focused on generalizing traditional definitions and intuitions from network science, primarily by adopting methods from algebraic topology \cite{giusti2016twos}. One notable approach, known as persistent homology, has allowed researchers to extrapolate conclusions about neural activity across scales, escape the problem of selecting appropriate thresholds for functional edge strengths \cite{giusti2015clique}, and extract principled mesoscale features of network organization \cite{sizemore2017cliques, reimann2017cliques}.

Efforts have also been made to expand traditional metrics of functional connectivity, which are typically based on correlation, to include more sophisticated notions of causality \cite{bettencourt2007functional}. Since causality reflects the flow of information in a network from one element to another, efforts which aim to uncover causal relationships between neurons and brain regions have naturally drawn inspiration from concepts in information theory (see Box 3) \cite{battaglia2012dynamic}. From mutual information to transfer entropy, information theoretic notions of functional connectivity are increasingly being used to quantify the flow of information in the brain \cite{zylberg2017robust, kirst2016dynamic, palmigiano2017flexible}. These measures of causality, in turn, have real-world implications for controlling brain networks and intervening to treat neurological disease and psychiatric disorders.

\begin{center}
[Box 3 here]
\end{center}

\section*{Perturbation experiments and the physics of brain network control}

Thus far, we have examined what is known about the structural circuitry connecting neural components in the brain as well as the dynamical laws governing the interactions between these components. An ultimate test of our understanding, however, lies in our ability to intervene and shift the brain's dynamics to facilitate desirable behaviors. An important implication of the brain's networked structure is that localized perturbations (e.g., targeted lesions or stimulation) do not just yield localized effects -- they also induce indirect effects that propagate along neural pathways \cite{mcintyre2004uncovering, lozano2013probing}. In this way, the task of controlling brain dynamics requires knowledge of how signals transmit along the brain's structural wires, making the problem inherently one of network control \cite{liu2016control}. Building upon targeted lesioning experiments in animals and clinical interventions in humans, efforts toward a theory of network control in the brain have recently taken shape, inspiring several fundamental questions \cite{schiff2012neural}. Are brain networks designed to facilitate control \cite{kim2018role}? What are the principles that allow brain networks to control themselves toward desired activity states \cite{gu2015controllability, jeganathan2018fronto}? Can we leverage these principles to inform stimulation-based therapies for neurological diseases and psychiatric disorders \cite{muldoon2016stimulation, taylor2015optimal, medaglia2018network, holt2016phasic}? To address these questions, here we review the current frontiers in the physics of brain network control.

\begin{figure}
\centering
\includegraphics[width = \textwidth]{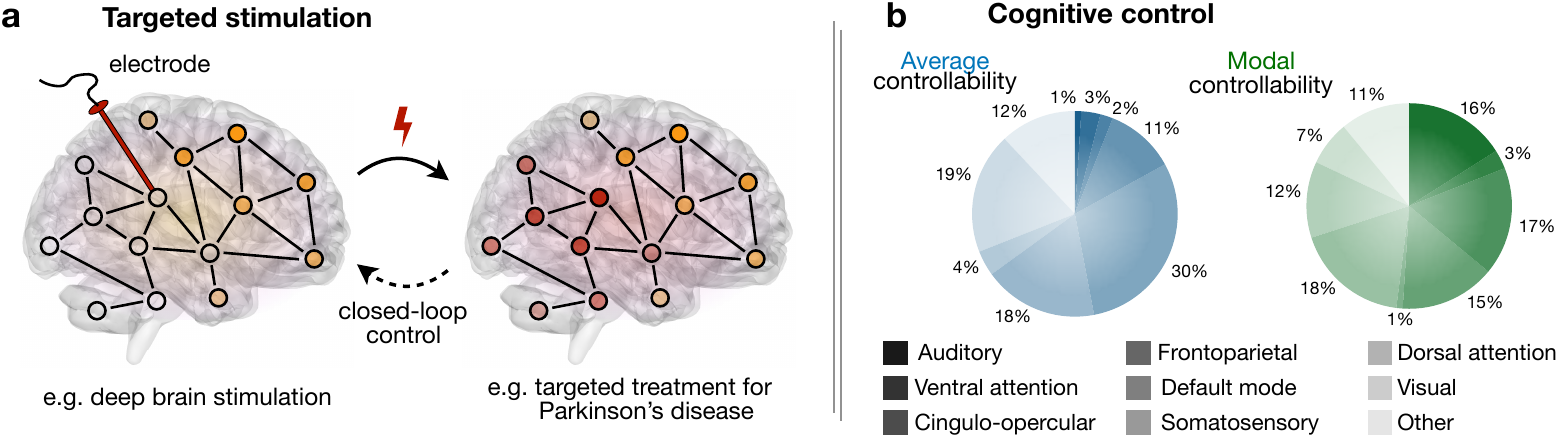} \\
\captionsetup{labelformat=empty}
{\spacing{1.25} \caption{\label{control} \myfont Figure 3 $\vert$ \textbf{Targeted perturbations and brain network control.} \textbf{a} $\vert$ Methods for targeted control are used in the study, design, and optimization of external control processes, such as transcranial magnetic stimulation and deep brain stimulation. These targeted perturbations of neural activity are being utilized in clinical settings to treat major depression, epilepsy, and Parkinson's disease. By simultaneously stimulating and measuring neural activity, researchers can now perform closed-loop control, continuously updating stimulation strategies in real time. \textbf{b} $\vert$ Controllability metrics provide summary statistics regarding the ease with which a given node can enact influence on the network. Two common metrics are the average controllability, which assesses the ease of moving the system to all nearby states, and the modal controllability, which assesses the ability to move the system to distant states (see Box 4). Notions of controllability have proven useful in the study of the brain's internal control processes, such as homeostatic regulation and cognitive control. For example, the human brain displays marked levels of both average and modal controllability, and the proportion of average and modal controllers differs across cognitive systems, suggesting the capacity for a diverse repertoire of dynamics \cite{gu2015controllability}. $\vert$}}
\end{figure}

\noindent \textbf{Targeted perturbations and clinical interventions.} The first attempts to systematically control brain dynamics date to the early 19${}^{\text{th}}$ century, when Marie-Jean-Pierre Flourens noticed that targeted lesions to the brain in living rabbits and pigeons yielded specific changes in the animals' perception, motor coordination, and behavior \cite{flourens1842recherches}. These efforts, in conjunction with other targeted lesioning experiments in animals \cite{panizza1855osservazioni, broca1861remarques}, supported the notion of functional localization -- the theory that specific cognitive functions are supported by specific parts of the brain. In humans, evidence for functional localization has typically relied on patients with localized brain damage (e.g., due to a stroke or head trauma). Historical studies of this kind have revealed, for instance, that damage to one half of the occipital lobe often induces blindness in the opposite field of vision \cite{holmes1918disturbances} and that lesions in the frontal lobe can result in memory loss and an increase in impulsivity and risk taking \cite{owen1990planning}. More recently, advances in non-invasive stimulation techniques such as transcranial magnetic stimulation (TMS) \cite{walsh2000transcranial}, which induces ``transient" lesions by disrupting the brain's normal electrical activity, have opened the door for the control of localized brain functions, including perception \cite{amassian1993measurement}, learning \cite{pascual1994modulation}, language processing \cite{pascual1991induction}, and attention \cite{walsh1998task}. These non-invasive transcranial techniques have been supplemented by more invasive deep brain stimulation (DBS) methods to provide targeted therapies for a number of psychiatric and neurological disorders \cite{walsh2000transcranial, kringelbach2007translational}. By focusing electromagnetic stimulation on the brain regions associated with specific disorders, both TMS and DBS have been used to treat Parkinson's disease, epilepsy, depression, and schizophrenia, among other disorders that are resistant to traditional therapies \cite{george1999transcranial, perlmutter2006deep} (Fig. \ref{control}a). Despite these therapeutic benefits, it remains unclear exactly how and why TMS and DBS are so effective \cite{kringelbach2007translational, mcintyre2004uncovering}; however, recent evidence suggests that the answers may rely on a deeper understanding of the indirect effects of stimulation that are mediated by the brain's physical circuitry \cite{tass1998detection, santaniello2015therapeutic}.

With the recent development of whole-brain neuroimaging methods such as fMRI, evidence continues to mount that brain regions are heavily interdependent on one another, often working in unison to process information and formulate responses \cite{van2010exploring, deco2015rethinking}. In a particularly clear demonstration of the brain's functional integration, Anthony Randall McIntosh and colleagues trained human subjects to associate an auditory stimulus with a visual event. Later, when the auditory stimulus was presented alone, the investigators observed increased activity in the occipital lobe, more traditionally thought of as being reserved for visual processing \cite{zeki1993vision}. Experiments such as these reveal how activity or stimulation in one part of the brain can propagate along neural pathways to induce activity in other distant parts. To understand the system-wide impacts of targeted stimulation, researchers have increasingly drawn upon network models of brain dynamics \cite{tass1998detection, santaniello2015therapeutic}. These efforts have resulted in the identification of neural circuits, rather than isolated regions, that are critical for reducing the symptoms of Parkinson's disease \cite{santaniello2015therapeutic, chiken2014disrupting}. Similar network-based approaches are also being used to suppress epileptic seizures using DBS \cite{berenyi2012closed}, non-invasively treat depression using TMS \cite{kedzior2016cognitive}, and modulate consciousness during surgery using anesthesia \cite{ching2013real}. Moreover, by stimulating and recording neural activity in several brain regions simultaneously, researchers have achieved closed-loop strategies for dynamically updating targeted treatments \cite{holt2014origins, heck2014two} (Fig. \ref{control}a). Meanwhile, clinical applications are increasingly being informed by detailed computational simulations of perturbations to specific brain regions, typically employing networked biophysical models such as those discussed in the previous section \cite{crinion2007spatial, santaniello2011closed}. Together, these real-world and computational studies of targeted stimulation have opened the door for sophisticated strategies that aim to shift neural activity with the ultimate goal of guiding healthy cognitive function.

\noindent \textbf{Network control in the brain.} To inform strategies for targeted stimulation and brain network control, it helps to draw upon existing tools from control theory in mathematics and intuitions from cognitive control in psychology. Given a mathematical model of a system, control theory seeks to understand how the system can be influenced such that it moves toward a desired state \cite{liu2016control, iudice2015structural} (see Box 4). Cognitive control, on the other hand, encompasses a broad class of processes by which the brain enacts control over itself, typically to achieve an abstract goal or desired response \cite{posner2004attention}. For example, dating to the early 1970s neurophysiological studies revealed that the act of holding an object in working memory induces a sustained neural response in the prefrontal cortex \cite{fuster1971neuron, goldman1970localization}. In fact, the prefrontal cortex is now believed to play a key role in many cognitive control processes, from the representation of complex goal-directed behaviors \cite{bechara1994insensitivity} to the support of flexible responses to changes in the environment \cite{dias1996dissociation}. But how do these notions of cognitive control (as defined by psychologists and cognitive neuroscientists) compare to theories of network control (as defined by physicists and engineers)? Furthermore, how can knowledge of the brain's intrinsic control processes inform targeted therapies for mental illness?

\begin{center}
[Box 4 here]
\end{center}

To address these questions, we begin by comparing cognitive notions of intrinsic control with theoretical measures of control and controllability in brain networks (see Box 4). It is interesting, for example, to ask which brain regions are most capable of inducing desired neural responses in other brain regions that are responsible for common functions such as vision, audition, and motor coordination. Toward this end, Gu \emph{et al.} used methods from control theory to demonstrate that the strongest driver nodes corresponded to brain regions with high communicability -- or many topological paths through the brain network -- to the target brain regions \cite{gu2017optimal}. In a related study, Betzel \emph{et al.} used the structural wiring of the brain to simulate transitions between commonly observed activity states \cite{betzel2016optimally}. They found that optimal control nodes tended to have high degree in the network, and that when this rich-club of hub regions was destroyed by simulated lesioning, the ability of the brain to make common transitions was significantly reduced.

In addition to studying the roles of specific control trajectories, complementary approaches have considered trajectory-independent metrics such as the average and modal controllabilities discussed in Box 4 \cite{pasqualetti2014controllability}. By comparing control theoretic measures of node controllability with the cognitive functions associated with each brain region, researchers have observed that different types of controllers are located in distinct areas of the brain (Fig. \ref{control}b) \cite{gu2015controllability}. For example, brain regions with strong average controllability are disproportionately located in the default mode system, which is associated with baseline neural activity; meanwhile, strong modal controllers are primarily located in cognitive control systems. These observations are particularly interesting because they suggest that regions associated with the default mode are optimally positioned to push the system into many easily reachable states, while regions associated with cognitive control are optimally positioned to steer the system toward distant states.

As a final layer of abstraction, rather than studying the controllabilities of specific brain regions, one could envision averaging over all regions to quantify the mean controllability of an entire brain network. Interestingly, by taking precisely this approach, Tang \emph{et al.} established that brain networks as a whole are finely tuned to maximize both average and modal controllability, thereby supporting a diverse range of possible control strategies \cite{tang2017developmental}. Furthermore, by comparing subjects in different stages of adolescence, the researchers found that brain network controllability increases with age, suggesting that neural circuitry evolves over time to support increasingly complex dynamics. In related studies, metrics of network controllability were found to differ by sex \cite{cornblath2018sex} and to be altered in individuals with high genetic risk for bipolar disorder \cite{jeganathan2018fronto}. Taken together, these results demonstrate that network measures of optimal control and controllability correspond closely to existing notions of intrinsic and cognitive control in neuroscience. This close correspondence, in turn, suggests that network control theory, by taking into account the complex wiring of the brain, has the promise to enrich our understanding of the brain's control principles \cite{tang2018control}.

\noindent \textbf{The future of brain network control.} Throughout this section, we have focused primarily on targeted therapies that rely on the coarse-grained stimulation of entire brain regions and simple control strategies that assume idealized linear dynamics. Emerging efforts in neuroscience and control theory, however, are opening the door for a number of significant improvements, including: (i) techniques for fine-scale control of neural activity \cite{adamantidis2007neural, deisseroth2011optogenetics, gunaydin2010ultrafast, grosenick2015closed}, even down to the level of individual neurons \cite{prakash2012two, rickgauer2014simultaneous}, (ii) systems identification approaches that allow for the incorporation of effective connectivity measurements to inform control, superseding solely structural explanations \cite{becker2018large}, and (iii) generalizations of linear control theory that include more realistic nonlinear dynamics \cite{coron2007control, klickstein2017locally}. Among recent advances in the manipulation of fine-scale neural activity, arguably the most promising tool is optogenetics, which offers millisecond-scale optical control of specific cell types within the brains of conscious animals \cite{adamantidis2007neural, deisseroth2011optogenetics}. Its striking precision \cite{gunaydin2010ultrafast}, in some cases even down to single-cell resolution \cite{prakash2012two, rickgauer2014simultaneous}, has enabled researchers to investigate the nature of causal signals between neurons and to study how these signals give rise to qualitative changes in animal behavior \cite{grosenick2015closed}.

While linear control theory continues to provide critical insights about how signals propagate along the brain's structural wiring \cite{gu2017optimal, betzel2016optimally, gu2015controllability, kim2018role}, interactions between neural components, from individual neurons to entire brain regions, are highly nonlinear (Fig. \ref{function}b) \cite{breakspear2017dynamic}. Initial efforts to develop a theory of nonlinear control, dating as early as the 1970s \cite{haynes1970nonlinear, sussmann1972controllability, hermann1977nonlinear}, quickly converged on the conclusion that results as strong and general as those derived for linear dynamics could not be obtained for a general nonlinear system \cite{liu2016control}. Fortunately, concerted theoretic efforts have led to weaker notions of nonlinear controllability \cite{cornelius2013realistic}, notable among which are techniques for linearizing nonlinear systems around stable equilibrium states \cite{coron2007control, klickstein2017locally} and methods for leveraging the symmetries of a system \cite{whalen2015observability} such as repeated network motifs to simplify control strategies \cite{isidori2013nonlinear}. Additional efforts have utilized advances in computing power to simulate the effects of external perturbations across a range of model systems, including networks of FitzHugh--Nagumo neurons \cite{whalen2015observability}, Wilson--Cowan neural masses \cite{muldoon2016stimulation}, and Kuramoto oscillators \cite{chopra2009exponential} as well as artificial neural networks such as the Ising model \cite{lynn2017statistical, lynn2018maximizing}. Together, recent advances in high-precision neural stimulation like optogenetics and our emerging understanding of the principles governing nonlinear control are pushing the boundaries of what is considered possible in the investigation of neural activity. Targeted control of the brain's complex behavior -- once considered a topic of science fiction -- now has the promise to shape targeted therapies for a range of psychiatric and neurological disorders.

\section*{Conclusions and future directions in the neurophysics of brain networks}

The intricate inner workings of the brain remains one of the greatest mysteries defying resolution by contemporary scientific inquiry. On the heels of decades of effort investigating the functions of the brain's individual components \cite{amunts2015architectonic}, from neurons to neuronal ensembles and large-scale brain regions, conclusive evidence points to the need for maps and models of the interactions between these components in order to fundamentally understand the brain's ensemble dynamics, circuit function, and emergent behavior \cite{cohen2011measuring, bassett2018nature}. Here we reviewed recent advances toward meeting this challenge with an eclectic array of curios from the physicist's cabinet: statistical mechanics of complex networks, thermodynamics, information theory, dynamical systems theory, and control theory. In the course of our exposition, we considered the principles of small-worldness \cite{bassett2016small}, interconnected high-degree hubs \cite{heuvel2011rich}, modularity \cite{sporns2016modular}, and spatial embedding \cite{stiso2018spatial} that provide useful explanations for the architecture of structural brain networks. We then saw these same principles reflected in the organization of long-range functional connectivity supporting information dissemination, and the computations that can result therefrom \cite{avena2017communication, palmigiano2017flexible}. As with any physical system, a natural next step is to probe the validity of our descriptive and explanatory models using perturbative approaches both in theory and experiment. Thus, we next summarized the utility of network control theory in offering insights into internal control processes such as homeostatic regulation and cognitive control, as well as external control processes such as neurostimulation, which are currently being used to treat multiple disorders of mental health \cite{tang2018control}.

Throughout the exposition, we described current frontiers in the investigation of brain network structure, function, and control. Although we will not reiterate those points here, we do wish to offer the sentiment that, while the empirical advances laying the foundation of the field have spanned several decades, the network physics of the brain is an incredibly young area, rich with opportunities for discovery. And perhaps -- with a bit of courage -- we may even begin to provide an empirical constitution to the deeper philosophical questions that humans have wrestled with for millennia: What makes us unique and different from non-human animals \cite{heuvel2016comparative,kim2018role}? How do we represent abstract concepts such as value to ourselves \cite{persichetti2015value} and others \cite{dore2018brain}? How are representations transmitted throughout the brain or reconfigured based on new knowledge \cite{constantinescu2016organizing}? What makes a mind from a brain? Physicists, the brain is calling you.

\newpage
\clearpage

\noindent \textbf{Box 1} $\vert$ \textbf{A simple primer on networks.} Here, we define what we mean by a network and describe tools for summarizing its architecture. Importantly, a network is agnostic to the system that it represents \cite{bassett2018nature}, whether it be a brain, a granular material \cite{papadopoulos2018network}, or a quantum lattice \cite{bianconi2015complex}. By far the simplest network model is represented by a binary undirected graph in which identical nodes represent system components and identical edges indicate relations or connections between pairs of nodes (see the figure). Such a network can be encoded in an adjacency matrix $\mathbf{A}$, where each element $A_{ij}$ indicates the strength of connectivity between nodes $i$ and $j$. When all edge strengths are unity, the network is said to be binary. When edges have a range of weights, the network represented by the adjacency matrix is said to be weighted. When $\mathbf{A}=\mathbf{A}^{\intercal}$, the network is undirected; otherwise, the network is directed.

One can extend this simple encoding to study multilayer, multislice, and multiplex networks \cite{kivela2014multilayer}; dynamic or temporal networks \cite{domenico2016physics, holme2012temporal}; annotated networks \cite{newman2016structure}; hypergraphs \cite{bassett2014cross}; and simplicial complexes \cite{giusti2016twos}. One can also calculate various statistics to quantify the architecture of a network and to infer the function thereof (see figure). Intuitively, these statistics range from measures of the local structure in the network, which depend solely on the links directly emanating from a given node (e.g., degree and clustering), to measures of the network's global structure, which depend on the complex pattern of interconnections between all nodes (e.g., path lengths and centrality) \cite{costa2006characterization}. Intermediate statistics exist to study network organization at the mesoscale, such as cavity structure and community structure, the latter of which describes the presence of communities of densely connected nodes \cite{porter2009communities, fortunato2010community, fortunato2016community}. As we will see, the encoding of a system as a network and the quantitative assessment of its architecture can provide important insights into its function \cite{newman2003structure, watts1998collective}.

\begin{figure}[h]
\centering
\includegraphics{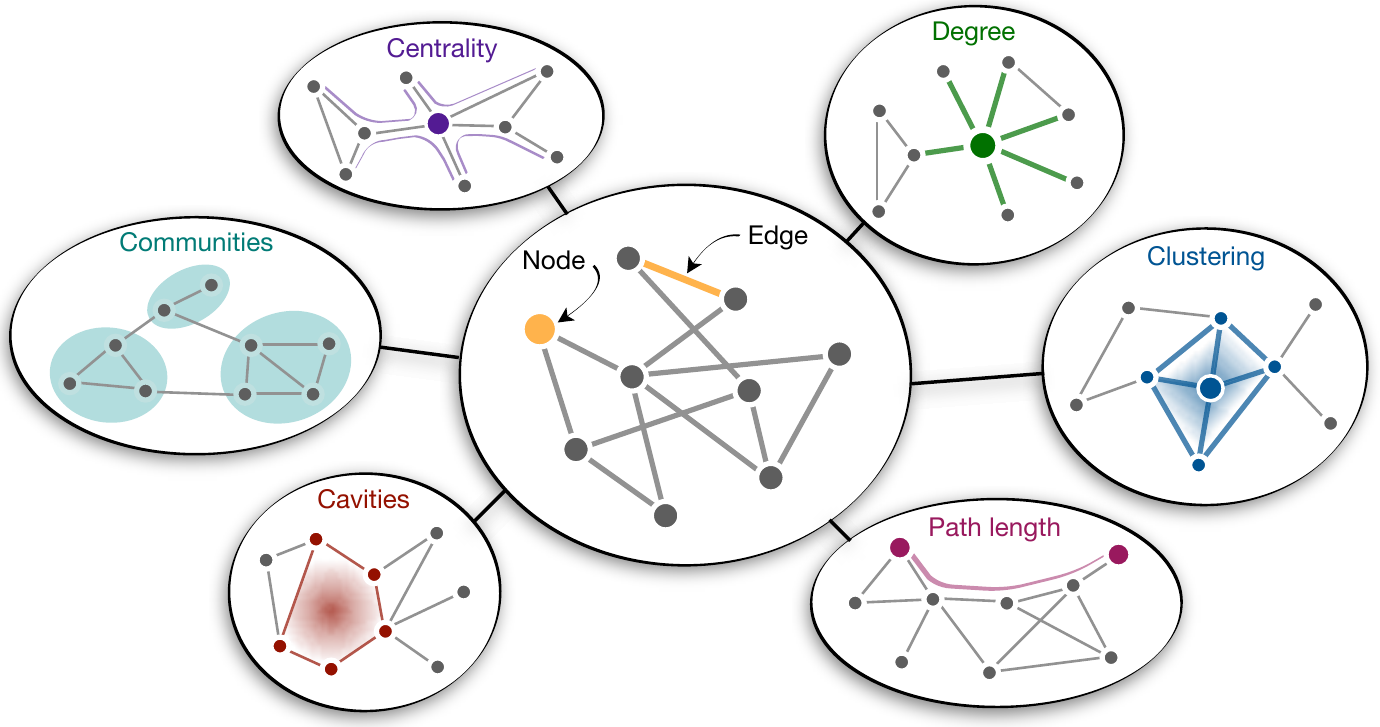} \\
\captionsetup{labelformat=empty}
\end{figure}

\newpage
\clearpage

\noindent \textbf{Box 2} $\vert$ \textbf{Bridging spatiotemporal scales.} In the context of complex systems generally and neural systems specifically, the cutting edge work relates to extending our tools, theories, and intuitions from a single network to so-called multiscale, multilayer, and multiplex networks \cite{kivela2014multilayer, gosak2018network}. Perhaps the most obvious context in which to make this extension is from regional networks to cellular-scale neuronal networks \cite{mejias2016feedforward}. Large-scale brain activity provides a coarse-grained encoding of neural processes, and the map from cellular dynamics to regional dynamics reflects rules of system function. By combining these two layers we can address questions like, ``How do cellular processes shape circuit behavior?'' The next logical extension is to move even further down the natural hierarchy of scales to understand how molecular networks -- including gene coexpression networks \cite{richiardi2015correlated, arnatkeviciute2018hub, romero2018structural, whitaker2016adolescence} -- shape the behavior of cells \cite{hardingham2018lineage}. Understanding how molecular mechanisms affect large-scale brain network function is critical for the development of effective pharmacological interventions \cite{luke2007network, braun2018from, stam2014modern}. By extending the network model from regions to cells to molecular drivers, we can ask questions like, ``How do genetic codes and epigenetic drivers shape circuit behavior across spatial scales?'' And in a final extension, it is time to move up in the natural hierarchy of scales to combine information from the connectivity within a single human brain to the connectivity between human brains in large-scale social networks \cite{schmalzle2017brain, dore2018brain, parkinson2018similar, parkinson2014common}. While brain activity and structure offer biological mechanisms for human behaviors, social networks offer external inducers or modulators of those behaviors \cite{falk2017brain}. By extending the network model to this larger scale, we can start to ask -- and potentially answer -- questions like, ``How do brains shape social networks? And how do social ties shape the brain?'' This extension will be important in understanding human behavior within the broader contexts of culture and society. 

\begin{figure}[h]
\centering
\includegraphics{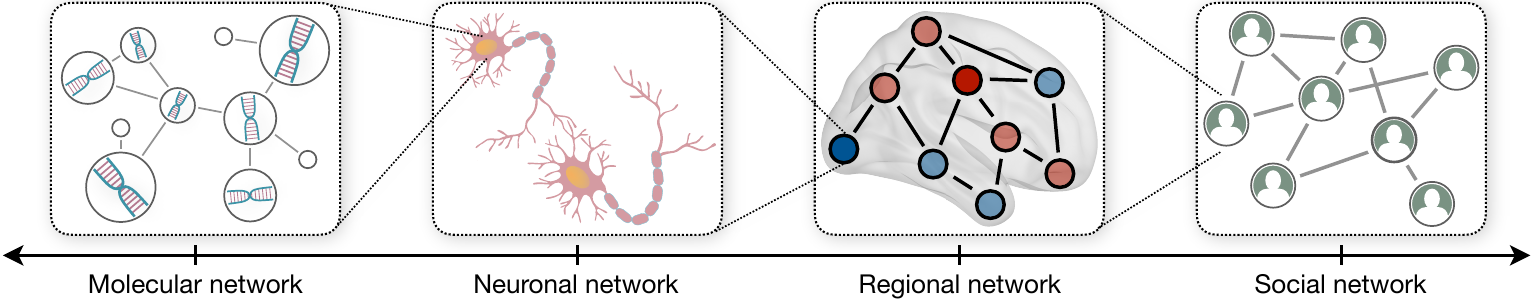} \\
\captionsetup{labelformat=empty}
\end{figure}

\newpage
\clearpage

\noindent \textbf{Box 3} $\vert$ \textbf{Information theory and network neuroscience.} At its core the brain is an information processing system, having evolved over millions of years to encode and manipulate a continuous stream of sensory signals \cite{rieke1999spikes}. As such, information theory -- the science of how signals are encoded and processed -- provides a compelling lens through which to study the brain's function \cite{cover2012elements}. Information theory began with the 1948 paper ``A Mathematical Theory of Communication,'' wherein Claude Shannon proposed the entropy of a signal as the natural measure of its information content and derived fundamental limits on the information capacity of a communication channel \cite{shannon1948mathematical}. Soon after, MacKay and McCulloch adapted the concept of channel capacity to obtain limits on the rate at which one neuron can transmit information to another \cite{mackay1952limiting}, sparking the study of information flow in the brain. Subsequent work by Attneave and Barlow proposed the idea that neural activity is optimized for the transmission of sensory information \cite{attneave1954some, barlow1961possible}, providing the foundation for future investigations of neural coding \cite{rieke1999spikes,abbott2001theoretical}.

Despite these initial efforts bridging information theory and neuroscience, progress slowed primarily due to difficulties obtaining unbiased information estimates from neural systems. Improvements in experimental techniques, however, eventually sparked renewed interest \cite{van1988real}, spurring the introduction of robust methods for estimating information theoretic quantities \cite{strong1998entropy, paninski2003estimation, nemenman2004entropy}. On the basis of these advancements, information theory has once again become a powerful tool for the network neuroscientist. Recent attempts, for instance, to uncover causal relationships between neural elements have successfully adapted notions of information flow, such as mutual information and transfer entropy \cite{schreiber2000measuring, vicente2011transfer}. At the same time, efforts to understand large-scale correlations within neuronal populations have utilized the principle of maximum entropy \cite{jaynes1957information}, resulting in Ising-like models of collective neural behavior \cite{schneidman2006weak, ganmor2011sparse}. As information theory becomes increasingly integrated into the fabric of neuroscience, physicists are uniquely positioned to pioneer exciting new techniques for investigating the nature of information processing in the brain.

\newpage
\clearpage

\noindent \textbf{Box 4} $\vert$ \textbf{Linear control and network controllability.} To investigate the principles of control in the brain, it is useful to understand the theory of network control generally. In network control, the system in question typically comprises a complex web of interacting components, and the goal is to drive this networked system toward a desired state by influencing a select number of input nodes \cite{liu2016control}. The starting point for most control theoretic problems is the linear time-invariant control system $\mathbf{x}(t+1) = \mathbf{A}\mathbf{x}(t) + \mathbf{u}(t)$, where $\mathbf{x}(t)$ defines the state of the system (e.g., the BOLD signal measured by fMRI), $\mathbf{A}$ is the interaction matrix (e.g., white matter tracts estimated using DTI), and $\mathbf{u}(t)$ defines the input signal (e.g., electromagnetic stimulation using TMS or DBS) \cite{kailath1980linear}. Such a system is said to be \textit{controllable} if it can be driven to any desired state. Often, however, many naturally occurring networks that are theoretically controllable cannot be steered to certain states due to limitations on control resources \cite{liu2011controllability, klickstein2017energy}, motivating the introduction of control strategies $\mathbf{u}^*(t)$ that minimize the so-called \textit{control energy} $E(\mathbf{u}) = \sum_{t = 0}^{\infty} |\mathbf{u}(t)|_2^2$.

By limiting the control input to a single node, we can quantify the ability of that node to steer the dynamics of the entire system.  For example, the \textit{average controllability} of a node represents its capacity to drive the network to many nearby states \cite{gu2015controllability}, while a node's \textit{modal controllability} quantifies its ability to push the network toward distant hard-to-reach states \cite{pasqualetti2014controllability} (see figure). Averaging these metrics over all nodes in a system, one can estimate the inherent controllability of an entire network itself. Control theoretic efforts such as these have only recently been applied to understand the locomotion of the nematode \cite{yan2017network} and the networked behavior of the brain more broadly \cite{schiff2012neural, lozano2013probing, tang2018control}, promising new strategies for stimulation-based therapies and fresh insights about the brain's capacity for intrinsic control.

\begin{figure}[h]
\centering
\includegraphics{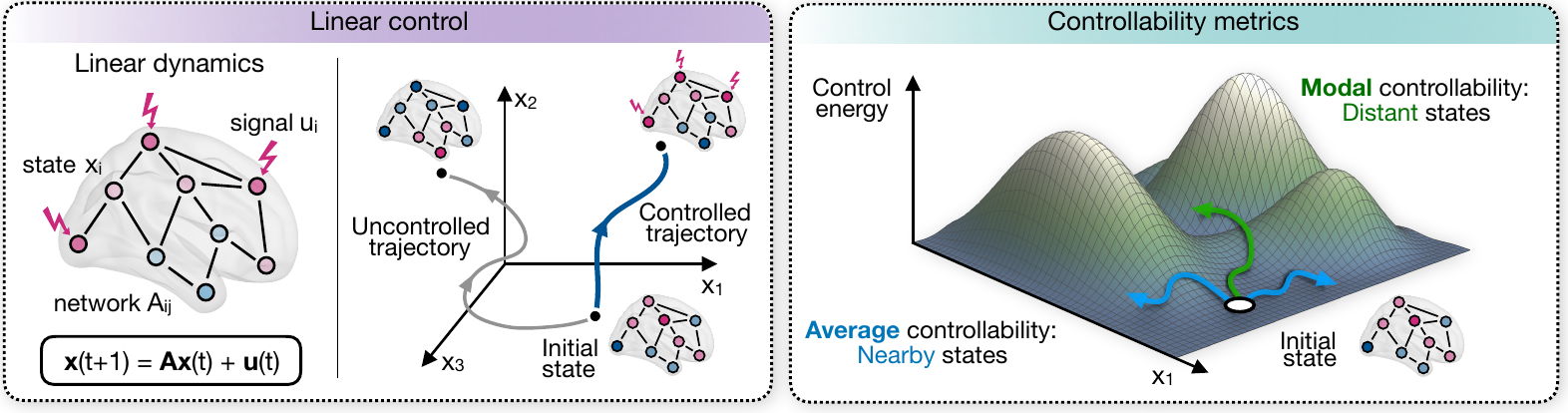} \\
\captionsetup{labelformat=empty}
\end{figure}

\newpage

%% Put the bibliography here, most people will use BiBTeX in
%% which case the environment below should be replaced with
%% the \bibliography{} command.

\section*{References}

\bibliographystyle{naturemag}
\bibliography{bibfile}

\begin{thebibliography}{100}
\expandafter\ifx\csname url\endcsname\relax
  \def\url#1{\texttt{#1}}\fi
\expandafter\ifx\csname urlprefix\endcsname\relax\def\urlprefix{URL }\fi
\providecommand{\bibinfo}[2]{#2}
\providecommand{\eprint}[2][]{\url{#2}}

\bibitem{lear1988desire}
\bibinfo{author}{Lear, J.}
\newblock \emph{\bibinfo{title}{Aristotle: The Desire to Understand}}
  (\bibinfo{publisher}{Cambridge University Press}, \bibinfo{year}{1988}).

\bibitem{metaphysics}
\bibinfo{author}{Aristotle}.
\newblock \emph{\bibinfo{title}{Metaphysics}}, vol. \bibinfo{volume}{VII.7}.

\bibitem{stenger2015physicists}
\bibinfo{author}{Stenger, V.~J.}, \bibinfo{author}{Lindsay, J.~A.} \&
  \bibinfo{author}{Boghossian, P.}
\newblock \bibinfo{title}{Physicists are philosophers, too}.
\newblock \emph{\bibinfo{journal}{Scientific American}}
  (\bibinfo{year}{2015}).

\bibitem{neumann1958computer}
\bibinfo{author}{von Neumann, J.}
\newblock \emph{\bibinfo{title}{The Computer and the Brain}}
  (\bibinfo{year}{1958}).

\bibitem{hopfield1982neural}
\bibinfo{author}{Hopfield, J.~J.}
\newblock \bibinfo{title}{Neural networks and physical systems with emergent
  collective computational abilities}.
\newblock \emph{\bibinfo{journal}{Proc. Nat. Acad. Sci. (USA)}}
  \textbf{\bibinfo{volume}{79}}, \bibinfo{pages}{2554--2558}
  (\bibinfo{year}{1982}).

\bibitem{scott1977neurophysics}
\bibinfo{author}{Scott, A.}
\newblock \emph{\bibinfo{title}{Neurophysics}} (\bibinfo{publisher}{Wiley},
  \bibinfo{year}{1977}).

\bibitem{koch1983theoretical}
\bibinfo{author}{Koch, C.} \& \bibinfo{author}{Poggio, T.}
\newblock \bibinfo{title}{A theoretical analysis of electrical properties of
  spines}.
\newblock \emph{\bibinfo{journal}{Proc R Soc Lond B Biol Sci}}
  \textbf{\bibinfo{volume}{218}}, \bibinfo{pages}{455--477}
  (\bibinfo{year}{1983}).

\bibitem{tyler2012mechanobiology}
\bibinfo{author}{Tyler, W.~J.}
\newblock \bibinfo{title}{The mechanobiology of brain function}.
\newblock \emph{\bibinfo{journal}{Nature Reviews Neuroscience}}
  \textbf{\bibinfo{volume}{13}}, \bibinfo{pages}{867--878}
  (\bibinfo{year}{2012}).

\bibitem{friston2006free}
\bibinfo{author}{Friston, K.}, \bibinfo{author}{Kilner, J.} \&
  \bibinfo{author}{Harrison, L.}
\newblock \bibinfo{title}{A free energy principle for the brain}.
\newblock \emph{\bibinfo{journal}{J Physiol Paris}}
  \textbf{\bibinfo{volume}{100}}, \bibinfo{pages}{70--87}
  (\bibinfo{year}{2006}).

\bibitem{plewes2012physics}
\bibinfo{author}{Plewes, D.~B.} \& \bibinfo{author}{Kucharczyk, W.}
\newblock \bibinfo{title}{Physics of {MRI}: a primer}.
\newblock \emph{\bibinfo{journal}{J Magn Reson Imaging}}
  \textbf{\bibinfo{volume}{35}}, \bibinfo{pages}{1038--1054}
  (\bibinfo{year}{2012}).

\bibitem{hari2012magnetoencephalography}
\bibinfo{author}{Hari, R.} \& \bibinfo{author}{Salmelin, R.}
\newblock \bibinfo{title}{{Magnetoencephalography: From SQUIDs to neuroscience.
  Neuroimage 20th anniversary special edition}}.
\newblock \emph{\bibinfo{journal}{Neuroimage}} \textbf{\bibinfo{volume}{61}},
  \bibinfo{pages}{386--396} (\bibinfo{year}{2012}).

\bibitem{boto2018moving}
\bibinfo{author}{Boto, E.} \emph{et~al.}
\newblock \bibinfo{title}{Moving magnetoencephalography towards real-world
  applications with a wearable system}.
\newblock \emph{\bibinfo{journal}{Nature}} \textbf{\bibinfo{volume}{555}},
  \bibinfo{pages}{657--661} (\bibinfo{year}{2018}).

\bibitem{alivisatos2013nanotools}
\bibinfo{author}{Alivisatos, A.~P.} \emph{et~al.}
\newblock \bibinfo{title}{Nanotools for neuroscience and brain activity
  mapping}.
\newblock \emph{\bibinfo{journal}{ACS Nano}} \textbf{\bibinfo{volume}{7}},
  \bibinfo{pages}{1850--1866} (\bibinfo{year}{2013}).

\bibitem{piazza2018enhanced}
\bibinfo{author}{Piazza, S.}, \bibinfo{author}{Bianchini, P.},
  \bibinfo{author}{Sheppard, C.}, \bibinfo{author}{Diaspro, A.} \&
  \bibinfo{author}{Duocastella, M.}
\newblock \bibinfo{title}{Enhanced volumetric imaging in 2-photon microscopy
  via acoustic lens beam shaping}.
\newblock \emph{\bibinfo{journal}{J Biophotonics}}
  \textbf{\bibinfo{volume}{11}} (\bibinfo{year}{2018}).

\bibitem{boyden2005millisecond}
\bibinfo{author}{Boyden, E.~S.}, \bibinfo{author}{Zhang, F.},
  \bibinfo{author}{Bamberg, E.}, \bibinfo{author}{Nagel, G.} \&
  \bibinfo{author}{Deisseroth, K.}
\newblock \bibinfo{title}{Millisecond-timescale, genetically targeted optical
  control of neural activity}.
\newblock \emph{\bibinfo{journal}{Nat Neurosci}} \textbf{\bibinfo{volume}{8}},
  \bibinfo{pages}{1263--1268} (\bibinfo{year}{2005}).

\bibitem{popkin2016physicists}
\bibinfo{author}{Popkin, G.}
\newblock \bibinfo{title}{Physicists, the brain is calling you}
  (\bibinfo{year}{2016}).

\bibitem{mcculloch1943logical}
\bibinfo{author}{McCulloch, W.~S.} \& \bibinfo{author}{Pitts, W.}
\newblock \bibinfo{title}{A logical calculus of the ideas immanent in nervous
  activity}.
\newblock \emph{\bibinfo{journal}{Bull Math Biol}}
  \textbf{\bibinfo{volume}{5}}, \bibinfo{pages}{115--133}
  (\bibinfo{year}{1943}).

\bibitem{fries2015rhythms}
\bibinfo{author}{Fries, P.}
\newblock \bibinfo{title}{Rhythms for cognition: Communication through
  coherence}.
\newblock \emph{\bibinfo{journal}{Neuron}} \textbf{\bibinfo{volume}{88}},
  \bibinfo{pages}{220--235} (\bibinfo{year}{2015}).

\bibitem{betzel2018specificity}
\bibinfo{author}{Betzel, R.~F.} \& \bibinfo{author}{Bassett, D.~S.}
\newblock \bibinfo{title}{Specificity and robustness of long-distance
  connections in weighted, interareal connectomes}.
\newblock \emph{\bibinfo{journal}{Proc Natl Acad Sci U S A}}
  \textbf{\bibinfo{volume}{115}}, \bibinfo{pages}{E4880--E4889}
  (\bibinfo{year}{2018}).

\bibitem{poincare1905science}
\bibinfo{author}{Poincare, H.}
\newblock \emph{\bibinfo{title}{Science and Hypothesis}}
  (\bibinfo{publisher}{London: Walter Scott Publishing}, \bibinfo{year}{1905}).

\bibitem{essen2013wuminn}
\bibinfo{author}{Van~Essen, D.~C.} \emph{et~al.}
\newblock \bibinfo{title}{{The WU-Minn Human Connectome Project: an overview}}.
\newblock \emph{\bibinfo{journal}{Neuroimage}} \textbf{\bibinfo{volume}{80}},
  \bibinfo{pages}{62--79} (\bibinfo{year}{2013}).

\bibitem{markram2015reconstruction}
\bibinfo{author}{Markram, H.} \emph{et~al.}
\newblock \bibinfo{title}{Reconstruction and simulation of neocortical
  microcircuitry}.
\newblock \emph{\bibinfo{journal}{Cell}} \textbf{\bibinfo{volume}{163}},
  \bibinfo{pages}{456--492} (\bibinfo{year}{2015}).

\bibitem{poo2016china}
\bibinfo{author}{Poo, M.~M.} \emph{et~al.}
\newblock \bibinfo{title}{{China Brain Project: Basic Neuroscience, Brain
  Diseases, and Brain-Inspired Computing}}.
\newblock \emph{\bibinfo{journal}{Neuron}} \textbf{\bibinfo{volume}{92}},
  \bibinfo{pages}{591--596} (\bibinfo{year}{2016}).

\bibitem{okano2015brain}
\bibinfo{author}{Okano, H.}, \bibinfo{author}{Miyawaki, A.} \&
  \bibinfo{author}{Kasai, K.}
\newblock \bibinfo{title}{{Brain/MINDS: brain-mapping project in Japan}}.
\newblock \emph{\bibinfo{journal}{Philos Trans R Soc Lond B Biol Sci}}
  \textbf{\bibinfo{volume}{370}}, \bibinfo{pages}{20140310}
  (\bibinfo{year}{2015}).

\bibitem{bassett2011understanding}
\bibinfo{author}{Bassett, D.~S.} \& \bibinfo{author}{Gazzaniga, M.~S.}
\newblock \bibinfo{title}{Understanding complexity in the human brain}.
\newblock \emph{\bibinfo{journal}{Trends Cogn Sci}}
  \textbf{\bibinfo{volume}{15}}, \bibinfo{pages}{200--209}
  (\bibinfo{year}{2011}).

\bibitem{sethna2006statistical}
\bibinfo{author}{Sethna, J.~P.}
\newblock \emph{\bibinfo{title}{Statistical Mechanics: Entropy, Order
  Parameters and Complexity}} (\bibinfo{publisher}{Oxford University Press},
  \bibinfo{year}{2006}).

\bibitem{bassett2016small}
\bibinfo{author}{Bassett, D.~S.} \& \bibinfo{author}{Bullmore, E.~T.}
\newblock \bibinfo{title}{Small-world brain networks revisited}.
\newblock \emph{\bibinfo{journal}{Neuroscientist}} \textbf{\bibinfo{volume}{Sep
  21}}, \bibinfo{pages}{1073858416667720} (\bibinfo{year}{2016}).

\bibitem{albert2002statistical}
\bibinfo{author}{Albert, E.} \& \bibinfo{author}{Barabasi, A.-L.}
\newblock \bibinfo{title}{Statistical mechanics of complex networks}.
\newblock \emph{\bibinfo{journal}{Rev. Mod. Phys.}}
  \textbf{\bibinfo{volume}{74}} (\bibinfo{year}{2002}).

\bibitem{butts2009revisiting}
\bibinfo{author}{Butts, C.~T.}
\newblock \bibinfo{title}{Revisiting the foundations of network analysis}.
\newblock \emph{\bibinfo{journal}{Science}} \textbf{\bibinfo{volume}{325}},
  \bibinfo{pages}{414--416} (\bibinfo{year}{2009}).

\bibitem{costa2006characterization}
\bibinfo{author}{Costa, L. d.~F.}, \bibinfo{author}{Rodrigues, F.~A.},
  \bibinfo{author}{Travieso, G.} \& \bibinfo{author}{Villas~Boas, P.~R.}
\newblock \bibinfo{title}{Characterization of complex networks: {A} survey of
  measurements}.
\newblock \emph{\bibinfo{journal}{Advances In Physics}}
  \textbf{\bibinfo{volume}{56}}, \bibinfo{pages}{167--242}
  (\bibinfo{year}{2006}).

\bibitem{gross2008adaptive}
\bibinfo{author}{Gross, T.} \& \bibinfo{author}{Blasius, B.}
\newblock \bibinfo{title}{Adaptive coevolutionary networks: a review}.
\newblock \emph{\bibinfo{journal}{J R Soc Interface}}
  \textbf{\bibinfo{volume}{5}}, \bibinfo{pages}{259--271}
  (\bibinfo{year}{2008}).

\bibitem{zhang2017random}
\bibinfo{author}{Zhang, X.}, \bibinfo{author}{Moore, C.} \&
  \bibinfo{author}{Newman, M. E.~J.}
\newblock \bibinfo{title}{Random graph models for dynamic networks}.
\newblock \emph{\bibinfo{journal}{Eur. Phys. J. B}}
  \textbf{\bibinfo{volume}{90}}, \bibinfo{pages}{200} (\bibinfo{year}{2017}).

\bibitem{hackett2011cascades}
\bibinfo{author}{Hackett, A.}, \bibinfo{author}{Melnik, s.} \&
  \bibinfo{author}{Gleeson, J.~P.}
\newblock \bibinfo{title}{Cascades on a class of clustered random networks}.
\newblock \emph{\bibinfo{journal}{Phys. Rev. E}} \textbf{\bibinfo{volume}{83}},
  \bibinfo{pages}{056107} (\bibinfo{year}{2011}).

\bibitem{newman2003structure}
\bibinfo{title}{The structure and function of complex networks}.
\newblock \emph{\bibinfo{journal}{SIAM REVIEW}} \textbf{\bibinfo{volume}{45}},
  \bibinfo{pages}{167--–256} (\bibinfo{year}{2003}).

\bibitem{motter2015networkcontrology}
\bibinfo{author}{Motter, A.~E.}
\newblock \bibinfo{title}{Networkcontrology}.
\newblock \emph{\bibinfo{journal}{Chaos}} \textbf{\bibinfo{volume}{25}},
  \bibinfo{pages}{097621} (\bibinfo{year}{2015}).

\bibitem{bassett2018nature}
\bibinfo{author}{Bassett, D.~S.}, \bibinfo{author}{Zurn, P.} \&
  \bibinfo{author}{Gold, J.~I.}
\newblock \bibinfo{title}{On the nature and use of models in network
  neuroscience}.
\newblock \emph{\bibinfo{journal}{Nat Rev Neurosci}}
  \textbf{\bibinfo{volume}{Epub ahead of print}} (\bibinfo{year}{2018}).

\bibitem{pereda2014electrical}
\bibinfo{author}{Pereda, A.~E.}
\newblock \bibinfo{title}{Electrical synapses and their functional interactions
  with chemical synapses}.
\newblock \emph{\bibinfo{journal}{Nat Rev Neurosci}}
  \textbf{\bibinfo{volume}{15}}, \bibinfo{pages}{250--263}
  (\bibinfo{year}{2014}).

\bibitem{avena2017communication}
\bibinfo{author}{Avena-Koenigsberger, A.}, \bibinfo{author}{Misic, B.} \&
  \bibinfo{author}{Sporns, O.}
\newblock \bibinfo{title}{Communication dynamics in complex brain networks}.
\newblock \emph{\bibinfo{journal}{Nat Rev Neurosci}}
  \textbf{\bibinfo{volume}{19}}, \bibinfo{pages}{17--33}
  (\bibinfo{year}{2017}).

\bibitem{ising1925beitrag}
\bibinfo{author}{Ising, E.}
\newblock \bibinfo{title}{Beitrag zur theorie des ferromagnetismus}.
\newblock \emph{\bibinfo{journal}{Zeitschrift f{\"u}r Physik}}
  \textbf{\bibinfo{volume}{31}}, \bibinfo{pages}{253--258}
  (\bibinfo{year}{1925}).

\bibitem{onsager1944crystal}
\bibinfo{author}{Onsager, L.}
\newblock \bibinfo{title}{Crystal statistics. i. a two-dimensional model with
  an order-disorder transition}.
\newblock \emph{\bibinfo{journal}{Phys. Rev.}} \textbf{\bibinfo{volume}{65}},
  \bibinfo{pages}{117} (\bibinfo{year}{1944}).

\bibitem{brush1967history}
\bibinfo{author}{Brush, S.~G.}
\newblock \bibinfo{title}{History of the lenz-ising model}.
\newblock \emph{\bibinfo{journal}{Rev. Mod. Phys.}}
  \textbf{\bibinfo{volume}{39}}, \bibinfo{pages}{883} (\bibinfo{year}{1967}).

\bibitem{sporns2004organization}
\bibinfo{author}{Sporns, O.}, \bibinfo{author}{Chialvo, D.~R.},
  \bibinfo{author}{Kaiser, M.} \& \bibinfo{author}{Hilgetag, C.~C.}
\newblock \bibinfo{title}{Organization, development and function of complex
  brain networks}.
\newblock \emph{\bibinfo{journal}{Trends Cogn Sci}}
  \textbf{\bibinfo{volume}{8}}, \bibinfo{pages}{418--425}
  (\bibinfo{year}{2004}).

\bibitem{medaglia2015cognitive}
\bibinfo{author}{Medaglia, J.~D.}, \bibinfo{author}{Lynall, M.~E.} \&
  \bibinfo{author}{Bassett, D.~S.}
\newblock \bibinfo{title}{Cognitive network neuroscience}.
\newblock \emph{\bibinfo{journal}{J Cogn Neurosci}}
  \textbf{\bibinfo{volume}{27}}, \bibinfo{pages}{1471--1491}
  (\bibinfo{year}{2015}).

\bibitem{sporns2014contributions}
\bibinfo{author}{Sporns, O.}
\newblock \bibinfo{title}{Contributions and challenges for network models in
  cognitive neuroscience}.
\newblock \emph{\bibinfo{journal}{Nat Neurosci}} \textbf{\bibinfo{volume}{17}},
  \bibinfo{pages}{652--660} (\bibinfo{year}{2014}).

\bibitem{petersen2015brain}
\bibinfo{author}{Petersen, S.~E.} \& \bibinfo{author}{Sporns, O.}
\newblock \bibinfo{title}{Brain networks and cognitive architectures}.
\newblock \emph{\bibinfo{journal}{Neuron}} \textbf{\bibinfo{volume}{88}},
  \bibinfo{pages}{207--219} (\bibinfo{year}{2015}).

\bibitem{misic2016from}
\bibinfo{author}{Misic, B.} \& \bibinfo{author}{Sporns, O.}
\newblock \bibinfo{title}{From regions to connections and networks: new bridges
  between brain and behavior}.
\newblock \emph{\bibinfo{journal}{Curr Opin Neurobiol}}
  \textbf{\bibinfo{volume}{40}}, \bibinfo{pages}{1--7} (\bibinfo{year}{2016}).

\bibitem{wallace2013randomly}
\bibinfo{author}{Wallace, E.}, \bibinfo{author}{Maei, H.~R.} \&
  \bibinfo{author}{Latham, P.~E.}
\newblock \bibinfo{title}{Randomly connected networks have short temporal
  memory}.
\newblock \emph{\bibinfo{journal}{Neural Comput}}
  \textbf{\bibinfo{volume}{25}}, \bibinfo{pages}{1408--1439}
  (\bibinfo{year}{2013}).

\bibitem{rajan2016recurrent}
\bibinfo{author}{Rajan, K.}, \bibinfo{author}{Harvey, C.~D.} \&
  \bibinfo{author}{Tank, D.~W.}
\newblock \bibinfo{title}{Recurrent network models of sequence generation and
  memory}.
\newblock \emph{\bibinfo{journal}{Neuron}} \textbf{\bibinfo{volume}{90}},
  \bibinfo{pages}{128--142} (\bibinfo{year}{2016}).

\bibitem{chaudhuri2016computational}
\bibinfo{author}{Chaudhuri, R.} \& \bibinfo{author}{Fiete, I.}
\newblock \bibinfo{title}{Computational principles of memory}.
\newblock \emph{\bibinfo{journal}{Nat Neurosci}} \textbf{\bibinfo{volume}{19}},
  \bibinfo{pages}{394--403} (\bibinfo{year}{2016}).

\bibitem{hermundstad2011learning}
\bibinfo{author}{Hermundstad, A.~M.}, \bibinfo{author}{Brown, K.~S.},
  \bibinfo{author}{Bassett, D.~S.} \& \bibinfo{author}{Carlson, J.~M.}
\newblock \bibinfo{title}{Learning, memory, and the role of neural network
  architecture}.
\newblock \emph{\bibinfo{journal}{PLoS Comput Biol}}
  \textbf{\bibinfo{volume}{7}}, \bibinfo{pages}{e1002063}
  (\bibinfo{year}{2011}).

\bibitem{tesileanu2017rules}
\bibinfo{author}{Teşileanu, T.}, \bibinfo{author}{Olveczky, B.} \&
  \bibinfo{author}{Balasubramanian, V.}
\newblock \bibinfo{title}{Rules and mechanisms for efficient two-stage learning
  in neural circuits}.
\newblock \emph{\bibinfo{journal}{Elife}} \textbf{\bibinfo{volume}{6}},
  \bibinfo{pages}{e20944} (\bibinfo{year}{2017}).

\bibitem{takemura2013visual}
\bibinfo{author}{Takemura, S.~Y.} \emph{et~al.}
\newblock \bibinfo{title}{A visual motion detection circuit suggested by
  drosophila connectomics}.
\newblock \emph{\bibinfo{journal}{Nature}} \textbf{\bibinfo{volume}{500}},
  \bibinfo{pages}{175--181} (\bibinfo{year}{2013}).

\bibitem{zhen2015locomotion}
\bibinfo{author}{Zhen, M.} \& \bibinfo{author}{Samuel, A.~D.}
\newblock \bibinfo{title}{{C}. elegans locomotion: small circuits, complex
  functions}.
\newblock \emph{\bibinfo{journal}{Curr Opin Neurobiol}}
  \textbf{\bibinfo{volume}{33}}, \bibinfo{pages}{117--126}
  (\bibinfo{year}{2015}).

\bibitem{golgi1885sulla}
\bibinfo{author}{Golgi, C.}
\newblock \emph{\bibinfo{title}{Sulla fina anatomia degli organi centrali del
  sistema nervoso}} (\bibinfo{publisher}{S. Calderini}, \bibinfo{year}{1885}).

\bibitem{y1888estructura}
\bibinfo{author}{y~Cajal, S.~R.}
\newblock \emph{\bibinfo{title}{Estructura de los centros nerviosos de las
  aves}} (\bibinfo{year}{1888}).

\bibitem{shepherd2015foundations}
\bibinfo{author}{Shepherd, G.~M.}
\newblock \emph{\bibinfo{title}{Foundations of the neuron doctrine}}
  (\bibinfo{publisher}{Oxford University Press}, \bibinfo{year}{2015}).

\bibitem{white1986structure}
\bibinfo{author}{White, J.~G.}, \bibinfo{author}{Southgate, E.},
  \bibinfo{author}{Thomson, J.~N.} \& \bibinfo{author}{Brenner, S.}
\newblock \bibinfo{title}{The structure of the nervous system of the nematode
  {C}aenorhabditis elegans}.
\newblock \emph{\bibinfo{journal}{Phil. Trans. R. Soc. Lond. B}}
  \textbf{\bibinfo{volume}{314}}, \bibinfo{pages}{1--340}
  (\bibinfo{year}{1986}).

\bibitem{helmstaedter2013connectomic}
\bibinfo{author}{Helmstaedter, M.} \emph{et~al.}
\newblock \bibinfo{title}{Connectomic reconstruction of the inner plexiform
  layer in the mouse retina}.
\newblock \emph{\bibinfo{journal}{Nature}} \textbf{\bibinfo{volume}{500}},
  \bibinfo{pages}{168--174} (\bibinfo{year}{2013}).

\bibitem{sporns2005human}
\bibinfo{author}{Sporns, O.}, \bibinfo{author}{Tononi, G.} \&
  \bibinfo{author}{K{\"o}tter, R.}
\newblock \bibinfo{title}{The human connectome: a structural description of the
  human brain}.
\newblock \emph{\bibinfo{journal}{PLoS Comput Biol}}
  \textbf{\bibinfo{volume}{1}}, \bibinfo{pages}{e42} (\bibinfo{year}{2005}).

\bibitem{stephan2001advanced}
\bibinfo{author}{Stephan, K.~E.} \emph{et~al.}
\newblock \bibinfo{title}{Advanced database methodology for the collation of
  connectivity data on the {M}acaque brain ({CoCoMac})}.
\newblock \emph{\bibinfo{journal}{Philos Trans R Soc Lond B Biol Sci}}
  \textbf{\bibinfo{volume}{356}}, \bibinfo{pages}{1159--1186}
  (\bibinfo{year}{2001}).

\bibitem{markov2014weighted}
\bibinfo{author}{Markov, N.~T.} \emph{et~al.}
\newblock \bibinfo{title}{A weighted and directed interareal connectivity
  matrix for macaque cerebral cortex}.
\newblock \emph{\bibinfo{journal}{Cereb Cortex}} \textbf{\bibinfo{volume}{24}},
  \bibinfo{pages}{17--36} (\bibinfo{year}{2014}).

\bibitem{young1994analysis}
\bibinfo{author}{Young, M.~P.}, \bibinfo{author}{Scannell, J.~W.},
  \bibinfo{author}{Burns, G.~A.} \& \bibinfo{author}{Blakemore, C.}
\newblock \bibinfo{title}{Analysis of connectivity: neural systems in the
  cerebral cortex}.
\newblock \emph{\bibinfo{journal}{Rev Neurosci}} \textbf{\bibinfo{volume}{5}},
  \bibinfo{pages}{227--250} (\bibinfo{year}{1994}).

\bibitem{oh2014mesoscale}
\bibinfo{author}{Oh, S.~W.} \emph{et~al.}
\newblock \bibinfo{title}{A mesoscale connectome of the mouse brain}.
\newblock \emph{\bibinfo{journal}{Nature}} \textbf{\bibinfo{volume}{508}},
  \bibinfo{pages}{207--214} (\bibinfo{year}{2014}).

\bibitem{shih2015connectomics}
\bibinfo{author}{Shih, C.~T.} \emph{et~al.}
\newblock \bibinfo{title}{Connectomics-based analysis of information flow in
  the {D}rosophila brain}.
\newblock \emph{\bibinfo{journal}{Curr Biol}} \textbf{\bibinfo{volume}{25}},
  \bibinfo{pages}{1249--1258} (\bibinfo{year}{2015}).

\bibitem{hsieh2009computed}
\bibinfo{author}{Hsieh, J.} \emph{et~al.}
\newblock \bibinfo{title}{Computed tomography: principles, design, artifacts,
  and recent advances} (\bibinfo{organization}{SPIE Bellingham, WA},
  \bibinfo{year}{2009}).

\bibitem{peirpaoli1996diffusion}
\bibinfo{author}{Pierpaoli, C.}, \bibinfo{author}{Jezzard, P.},
  \bibinfo{author}{Basser, P.~J.}, \bibinfo{author}{Barnett, A.} \&
  \bibinfo{author}{Di~Chiro, G.}
\newblock \bibinfo{title}{Diffusion tensor {MR} imaging of the human brain}.
\newblock \emph{\bibinfo{journal}{Radiology}} \textbf{\bibinfo{volume}{201}},
  \bibinfo{pages}{637--648} (\bibinfo{year}{1996}).

\bibitem{basser2000invivo}
\bibinfo{author}{Basser, P.~J.}, \bibinfo{author}{Pajevic, S.},
  \bibinfo{author}{Pierpaoli, C.}, \bibinfo{author}{Duda, J.} \&
  \bibinfo{author}{Aldroubi, A.}
\newblock \bibinfo{title}{In vivo fiber tractography using {DT-MRI} data}.
\newblock \emph{\bibinfo{journal}{Magn Reson Med}}
  \textbf{\bibinfo{volume}{44}}, \bibinfo{pages}{625--632}
  (\bibinfo{year}{2000}).

\bibitem{behrens2005relating}
\bibinfo{author}{Behrens, T.~E.} \& \bibinfo{author}{Johansen-Berg, H.}
\newblock \bibinfo{title}{Relating connectional architecture to grey matter
  function using diffusion imaging}.
\newblock \emph{\bibinfo{journal}{Philos Trans R Soc Lond B Biol Sci}}
  \textbf{\bibinfo{volume}{360}}, \bibinfo{pages}{903--911}
  (\bibinfo{year}{2005}).

\bibitem{bullmore2012economy}
\bibinfo{author}{Bullmore, E.} \& \bibinfo{author}{Sporns, O.}
\newblock \bibinfo{title}{The economy of brain network organization}.
\newblock \emph{\bibinfo{journal}{Nat. Rev. Neurosci.}}
  \textbf{\bibinfo{volume}{13}}, \bibinfo{pages}{336--349}
  (\bibinfo{year}{2012}).

\bibitem{betzel2017generative}
\bibinfo{author}{Betzel, R.~F.} \& \bibinfo{author}{Bassett, D.~S.}
\newblock \bibinfo{title}{Generative models for network neuroscience: prospects
  and promise}.
\newblock \emph{\bibinfo{journal}{J R Soc Interface}}
  \textbf{\bibinfo{volume}{14}}, \bibinfo{pages}{20170623}
  (\bibinfo{year}{2017}).

\bibitem{bassett2017network}
\bibinfo{author}{Bassett, D.~S.} \& \bibinfo{author}{Sporns, O.}
\newblock \bibinfo{title}{Network neuroscience}.
\newblock \emph{\bibinfo{journal}{Nat Neurosci}} \textbf{\bibinfo{volume}{20}},
  \bibinfo{pages}{353--364} (\bibinfo{year}{2017}).

\bibitem{nicosia2013phase}
\bibinfo{author}{Nicosia, V.}, \bibinfo{author}{V{\'e}rtes, P.~E.},
  \bibinfo{author}{Schafer, W.~R.}, \bibinfo{author}{Latora, V.} \&
  \bibinfo{author}{Bullmore, E.~T.}
\newblock \bibinfo{title}{Phase transition in the economically modeled growth
  of a cellular nervous system}.
\newblock \emph{\bibinfo{journal}{Proceedings of the National Academy of
  Sciences USA}} \textbf{\bibinfo{volume}{110}}, \bibinfo{pages}{7880--7885}
  (\bibinfo{year}{2013}).

\bibitem{henriksen2016simple}
\bibinfo{author}{Henriksen, S.}, \bibinfo{author}{Pang, R.} \&
  \bibinfo{author}{Wronkiewicz, M.}
\newblock \bibinfo{title}{A simple generative model of the mouse mesoscale
  connectome}.
\newblock \emph{\bibinfo{journal}{Elife}} \textbf{\bibinfo{volume}{5}},
  \bibinfo{pages}{e12366} (\bibinfo{year}{2016}).

\bibitem{beul2015predictive}
\bibinfo{author}{Beul, S.~F.}, \bibinfo{author}{Grant, S.} \&
  \bibinfo{author}{Hilgetag, C.~C.}
\newblock \bibinfo{title}{A predictive model of the cat cortical connectome
  based on cytoarchitecture and distance}.
\newblock \emph{\bibinfo{journal}{Brain Struct Funct}}
  \textbf{\bibinfo{volume}{220}}, \bibinfo{pages}{3167--3184}
  (\bibinfo{year}{2015}).

\bibitem{ercsey2013predictive}
\bibinfo{author}{Ercsey-Ravasz, M.} \emph{et~al.}
\newblock \bibinfo{title}{A predictive network model of cerebral cortical
  connectivity based on a distance rule}.
\newblock \emph{\bibinfo{journal}{Neuron}} \textbf{\bibinfo{volume}{80}},
  \bibinfo{pages}{184--197} (\bibinfo{year}{2013}).

\bibitem{beul2017predictive}
\bibinfo{author}{Beul, S.~F.}, \bibinfo{author}{Barbas, H.} \&
  \bibinfo{author}{Hilgetag, C.~C.}
\newblock \bibinfo{title}{A predictive structural model of the primate
  connectome}.
\newblock \emph{\bibinfo{journal}{Sci Rep}} \textbf{\bibinfo{volume}{7}},
  \bibinfo{pages}{43176} (\bibinfo{year}{2017}).

\bibitem{betzel2016generative}
\bibinfo{author}{Betzel, R.~F.} \emph{et~al.}
\newblock \bibinfo{title}{Generative models of the human connectome}.
\newblock \emph{\bibinfo{journal}{Neuroimage}} \textbf{\bibinfo{volume}{124}},
  \bibinfo{pages}{1054--1064} (\bibinfo{year}{2016}).

\bibitem{thompson2001genetic}
\bibinfo{author}{Thompson, P.~M.} \emph{et~al.}
\newblock \bibinfo{title}{Genetic influences on brain structure}.
\newblock \emph{\bibinfo{journal}{Nat. Neurosci.}}
  \textbf{\bibinfo{volume}{4}}, \bibinfo{pages}{1253} (\bibinfo{year}{2001}).

\bibitem{raz2005regional}
\bibinfo{author}{Raz, N.} \emph{et~al.}
\newblock \bibinfo{title}{Regional brain changes in aging healthy adults:
  general trends, individual differences and modifiers}.
\newblock \emph{\bibinfo{journal}{Cereb. Cortex}}
  \textbf{\bibinfo{volume}{15}}, \bibinfo{pages}{1676--1689}
  (\bibinfo{year}{2005}).

\bibitem{gong2009age}
\bibinfo{author}{Gong, G.} \emph{et~al.}
\newblock \bibinfo{title}{Age-and gender-related differences in the cortical
  anatomical network}.
\newblock \emph{\bibinfo{journal}{J. Neurosci.}} \textbf{\bibinfo{volume}{29}},
  \bibinfo{pages}{15684--15693} (\bibinfo{year}{2009}).

\bibitem{kanai2011structural}
\bibinfo{author}{Kanai, R.} \& \bibinfo{author}{Rees, G.}
\newblock \bibinfo{title}{The structural basis of inter-individual differences
  in human behaviour and cognition}.
\newblock \emph{\bibinfo{journal}{Nat. Rev. Neurosci.}}
  \textbf{\bibinfo{volume}{12}}, \bibinfo{pages}{231} (\bibinfo{year}{2011}).

\bibitem{banissy2012inter}
\bibinfo{author}{Banissy, M.~J.}, \bibinfo{author}{Kanai, R.},
  \bibinfo{author}{Walsh, V.} \& \bibinfo{author}{Rees, G.}
\newblock \bibinfo{title}{Inter-individual differences in empathy are reflected
  in human brain structure}.
\newblock \emph{\bibinfo{journal}{Neuroimage}} \textbf{\bibinfo{volume}{62}},
  \bibinfo{pages}{2034--2039} (\bibinfo{year}{2012}).

\bibitem{fleming2010relating}
\bibinfo{author}{Fleming, S.~M.}, \bibinfo{author}{Weil, R.~S.},
  \bibinfo{author}{Nagy, Z.}, \bibinfo{author}{Dolan, R.~J.} \&
  \bibinfo{author}{Rees, G.}
\newblock \bibinfo{title}{Relating introspective accuracy to individual
  differences in brain structure}.
\newblock \emph{\bibinfo{journal}{Science}} \textbf{\bibinfo{volume}{329}},
  \bibinfo{pages}{1541--1543} (\bibinfo{year}{2010}).

\bibitem{hartley2011brain}
\bibinfo{author}{Hartley, C.~A.}, \bibinfo{author}{Fischl, B.} \&
  \bibinfo{author}{Phelps, E.~A.}
\newblock \bibinfo{title}{Brain structure correlates of individual differences
  in the acquisition and inhibition of conditioned fear}.
\newblock \emph{\bibinfo{journal}{Cereb. Cortex}}
  \textbf{\bibinfo{volume}{21}}, \bibinfo{pages}{1954--1962}
  (\bibinfo{year}{2011}).

\bibitem{kanai2011political}
\bibinfo{author}{Kanai, R.}, \bibinfo{author}{Feilden, T.},
  \bibinfo{author}{Firth, C.} \& \bibinfo{author}{Rees, G.}
\newblock \bibinfo{title}{Political orientations are correlated with brain
  structure in young adults}.
\newblock \emph{\bibinfo{journal}{Curr. Biol.}} \textbf{\bibinfo{volume}{21}},
  \bibinfo{pages}{677--680} (\bibinfo{year}{2011}).

\bibitem{erdos1960evolution}
\bibinfo{author}{Erd\"{o}s, P.} \& \bibinfo{author}{R{\'e}nyi, A.}
\newblock \bibinfo{title}{On the evolution of random graphs}.
\newblock \emph{\bibinfo{journal}{Publ. Math. Inst. Hung. Acad. Sci}}
  \textbf{\bibinfo{volume}{5}}, \bibinfo{pages}{17--60} (\bibinfo{year}{1960}).

\bibitem{sherrington1906integrative}
\bibinfo{author}{Sherrington, C.~S.}
\newblock \emph{\bibinfo{title}{The Integrative Action of the Nervous System}}
  (\bibinfo{publisher}{Yale University Press}, \bibinfo{year}{1906}).

\bibitem{sporns2000theoretical}
\bibinfo{author}{Sporns, O.}, \bibinfo{author}{Tononi, G.} \&
  \bibinfo{author}{Edelman, G.~M.}
\newblock \bibinfo{title}{Theoretical neuroanatomy: relating anatomical and
  functional connectivity in graphs and cortical connection matrices}.
\newblock \emph{\bibinfo{journal}{Cereb. cortex}}
  \textbf{\bibinfo{volume}{10}}, \bibinfo{pages}{127--141}
  (\bibinfo{year}{2000}).

\bibitem{hilgetag2000anatomical}
\bibinfo{author}{Hilgetag, C.-C.}, \bibinfo{author}{Burns, G.~A.},
  \bibinfo{author}{O'Neill, M.~A.}, \bibinfo{author}{Scannell, J.~W.} \&
  \bibinfo{author}{Young, M.~P.}
\newblock \bibinfo{title}{Anatomical connectivity defines the organization of
  clusters of cortical areas in the macaque and the cat}.
\newblock \emph{\bibinfo{journal}{Phil. Trans. R. Soc. Lon. B}}
  \textbf{\bibinfo{volume}{355}}, \bibinfo{pages}{91--110}
  (\bibinfo{year}{2000}).

\bibitem{sporns2004small}
\bibinfo{author}{Sporns, O.} \& \bibinfo{author}{Zwi, J.~D.}
\newblock \bibinfo{title}{The small world of the cerebral cortex}.
\newblock \emph{\bibinfo{journal}{Neuroinformatics}}
  \textbf{\bibinfo{volume}{2}}, \bibinfo{pages}{145--162}
  (\bibinfo{year}{2004}).

\bibitem{sporns2016modular}
\bibinfo{author}{Sporns, O.} \& \bibinfo{author}{Betzel, R.~F.}
\newblock \bibinfo{title}{Modular brain networks}.
\newblock \emph{\bibinfo{journal}{Annu Rev Psychol}}
  \textbf{\bibinfo{volume}{67}}, \bibinfo{pages}{613--640}
  (\bibinfo{year}{2016}).

\bibitem{bassett2010efficient}
\bibinfo{author}{Bassett, D.~S.} \emph{et~al.}
\newblock \bibinfo{title}{Efficient physical embedding of topologically complex
  information processing networks in brains and computer circuits}.
\newblock \emph{\bibinfo{journal}{PLoS Comput. Biol.}}
  \textbf{\bibinfo{volume}{6}}, \bibinfo{pages}{e1000748}
  (\bibinfo{year}{2010}).

\bibitem{taylor2017within}
\bibinfo{author}{Taylor, P.~N.}, \bibinfo{author}{Wang, Y.} \&
  \bibinfo{author}{Kaiser, M.}
\newblock \bibinfo{title}{Within brain area tractography suggests local
  modularity using high resolution connectomics}.
\newblock \emph{\bibinfo{journal}{Sci Rep}} \textbf{\bibinfo{volume}{7}},
  \bibinfo{pages}{39859} (\bibinfo{year}{2017}).

\bibitem{lesicko2016connectional}
\bibinfo{author}{Lesicko, A.~M.}, \bibinfo{author}{Hristova, T.~S.},
  \bibinfo{author}{Maigler, K.~C.} \& \bibinfo{author}{Llano, D.~A.}
\newblock \bibinfo{title}{Connectional modularity of top-down and bottom-up
  multimodal inputs to the lateral cortex of the mouse inferior colliculus}.
\newblock \emph{\bibinfo{journal}{J Neurosci}} \textbf{\bibinfo{volume}{36}},
  \bibinfo{pages}{11037--11050} (\bibinfo{year}{2016}).

\bibitem{sohn2011topological}
\bibinfo{author}{Sohn, Y.}, \bibinfo{author}{Choi, M.~K.},
  \bibinfo{author}{Ahn, Y.~Y.}, \bibinfo{author}{Lee, J.} \&
  \bibinfo{author}{Jeong, J.}
\newblock \bibinfo{title}{Topological cluster analysis reveals the systemic
  organization of the {C}aenorhabditis elegans connectome}.
\newblock \emph{\bibinfo{journal}{PLoS Comput Biol}}
  \textbf{\bibinfo{volume}{7}}, \bibinfo{pages}{e1001139}
  (\bibinfo{year}{2011}).

\bibitem{azulay2016elegans}
\bibinfo{author}{Azulay, A.}, \bibinfo{author}{Itskovits, E.} \&
  \bibinfo{author}{Zaslaver, A.}
\newblock \bibinfo{title}{The {C}. elegans connectome consists of homogenous
  circuits with defined functional roles}.
\newblock \emph{\bibinfo{journal}{PLoS Comput Biol}}
  \textbf{\bibinfo{volume}{12}}, \bibinfo{pages}{e1005021}
  (\bibinfo{year}{2016}).

\bibitem{betzel2017multi}
\bibinfo{author}{Betzel, R.~F.} \& \bibinfo{author}{Bassett, D.~S.}
\newblock \bibinfo{title}{Multi-scale brain networks}.
\newblock \emph{\bibinfo{journal}{Neuroimage}} \textbf{\bibinfo{volume}{160}},
  \bibinfo{pages}{73--83} (\bibinfo{year}{2017}).

\bibitem{khambhati2017modeling}
\bibinfo{author}{Khambhati, A.~N.}, \bibinfo{author}{Sizemore, A.~E.},
  \bibinfo{author}{Betzel, R.~F.} \& \bibinfo{author}{Bassett, D.~S.}
\newblock \bibinfo{title}{Modeling and interpreting mesoscale network
  dynamics}.
\newblock \emph{\bibinfo{journal}{Neuroimage}}
  \textbf{\bibinfo{volume}{S1053-8119}}, \bibinfo{pages}{30500--1}
  (\bibinfo{year}{2017}).

\bibitem{aicher2015learning}
\bibinfo{author}{Aicher, C.}, \bibinfo{author}{Jacobs, A.~Z.} \&
  \bibinfo{author}{Clauset, A.}
\newblock \bibinfo{title}{Learning latent block structure in weighted
  networks}.
\newblock \emph{\bibinfo{journal}{Journal of Complex Networks}}
  \textbf{\bibinfo{volume}{3}}, \bibinfo{pages}{221--248}
  (\bibinfo{year}{2015}).

\bibitem{betzel2018diversity}
\bibinfo{author}{Betzel, R.~F.}, \bibinfo{author}{Medaglia, J.~D.} \&
  \bibinfo{author}{Bassett, D.~S.}
\newblock \bibinfo{title}{Diversity of meso-scale architecture in human and
  non-human connectomes}.
\newblock \emph{\bibinfo{journal}{Nature Communications}}
  \textbf{\bibinfo{volume}{9}}, \bibinfo{pages}{346} (\bibinfo{year}{2018}).

\bibitem{van2013network}
\bibinfo{author}{van~den Heuvel, M.~P.} \& \bibinfo{author}{Sporns, O.}
\newblock \bibinfo{title}{Network hubs in the human brain}.
\newblock \emph{\bibinfo{journal}{Trends Cogn. Sci.}}
  \textbf{\bibinfo{volume}{17}}, \bibinfo{pages}{683--696}
  (\bibinfo{year}{2013}).

\bibitem{liao2017small}
\bibinfo{author}{Liao, X.}, \bibinfo{author}{Vasilakos, A.~V.} \&
  \bibinfo{author}{He, Y.}
\newblock \bibinfo{title}{Small-world human brain networks: {P}erspectives and
  challenges}.
\newblock \emph{\bibinfo{journal}{Neurosci Biobehav Rev}}
  \textbf{\bibinfo{volume}{77}}, \bibinfo{pages}{286--300}
  (\bibinfo{year}{2017}).

\bibitem{deco2015rethinking}
\bibinfo{author}{Deco, G.}, \bibinfo{author}{Tononi, G.},
  \bibinfo{author}{Boly, M.} \& \bibinfo{author}{Kringelbach, M.~L.}
\newblock \bibinfo{title}{Rethinking segregation and integration: contributions
  of whole-brain modelling}.
\newblock \emph{\bibinfo{journal}{Nat. Rev. Neurosci.}}
  \textbf{\bibinfo{volume}{16}}, \bibinfo{pages}{430} (\bibinfo{year}{2015}).

\bibitem{latora2001efficient}
\bibinfo{author}{Latora, V.} \& \bibinfo{author}{Marchiori, M.}
\newblock \bibinfo{title}{Efficient behavior of small-world networks}.
\newblock \emph{\bibinfo{journal}{Phys. Rev. Lett.}}
  \textbf{\bibinfo{volume}{87}}, \bibinfo{pages}{198701}
  (\bibinfo{year}{2001}).

\bibitem{kaiser2006nonoptimal}
\bibinfo{author}{Kaiser, M.} \& \bibinfo{author}{Hilgetag, C.~C.}
\newblock \bibinfo{title}{Nonoptimal component placement, but short processing
  paths, due to long-distance projections in neural systems}.
\newblock \emph{\bibinfo{journal}{PLOS Comput. Biol.}}
  \textbf{\bibinfo{volume}{2}}, \bibinfo{pages}{e95} (\bibinfo{year}{2006}).

\bibitem{travers1967small}
\bibinfo{author}{Travers, J.} \& \bibinfo{author}{Milgram, S.}
\newblock \bibinfo{title}{The small world problem}.
\newblock \emph{\bibinfo{journal}{Phychology Today}}
  \textbf{\bibinfo{volume}{1}}, \bibinfo{pages}{61--67} (\bibinfo{year}{1967}).

\bibitem{watts1998collective}
\bibinfo{author}{Watts, D.~J.} \& \bibinfo{author}{Strogatz, S.~H.}
\newblock \bibinfo{title}{Collective dynamics of 'small-world' networks}.
\newblock \emph{\bibinfo{journal}{Nature}} \textbf{\bibinfo{volume}{393}},
  \bibinfo{pages}{440--442} (\bibinfo{year}{1998}).

\bibitem{gong2008mapping}
\bibinfo{author}{Gong, G.} \emph{et~al.}
\newblock \bibinfo{title}{Mapping anatomical connectivity patterns of human
  cerebral cortex using in vivo diffusion tensor imaging tractography}.
\newblock \emph{\bibinfo{journal}{Cereb. cortex}}
  \textbf{\bibinfo{volume}{19}}, \bibinfo{pages}{524--536}
  (\bibinfo{year}{2008}).

\bibitem{wedeen2005mapping}
\bibinfo{author}{Wedeen, V.~J.}, \bibinfo{author}{Hagmann, P.},
  \bibinfo{author}{Tseng, W.-Y.~I.}, \bibinfo{author}{Reese, T.~G.} \&
  \bibinfo{author}{Weisskoff, R.~M.}
\newblock \bibinfo{title}{Mapping complex tissue architecture with diffusion
  spectrum magnetic resonance imaging}.
\newblock \emph{\bibinfo{journal}{Magn. Reson. Med.}}
  \textbf{\bibinfo{volume}{54}}, \bibinfo{pages}{1377--1386}
  (\bibinfo{year}{2005}).

\bibitem{price1965networks}
\bibinfo{author}{de~Solla~Price, D.~J.}
\newblock \bibinfo{title}{Networks of scientific papers}.
\newblock \emph{\bibinfo{journal}{Science}} \textbf{\bibinfo{volume}{149}},
  \bibinfo{pages}{510--–515} (\bibinfo{year}{1965}).

\bibitem{barabasi1999emergence}
\bibinfo{author}{Barabasi, A.~L.} \& \bibinfo{author}{Albert, R.}
\newblock \bibinfo{title}{Emergence of scaling in random networks}.
\newblock \emph{\bibinfo{journal}{Science}} \textbf{\bibinfo{volume}{286}},
  \bibinfo{pages}{509--512} (\bibinfo{year}{1999}).

\bibitem{dall2002random}
\bibinfo{author}{Dall, J.} \& \bibinfo{author}{Christensen, M.}
\newblock \bibinfo{title}{Random geometric graphs}.
\newblock \emph{\bibinfo{journal}{Physical Review E}}
  \textbf{\bibinfo{volume}{66}}, \bibinfo{pages}{016121}
  (\bibinfo{year}{2002}).

\bibitem{vertes2012simple}
\bibinfo{author}{Vertes, P.~E.} \emph{et~al.}
\newblock \bibinfo{title}{Simple models of human brain functional networks}.
\newblock \emph{\bibinfo{journal}{Proc Natl Acad Sci U S A}}
  \textbf{\bibinfo{volume}{109}}, \bibinfo{pages}{5868--5873}
  (\bibinfo{year}{2012}).

\bibitem{rubinov2015wiring}
\bibinfo{author}{Rubinov, M.}, \bibinfo{author}{Ypma, R.},
  \bibinfo{author}{Watson, C.} \& \bibinfo{author}{Bullmore, E.}
\newblock \bibinfo{title}{Wiring cost and topological participation of the
  mouse brain connectome.}
\newblock \emph{\bibinfo{journal}{Proceedings of the National Academy of
  Sciences of the USA}} \bibinfo{pages}{doi/10.1073/pnas.1420315112}
  (\bibinfo{year}{2015}).

\bibitem{kaiser2017mechanisms}
\bibinfo{author}{Kaiser, M.}
\newblock \bibinfo{title}{Mechanisms of connectome development}.
\newblock \emph{\bibinfo{journal}{Trends Cogn Sci}}
  \textbf{\bibinfo{volume}{21}}, \bibinfo{pages}{703--717}
  (\bibinfo{year}{2017}).

\bibitem{stam2014modern}
\bibinfo{author}{Stam, C.~J.}
\newblock \bibinfo{title}{Modern network science of neurological disorders}.
\newblock \emph{\bibinfo{journal}{Nat Rev Neurosci}}
  \textbf{\bibinfo{volume}{15}}, \bibinfo{pages}{683--695}
  (\bibinfo{year}{2014}).

\bibitem{scholtens2014linking}
\bibinfo{author}{Scholtens, L.~H.}, \bibinfo{author}{Schmidt, R.},
  \bibinfo{author}{de~Reus, M.~A.} \& \bibinfo{author}{van~den Heuvel, M.~P.}
\newblock \bibinfo{title}{Linking macroscale graph analytical organization to
  microscale neuroarchitectonics in the macaque connectome}.
\newblock \emph{\bibinfo{journal}{J Neurosci}} \textbf{\bibinfo{volume}{34}},
  \bibinfo{pages}{12192--12205} (\bibinfo{year}{2014}).

\bibitem{chaudhuri2015large}
\bibinfo{author}{Chaudhuri, R.}, \bibinfo{author}{Knoblauch, K.},
  \bibinfo{author}{Gariel, M.~A.}, \bibinfo{author}{Kennedy, H.} \&
  \bibinfo{author}{Wang, X.~J.}
\newblock \bibinfo{title}{A large-scale circuit mechanism for hierarchical
  dynamical processing in the primate cortex}.
\newblock \emph{\bibinfo{journal}{Neuron}} \textbf{\bibinfo{volume}{88}},
  \bibinfo{pages}{419--431} (\bibinfo{year}{2015}).

\bibitem{breakspear2017dynamic}
\bibinfo{author}{Breakspear, M.}
\newblock \bibinfo{title}{Dynamic models of large-scale brain activity}.
\newblock \emph{\bibinfo{journal}{Nat Neurosci}} \textbf{\bibinfo{volume}{20}},
  \bibinfo{pages}{340--352} (\bibinfo{year}{2017}).

\bibitem{bentley2016multilayer}
\bibinfo{author}{Bentley, B.} \emph{et~al.}
\newblock \bibinfo{title}{The multilayer connectome of {C}aenorhabditis
  elegans}.
\newblock \emph{\bibinfo{journal}{PLoS Comput Biol}}
  \textbf{\bibinfo{volume}{12}}, \bibinfo{pages}{e1005283}
  (\bibinfo{year}{2016}).

\bibitem{mejias2016feedforward}
\bibinfo{author}{Mejias, J.~F.}, \bibinfo{author}{Murray, J.~D.},
  \bibinfo{author}{Kennedy, H.} \& \bibinfo{author}{Wang, X.~J.}
\newblock \bibinfo{title}{Feedforward and feedback frequency-dependent
  interactions in a large-scale laminar network of the primate cortex}.
\newblock \emph{\bibinfo{journal}{Sci Adv.}} \textbf{\bibinfo{volume}{2}},
  \bibinfo{pages}{e1601335} (\bibinfo{year}{2016}).

\bibitem{seung2014neuronal}
\bibinfo{author}{Seung, H.~S.} \& \bibinfo{author}{Sumbul, U.}
\newblock \bibinfo{title}{Neuronal cell types and connectivity: lessons from
  the retina}.
\newblock \emph{\bibinfo{journal}{Neuron}} \textbf{\bibinfo{volume}{83}},
  \bibinfo{pages}{1262--1272} (\bibinfo{year}{2014}).

\bibitem{arnatkeviciute2018hub}
\bibinfo{author}{Arnatkeviciute, A.}, \bibinfo{author}{Fulcher, B.~D.},
  \bibinfo{author}{Pocock, R.} \& \bibinfo{author}{Fornito, A.}
\newblock \bibinfo{title}{Hub connectivity, neuronal diversity, and gene
  expression in the {C}aenorhabditis elegans connectome}.
\newblock \emph{\bibinfo{journal}{PLoS Comput Biol}}
  \textbf{\bibinfo{volume}{14}}, \bibinfo{pages}{e1005989}
  (\bibinfo{year}{2018}).

\bibitem{scholz2009training}
\bibinfo{author}{Scholz, J.}, \bibinfo{author}{Klein, M.~C.},
  \bibinfo{author}{Behrens, T.~E.} \& \bibinfo{author}{Johansen-Berg, H.}
\newblock \bibinfo{title}{Training induces changes in white-matter
  architecture}.
\newblock \emph{\bibinfo{journal}{Nat Neurosci}} \textbf{\bibinfo{volume}{12}},
  \bibinfo{pages}{1370--1371} (\bibinfo{year}{2009}).

\bibitem{baum2017modular}
\bibinfo{author}{Baum, G.~L.} \emph{et~al.}
\newblock \bibinfo{title}{Modular segregation of structural brain networks
  supports the development of executive function in youth}.
\newblock \emph{\bibinfo{journal}{Curr Biol}} \textbf{\bibinfo{volume}{27}},
  \bibinfo{pages}{1561--1572.e8} (\bibinfo{year}{2017}).

\bibitem{zuo2017human}
\bibinfo{author}{Zuo, X.~N.} \emph{et~al.}
\newblock \bibinfo{title}{Human connectomics across the life span}.
\newblock \emph{\bibinfo{journal}{Trends Cogn Sci}}
  \textbf{\bibinfo{volume}{21}}, \bibinfo{pages}{32--45}
  (\bibinfo{year}{2017}).

\bibitem{holme2012temporal}
\bibinfo{author}{Holme, P.} \& \bibinfo{author}{Saramaki, J.}
\newblock \bibinfo{title}{Temporal networks}.
\newblock \emph{\bibinfo{journal}{Phys. Rep.}} \textbf{\bibinfo{volume}{519}},
  \bibinfo{pages}{97--125} (\bibinfo{year}{2012}).

\bibitem{li2017fundamental}
\bibinfo{author}{Li, A.}, \bibinfo{author}{Cornelius, S.~P.},
  \bibinfo{author}{Liu, Y.-Y.}, \bibinfo{author}{Wang, L.} \&
  \bibinfo{author}{Barab{\'a}si, A.-L.}
\newblock \bibinfo{title}{The fundamental advantages of temporal networks}.
\newblock \emph{\bibinfo{journal}{Science}} \textbf{\bibinfo{volume}{358}},
  \bibinfo{pages}{1042--1046} (\bibinfo{year}{2017}).

\bibitem{hebb1949organization}
\bibinfo{author}{Hebb, D.}
\newblock \emph{\bibinfo{title}{The Organization of Behavior}}
  (\bibinfo{publisher}{Wiley}, \bibinfo{year}{1949}).

\bibitem{magee1997synaptically}
\bibinfo{author}{Magee, J.~C.} \& \bibinfo{author}{Johnston, D.}
\newblock \bibinfo{title}{A synaptically controlled, associative signal for
  hebbian plasticity in hippocampal neurons}.
\newblock \emph{\bibinfo{journal}{Science}} \textbf{\bibinfo{volume}{275}},
  \bibinfo{pages}{209--213} (\bibinfo{year}{1997}).

\bibitem{montague1996framework}
\bibinfo{author}{Montague, P.~R.}, \bibinfo{author}{Dayan, P.} \&
  \bibinfo{author}{Sejnowski, T.~J.}
\newblock \bibinfo{title}{A framework for mesencephalic dopamine systems based
  on predictive hebbian learning}.
\newblock \emph{\bibinfo{journal}{J. Neurosci.}} \textbf{\bibinfo{volume}{16}},
  \bibinfo{pages}{1936--1947} (\bibinfo{year}{1996}).

\bibitem{song2000competitive}
\bibinfo{author}{Song, S.}, \bibinfo{author}{Miller, K.~D.} \&
  \bibinfo{author}{Abbott, L.~F.}
\newblock \bibinfo{title}{Competitive hebbian learning through
  spike-timing-dependent synaptic plasticity}.
\newblock \emph{\bibinfo{journal}{Nat. Neurosci.}}
  \textbf{\bibinfo{volume}{3}}, \bibinfo{pages}{919} (\bibinfo{year}{2000}).

\bibitem{chialvo2010emergent}
\bibinfo{author}{Chialvo, D.~R.}
\newblock \bibinfo{title}{Emergent complex neural dynamics}.
\newblock \emph{\bibinfo{journal}{Nat. Phys.}} \textbf{\bibinfo{volume}{6}},
  \bibinfo{pages}{744} (\bibinfo{year}{2010}).

\bibitem{tononi2016integrated}
\bibinfo{author}{Tononi, G.}, \bibinfo{author}{Boly, M.},
  \bibinfo{author}{Massimini, M.} \& \bibinfo{author}{Koch, C.}
\newblock \bibinfo{title}{Integrated information theory: from consciousness to
  its physical substrate}.
\newblock \emph{\bibinfo{journal}{Nat Rev Neurosci}}
  \textbf{\bibinfo{volume}{17}}, \bibinfo{pages}{450--461}
  (\bibinfo{year}{2016}).

\bibitem{abbott2001theoretical}
\bibinfo{author}{Abbott, L.~F.} \& \bibinfo{author}{Dayan, P.}
\newblock \emph{\bibinfo{title}{Theoretical Neuroscience}}
  (\bibinfo{publisher}{MIT Press}, \bibinfo{year}{2001}).

\bibitem{dechery2017emergent}
\bibinfo{author}{Dechery, J.~B.} \& \bibinfo{author}{MacLean, J.~N.}
\newblock \bibinfo{title}{Emergent cortical circuit dynamics contain dense,
  interwoven ensembles of spike sequences}.
\newblock \emph{\bibinfo{journal}{J Neurophysiol}}
  \textbf{\bibinfo{volume}{118}}, \bibinfo{pages}{1914--1925}
  (\bibinfo{year}{2017}).

\bibitem{reif2009fundamentals}
\bibinfo{author}{Reif, F.}
\newblock \emph{\bibinfo{title}{Fundamentals of statistical and thermal
  physics}} (\bibinfo{publisher}{Waveland Press}, \bibinfo{year}{2009}).

\bibitem{brody1999correlations}
\bibinfo{author}{Brody, C.~D.}
\newblock \bibinfo{title}{Correlations without synchrony}.
\newblock \emph{\bibinfo{journal}{Neural Comput}}
  \textbf{\bibinfo{volume}{11}}, \bibinfo{pages}{1537--1551}
  (\bibinfo{year}{1999}).

\bibitem{brody1999disambiguating}
\bibinfo{author}{Brody, C.~D.}
\newblock \bibinfo{title}{Disambiguating different covariation types}.
\newblock \emph{\bibinfo{journal}{Neural Comput}}
  \textbf{\bibinfo{volume}{11}}, \bibinfo{pages}{1527--1535}
  (\bibinfo{year}{1999}).

\bibitem{sporns2000connectivity}
\bibinfo{author}{Sporns, O.}, \bibinfo{author}{Tononi, G.} \&
  \bibinfo{author}{Edelman, G.~M.}
\newblock \bibinfo{title}{Connectivity and complexity: the relationship between
  neuroanatomy and brain dynamics}.
\newblock \emph{\bibinfo{journal}{Neural Netw}} \textbf{\bibinfo{volume}{13}},
  \bibinfo{pages}{909--922} (\bibinfo{year}{2000}).

\bibitem{schneidman2006weak}
\bibinfo{author}{Schneidman, E.}, \bibinfo{author}{Berry~II, M.~J.},
  \bibinfo{author}{Segev, R.} \& \bibinfo{author}{Bialek, W.}
\newblock \bibinfo{title}{Weak pairwise correlations imply strongly correlated
  network states in a neural population}.
\newblock \emph{\bibinfo{journal}{Nature}} \textbf{\bibinfo{volume}{440}},
  \bibinfo{pages}{1007} (\bibinfo{year}{2006}).

\bibitem{levina2007dynamical}
\bibinfo{author}{Levina, A.}, \bibinfo{author}{Herrmann, J.~M.} \&
  \bibinfo{author}{Geisel, T.}
\newblock \bibinfo{title}{Dynamical synapses causing self-organized criticality
  in neural networks}.
\newblock \emph{\bibinfo{journal}{Nat. Phys.}} \textbf{\bibinfo{volume}{3}},
  \bibinfo{pages}{857} (\bibinfo{year}{2007}).

\bibitem{hoevel2014functional}
\bibinfo{author}{Vuksanovic, V.} \& \bibinfo{author}{Hovel, P.}
\newblock \bibinfo{title}{Functional connectivity of distant cortical regions:
  role of remote synchronization and symmetry in interactions}.
\newblock \emph{\bibinfo{journal}{Neuroimage}} \textbf{\bibinfo{volume}{97}},
  \bibinfo{pages}{1--8} (\bibinfo{year}{2014}).

\bibitem{flourens1842recherches}
\bibinfo{author}{Flourens, P.}
\newblock \emph{\bibinfo{title}{Recherches exp{\'e}rimentales sur les
  propri{\'e}t{\'e}s et les fonctions du syst{\`e}me nerveux dans les animaux
  vert{\'e}br{\'e}s}} (\bibinfo{publisher}{Balli{\`e}re},
  \bibinfo{year}{1842}).

\bibitem{panizza1855osservazioni}
\bibinfo{author}{Panizza, B.}
\newblock \emph{\bibinfo{title}{Osservazioni sul nervo ottico}}
  (\bibinfo{publisher}{Bernardoni}, \bibinfo{year}{1855}).

\bibitem{broca1861remarques}
\bibinfo{author}{Broca, P.}
\newblock \bibinfo{title}{Remarques sur le si{\`e}ge de la facult{\'e} du
  langage articul{\'e}, suivies d’une observation d’aph{\'e}mie (perte de
  la parole)}.
\newblock \emph{\bibinfo{journal}{Bulletin et Memoires de la Societe anatomique
  de Paris}} \textbf{\bibinfo{volume}{6}}, \bibinfo{pages}{330--357}
  (\bibinfo{year}{1861}).

\bibitem{helmholtz1850Vorlaufiger}
\bibinfo{author}{von Helmholtz, H.}
\newblock \bibinfo{title}{Vorl\"{a}ufiger bericht \"{u}ber die
  fortpflanzungs-geschwindigkeit der nervenreizung}.
\newblock \emph{\bibinfo{journal}{Archiv f\"{u}r Anatomie, Physiologie und
  wissenschaftliche Medicin}} \bibinfo{pages}{71--73} (\bibinfo{year}{1850}).

\bibitem{haas2003hans}
\bibinfo{author}{Haas, L.~F.}
\newblock \bibinfo{title}{Hans berger (1873--1941), richard caton (1842--1926),
  and electroencephalography}.
\newblock \emph{\bibinfo{journal}{J. Neurol. Neurosurg. Psychiatry}}
  \textbf{\bibinfo{volume}{74}}, \bibinfo{pages}{9--9} (\bibinfo{year}{2003}).

\bibitem{caton1875electrical}
\bibinfo{author}{Caton, R.}
\newblock \bibinfo{title}{Electrical currents of the brain}.
\newblock \emph{\bibinfo{journal}{J. Nerv. Ment. Dis.}}
  \textbf{\bibinfo{volume}{2}}, \bibinfo{pages}{610} (\bibinfo{year}{1875}).

\bibitem{beck1890strome}
\bibinfo{author}{Beck, A.}
\newblock \bibinfo{title}{Die str{\"o}me der nervencentren}.
\newblock \emph{\bibinfo{journal}{Centralbl Physiol}}
  \textbf{\bibinfo{volume}{4}}, \bibinfo{pages}{572--573}
  (\bibinfo{year}{1890}).

\bibitem{lorente1934studies}
\bibinfo{author}{Lorente~de N{\'o}, R.}
\newblock \bibinfo{title}{Studies on the structure of the cerebral cortex. ii.
  continuation of the study of the ammonic system.}
\newblock \emph{\bibinfo{journal}{Journal f{\"u}r Psychologie und Neurologie}}
  (\bibinfo{year}{1934}).

\bibitem{green1982circadian}
\bibinfo{author}{Green, D.~J.} \& \bibinfo{author}{Gillette, R.}
\newblock \bibinfo{title}{Circadian rhythm of firing rate recorded from single
  cells in the rat suprachiasmatic brain slice}.
\newblock \emph{\bibinfo{journal}{Brain Res.}} \textbf{\bibinfo{volume}{245}},
  \bibinfo{pages}{198--200} (\bibinfo{year}{1982}).

\bibitem{edwards1989thin}
\bibinfo{author}{Edwards, F.~A.}, \bibinfo{author}{Konnerth, A.},
  \bibinfo{author}{Sakmann, B.} \& \bibinfo{author}{Takahashi, T.}
\newblock \bibinfo{title}{A thin slice preparation for patch clamp recordings
  from neurones of the mammalian central nervous system}.
\newblock \emph{\bibinfo{journal}{Pfl{\"u}gers Archiv}}
  \textbf{\bibinfo{volume}{414}}, \bibinfo{pages}{600--612}
  (\bibinfo{year}{1989}).

\bibitem{stosiek2003vivo}
\bibinfo{author}{Stosiek, C.}, \bibinfo{author}{Garaschuk, O.},
  \bibinfo{author}{Holthoff, K.} \& \bibinfo{author}{Konnerth, A.}
\newblock \bibinfo{title}{In vivo two-photon calcium imaging of neuronal
  networks}.
\newblock \emph{\bibinfo{journal}{Proc Natl Acad Sci U S A}}
  \textbf{\bibinfo{volume}{100}}, \bibinfo{pages}{7319--7324}
  (\bibinfo{year}{2003}).

\bibitem{grewe2010high}
\bibinfo{author}{Grewe, B.~F.}, \bibinfo{author}{Langer, D.},
  \bibinfo{author}{Kasper, H.}, \bibinfo{author}{Kampa, B.~M.} \&
  \bibinfo{author}{Helmchen, F.}
\newblock \bibinfo{title}{High-speed in vivo calcium imaging reveals neuronal
  network activity with near-millisecond precision}.
\newblock \emph{\bibinfo{journal}{Nat. Methods}} \textbf{\bibinfo{volume}{7}},
  \bibinfo{pages}{399} (\bibinfo{year}{2010}).

\bibitem{penny2011statistical}
\bibinfo{author}{Penny, W.~D.}, \bibinfo{author}{Friston, K.~J.},
  \bibinfo{author}{Ashburner, J.~T.}, \bibinfo{author}{Kiebel, S.~J.} \&
  \bibinfo{author}{Nichols, T.~E.}
\newblock \emph{\bibinfo{title}{Statistical parametric mapping: the analysis of
  functional brain images}} (\bibinfo{publisher}{Elsevier},
  \bibinfo{year}{2011}).

\bibitem{hamalainen1993magnetoencephalography}
\bibinfo{author}{H{\"a}m{\"a}l{\"a}inen, M.}, \bibinfo{author}{Hari, R.},
  \bibinfo{author}{Ilmoniemi, R.~J.}, \bibinfo{author}{Knuutila, J.} \&
  \bibinfo{author}{Lounasmaa, O.~V.}
\newblock \bibinfo{title}{Magnetoencephalography—theory, instrumentation, and
  applications to noninvasive studies of the working human brain}.
\newblock \emph{\bibinfo{journal}{Rev. Mod. Phys.}}
  \textbf{\bibinfo{volume}{65}}, \bibinfo{pages}{413} (\bibinfo{year}{1993}).

\bibitem{bailey2005positron}
\bibinfo{author}{Bailey, D.~L.}, \bibinfo{author}{Maisey, M.~N.},
  \bibinfo{author}{Townsend, D.~W.} \& \bibinfo{author}{Valk, P.~E.}
\newblock \emph{\bibinfo{title}{Positron emission tomography}}
  (\bibinfo{publisher}{Springer}, \bibinfo{year}{2005}).

\bibitem{raichle1998behind}
\bibinfo{author}{Raichle, M.~E.}
\newblock \bibinfo{title}{Behind the scenes of functional brain imaging: a
  historical and physiological perspective}.
\newblock \emph{\bibinfo{journal}{Proc Natl Acad Sci U S A}}
  \textbf{\bibinfo{volume}{95}}, \bibinfo{pages}{765--772}
  (\bibinfo{year}{1998}).

\bibitem{zarahn1997empirical}
\bibinfo{author}{Zarahn, E.}, \bibinfo{author}{Aguirre, G.~K.} \&
  \bibinfo{author}{D'Esposito, M.}
\newblock \bibinfo{title}{Empirical analyses of bold fmri statistics}.
\newblock \emph{\bibinfo{journal}{Neuroimage}} \textbf{\bibinfo{volume}{5}},
  \bibinfo{pages}{179--197} (\bibinfo{year}{1997}).

\bibitem{van2010exploring}
\bibinfo{author}{Van Den~Heuvel, M.~P.} \& \bibinfo{author}{Pol, H. E.~H.}
\newblock \bibinfo{title}{Exploring the brain network: a review on
  resting-state fmri functional connectivity}.
\newblock \emph{\bibinfo{journal}{Eur Neuropsychopharmacol}}
  \textbf{\bibinfo{volume}{20}}, \bibinfo{pages}{519--534}
  (\bibinfo{year}{2010}).

\bibitem{bullmore2009complex}
\bibinfo{author}{Bullmore, E.} \& \bibinfo{author}{Sporns, O.}
\newblock \bibinfo{title}{Complex brain networks: graph theoretical analysis of
  structural and functional systems}.
\newblock \emph{\bibinfo{journal}{Nat Rev Neurosci}}
  \textbf{\bibinfo{volume}{10}}, \bibinfo{pages}{186--198}
  (\bibinfo{year}{2009}).

\bibitem{zalesky2012correlation}
\bibinfo{author}{Zalesky, A.}, \bibinfo{author}{Fornito, A.} \&
  \bibinfo{author}{Bullmore, E.}
\newblock \bibinfo{title}{On the use of correlation as a measure of network
  connectivity}.
\newblock \emph{\bibinfo{journal}{Neuroimage}} \textbf{\bibinfo{volume}{60}},
  \bibinfo{pages}{2096--2106} (\bibinfo{year}{2012}).

\bibitem{he2009uncovering}
\bibinfo{author}{He, Y.} \emph{et~al.}
\newblock \bibinfo{title}{Uncovering intrinsic modular organization of
  spontaneous brain activity in humans}.
\newblock \emph{\bibinfo{journal}{PloS one}} \textbf{\bibinfo{volume}{4}},
  \bibinfo{pages}{e5226} (\bibinfo{year}{2009}).

\bibitem{salvador2005neurophysiological}
\bibinfo{author}{Salvador, R.} \emph{et~al.}
\newblock \bibinfo{title}{Neurophysiological architecture of functional
  magnetic resonance images of human brain}.
\newblock \emph{\bibinfo{journal}{Cerebral cortex}}
  \textbf{\bibinfo{volume}{15}}, \bibinfo{pages}{1332--1342}
  (\bibinfo{year}{2005}).

\bibitem{achard2006resilient}
\bibinfo{author}{Achard, S.}, \bibinfo{author}{Salvador, R.},
  \bibinfo{author}{Whitcher, B.}, \bibinfo{author}{Suckling, J.} \&
  \bibinfo{author}{Bullmore, E.}
\newblock \bibinfo{title}{A resilient, low-frequency, small-world human brain
  functional network with highly connected association cortical hubs}.
\newblock \emph{\bibinfo{journal}{J Neurosci}} \textbf{\bibinfo{volume}{26}},
  \bibinfo{pages}{63--72} (\bibinfo{year}{2006}).

\bibitem{bettencourt2007functional}
\bibinfo{author}{Bettencourt, L.~M.}, \bibinfo{author}{Stephens, G.~J.},
  \bibinfo{author}{Ham, M.~I.} \& \bibinfo{author}{Gross, G.~W.}
\newblock \bibinfo{title}{Functional structure of cortical neuronal networks
  grown in vitro}.
\newblock \emph{\bibinfo{journal}{Phys Rev E}} \textbf{\bibinfo{volume}{75}},
  \bibinfo{pages}{021915} (\bibinfo{year}{2007}).

\bibitem{sadovsky2013scaling}
\bibinfo{author}{Sadovsky, A.~J.} \& \bibinfo{author}{MacLean, J.~N.}
\newblock \bibinfo{title}{Scaling of topologically similar functional modules
  defines mouse primary auditory and somatosensory microcircuitry}.
\newblock \emph{\bibinfo{journal}{J Neurosci}} \textbf{\bibinfo{volume}{33}},
  \bibinfo{pages}{14048--14060} (\bibinfo{year}{2013}).

\bibitem{yue2017brain}
\bibinfo{author}{Yue, Q.} \emph{et~al.}
\newblock \bibinfo{title}{Brain modularity mediates the relation between task
  complexity and performance}.
\newblock \emph{\bibinfo{journal}{J Cogn Neurosci}}
  \textbf{\bibinfo{volume}{29}}, \bibinfo{pages}{1532--1546}
  (\bibinfo{year}{2017}).

\bibitem{bassett2006small}
\bibinfo{author}{Bassett, D.~S.} \& \bibinfo{author}{Bullmore, E.}
\newblock \bibinfo{title}{Small-world brain networks}.
\newblock \emph{\bibinfo{journal}{Neuroscientist}}
  \textbf{\bibinfo{volume}{12}}, \bibinfo{pages}{512--523}
  (\bibinfo{year}{2006}).

\bibitem{rosenbaum2017spatial}
\bibinfo{author}{Rosenbaum, R.}, \bibinfo{author}{Smith, M.~A.},
  \bibinfo{author}{Kohn, A.}, \bibinfo{author}{Rubin, J.~E.} \&
  \bibinfo{author}{Doiron, B.}
\newblock \bibinfo{title}{The spatial structure of correlated neuronal
  variability}.
\newblock \emph{\bibinfo{journal}{Nat Neurosci}} \textbf{\bibinfo{volume}{20}},
  \bibinfo{pages}{107--114} (\bibinfo{year}{2017}).

\bibitem{goni2014resting}
\bibinfo{author}{Go{\~n}i, J.} \emph{et~al.}
\newblock \bibinfo{title}{Resting-brain functional connectivity predicted by
  analytic measures of network communication}.
\newblock \emph{\bibinfo{journal}{Proceedings of the National Academy of
  Sciences}} \textbf{\bibinfo{volume}{111}}, \bibinfo{pages}{833--838}
  (\bibinfo{year}{2014}).

\bibitem{honey2009predicting}
\bibinfo{author}{Honey, C.} \emph{et~al.}
\newblock \bibinfo{title}{Predicting human resting-state functional
  connectivity from structural connectivity}.
\newblock \emph{\bibinfo{journal}{Proceedings of the National Academy of
  Sciences}} \textbf{\bibinfo{volume}{106}}, \bibinfo{pages}{2035--2040}
  (\bibinfo{year}{2009}).

\bibitem{medaglia2018functional}
\bibinfo{author}{Medaglia, J.~D.} \emph{et~al.}
\newblock \bibinfo{title}{Functional alignment with anatomical networks is
  associated with cognitive flexibility}.
\newblock \emph{\bibinfo{journal}{Nature Human Behaviour}}
  \textbf{\bibinfo{volume}{2}}, \bibinfo{pages}{156--164}
  (\bibinfo{year}{2018}).

\bibitem{park2013structural}
\bibinfo{author}{Park, H.-J.} \& \bibinfo{author}{Friston, K.}
\newblock \bibinfo{title}{Structural and functional brain networks: from
  connections to cognition}.
\newblock \emph{\bibinfo{journal}{Science}} \textbf{\bibinfo{volume}{342}},
  \bibinfo{pages}{1238411} (\bibinfo{year}{2013}).

\bibitem{hodgkin1952quantitative}
\bibinfo{author}{Hodgkin, A.~L.} \& \bibinfo{author}{Huxley, A.~F.}
\newblock \bibinfo{title}{A quantitative description of membrane current and
  its application to conduction and excitation in nerve}.
\newblock \emph{\bibinfo{journal}{J. Physiol.}} \textbf{\bibinfo{volume}{117}},
  \bibinfo{pages}{500--544} (\bibinfo{year}{1952}).

\bibitem{fitzhugh1961impulses}
\bibinfo{author}{FitzHugh, R.}
\newblock \bibinfo{title}{Impulses and physiological states in theoretical
  models of nerve membrane}.
\newblock \emph{\bibinfo{journal}{Biophys. J.}} \textbf{\bibinfo{volume}{1}},
  \bibinfo{pages}{445--466} (\bibinfo{year}{1961}).

\bibitem{beurle1956properties}
\bibinfo{author}{Beurle, R.~L.}
\newblock \bibinfo{title}{Properties of a mass of cells capable of regenerating
  pulses}.
\newblock \emph{\bibinfo{journal}{Phil. Trans. R. Soc. Lond. B}}
  \textbf{\bibinfo{volume}{240}}, \bibinfo{pages}{55--94}
  (\bibinfo{year}{1956}).

\bibitem{wilson1972excitatory}
\bibinfo{author}{Wilson, H.~R.} \& \bibinfo{author}{Cowan, J.~D.}
\newblock \bibinfo{title}{Excitatory and inhibitory interactions in localized
  populations of model neurons}.
\newblock \emph{\bibinfo{journal}{Biophys. J.}} \textbf{\bibinfo{volume}{12}},
  \bibinfo{pages}{1--24} (\bibinfo{year}{1972}).

\bibitem{kuramoto2012chemical}
\bibinfo{author}{Kuramoto, Y.}
\newblock \emph{\bibinfo{title}{Chemical oscillations, waves, and turbulence}},
  vol.~\bibinfo{volume}{19} (\bibinfo{publisher}{Springer Science \& Business
  Media}, \bibinfo{year}{2012}).

\bibitem{cash1999linear}
\bibinfo{author}{Cash, S.} \& \bibinfo{author}{Yuste, R.}
\newblock \bibinfo{title}{Linear summation of excitatory inputs by ca1
  pyramidal neurons}.
\newblock \emph{\bibinfo{journal}{Neuron}} \textbf{\bibinfo{volume}{22}},
  \bibinfo{pages}{383--394} (\bibinfo{year}{1999}).

\bibitem{ferrell1998biochemical}
\bibinfo{author}{Ferrell, J.~E.} \& \bibinfo{author}{Machleder, E.~M.}
\newblock \bibinfo{title}{The biochemical basis of an all-or-none cell fate
  switch in xenopus oocytes}.
\newblock \emph{\bibinfo{journal}{Science}} \textbf{\bibinfo{volume}{280}},
  \bibinfo{pages}{895--898} (\bibinfo{year}{1998}).

\bibitem{hearst1998support}
\bibinfo{author}{Hearst, M.~A.}, \bibinfo{author}{Dumais, S.~T.},
  \bibinfo{author}{Osuna, E.}, \bibinfo{author}{Platt, J.} \&
  \bibinfo{author}{Scholkopf, B.}
\newblock \bibinfo{title}{Support vector machines}.
\newblock \emph{\bibinfo{journal}{IEEE Intell. Syst.}}
  \textbf{\bibinfo{volume}{13}}, \bibinfo{pages}{18--28}
  (\bibinfo{year}{1998}).

\bibitem{kleene1951representation}
\bibinfo{author}{Kleene, S.~C.}
\newblock \bibinfo{title}{Representation of events in nerve nets and finite
  automata}.
\newblock \bibinfo{type}{Tech. Rep.}, \bibinfo{institution}{RAND PROJECT AIR
  FORCE SANTA MONICA CA} (\bibinfo{year}{1951}).

\bibitem{schmidhuber2015deep}
\bibinfo{author}{Schmidhuber, J.}
\newblock \bibinfo{title}{Deep learning in neural networks: An overview}.
\newblock \emph{\bibinfo{journal}{Neural Netw.}} \textbf{\bibinfo{volume}{61}},
  \bibinfo{pages}{85--117} (\bibinfo{year}{2015}).

\bibitem{egmont2002image}
\bibinfo{author}{Egmont-Petersen, M.}, \bibinfo{author}{de~Ridder, D.} \&
  \bibinfo{author}{Handels, H.}
\newblock \bibinfo{title}{Image processing with neural networks—a review}.
\newblock \emph{\bibinfo{journal}{Pattern Recognit.}}
  \textbf{\bibinfo{volume}{35}}, \bibinfo{pages}{2279--2301}
  (\bibinfo{year}{2002}).

\bibitem{hinton2012deep}
\bibinfo{author}{Hinton, G.} \emph{et~al.}
\newblock \bibinfo{title}{Deep neural networks for acoustic modeling in speech
  recognition: The shared views of four research groups}.
\newblock \emph{\bibinfo{journal}{IEEE Signal Process. Mag.}}
  \textbf{\bibinfo{volume}{29}}, \bibinfo{pages}{82--97}
  (\bibinfo{year}{2012}).

\bibitem{silver2016mastering}
\bibinfo{author}{Silver, D.} \emph{et~al.}
\newblock \bibinfo{title}{Mastering the game of go with deep neural networks
  and tree search}.
\newblock \emph{\bibinfo{journal}{Nature}} \textbf{\bibinfo{volume}{529}},
  \bibinfo{pages}{484} (\bibinfo{year}{2016}).

\bibitem{newman1988memory}
\bibinfo{author}{Newman, C.~M.}
\newblock \bibinfo{title}{Memory capacity in neural network models: Rigorous
  lower bounds}.
\newblock \emph{\bibinfo{journal}{Neural Netw.}} \textbf{\bibinfo{volume}{1}},
  \bibinfo{pages}{223--238} (\bibinfo{year}{1988}).

\bibitem{hertz1991introduction}
\bibinfo{author}{Hertz, J.}, \bibinfo{author}{Krogh, A.} \&
  \bibinfo{author}{Palmer, R.~G.}
\newblock \emph{\bibinfo{title}{Introduction to the theory of neural
  computation.}} (\bibinfo{publisher}{Addison-Wesley/Addison Wesley Longman},
  \bibinfo{year}{1991}).

\bibitem{moosavi2015structural}
\bibinfo{author}{Moosavi, S.~A.} \& \bibinfo{author}{Montakhab, A.}
\newblock \bibinfo{title}{Structural versus dynamical origins of mean-field
  behavior in a self-organized critical model of neuronal avalanches.}
\newblock \emph{\bibinfo{journal}{Phys Rev E Stat Nonlin Soft Matter Phys}}
  \textbf{\bibinfo{volume}{92}}, \bibinfo{pages}{052804}
  (\bibinfo{year}{2015}).

\bibitem{karimipanah2015adaptation}
\bibinfo{author}{Woodrow, W.~L.} \emph{et~al.}
\newblock \bibinfo{title}{Adaptation to sensory input tunes visual cortex to
  criticality}.
\newblock \emph{\bibinfo{journal}{Nature Physics}}
  \textbf{\bibinfo{volume}{11}}, \bibinfo{pages}{659--663}
  (\bibinfo{year}{2015}).

\bibitem{haldeman2005}
\bibinfo{author}{Haldeman, C.} \& \bibinfo{author}{Beggs, J.~M.}
\newblock \bibinfo{title}{Critical branching captures activity in living neural
  networks and maximizes the number of metastable states}.
\newblock \emph{\bibinfo{journal}{Phys. Rev. Lett.}}
  \textbf{\bibinfo{volume}{94}}, \bibinfo{pages}{058101}
  (\bibinfo{year}{2005}).
\newblock
  \urlprefix\url{https://link.aps.org/doi/10.1103/PhysRevLett.94.058101}.

\bibitem{beggs2003}
\bibinfo{author}{Beggs, J.~M.} \& \bibinfo{author}{Plenz, D.}
\newblock \bibinfo{title}{Neuronal avalanches in neocortical circuits}.
\newblock \emph{\bibinfo{journal}{Journal of Neuroscience}}
  \textbf{\bibinfo{volume}{23}}, \bibinfo{pages}{11167--11177}
  (\bibinfo{year}{2003}).
\newblock \urlprefix\url{http://www.jneurosci.org/content/23/35/11167}.
\newblock \eprint{http://www.jneurosci.org/content/23/35/11167.full.pdf}.

\bibitem{kinouchi2006}
\bibinfo{author}{Kinouchi, O.} \& \bibinfo{author}{Copelli, M.}
\newblock \bibinfo{title}{Optimal dynamical range of excitable networks at
  criticality}.
\newblock \emph{\bibinfo{journal}{Nature Physics}}
  \textbf{\bibinfo{volume}{2}}, \bibinfo{pages}{348 EP --}
  (\bibinfo{year}{2006}).
\newblock \urlprefix\url{http://dx.doi.org/10.1038/nphys289}.

\bibitem{shew2009}
\bibinfo{author}{Shew, W.~L.}, \bibinfo{author}{Yang, H.},
  \bibinfo{author}{Petermann, T.}, \bibinfo{author}{Roy, R.} \&
  \bibinfo{author}{Plenz, D.}
\newblock \bibinfo{title}{Neuronal avalanches imply maximum dynamic range in
  cortical networks at criticality}.
\newblock \emph{\bibinfo{journal}{Journal of Neuroscience}}
  \textbf{\bibinfo{volume}{29}}, \bibinfo{pages}{15595--15600}
  (\bibinfo{year}{2009}).
\newblock \urlprefix\url{http://www.jneurosci.org/content/29/49/15595}.
\newblock \eprint{http://www.jneurosci.org/content/29/49/15595.full.pdf}.

\bibitem{bertschinger2004}
\bibinfo{author}{Bertschinger, N.} \& \bibinfo{author}{Natschl{\"a}ger, T.}
\newblock \bibinfo{title}{Real-time computation at the edge of chaos in
  recurrent neural networks}.
\newblock \emph{\bibinfo{journal}{Neural Computation}}
  \textbf{\bibinfo{volume}{16}}, \bibinfo{pages}{1413--1436}
  (\bibinfo{year}{2004}).
\newblock \urlprefix\url{https://doi.org/10.1162/089976604323057443}.
\newblock \eprint{https://doi.org/10.1162/089976604323057443}.

\bibitem{ganmor2011sparse}
\bibinfo{author}{Ganmor, E.}, \bibinfo{author}{Segev, R.} \&
  \bibinfo{author}{Schneidman, E.}
\newblock \bibinfo{title}{Sparse low-order interaction network underlies a
  highly correlated and learnable neural population code}.
\newblock \emph{\bibinfo{journal}{Proc Natl Acad Sci U S A}}
  \textbf{\bibinfo{volume}{108}}, \bibinfo{pages}{9679--9684}
  (\bibinfo{year}{2011}).

\bibitem{lee1998coherence}
\bibinfo{author}{Lee, S.-G.}, \bibinfo{author}{Neiman, A.} \&
  \bibinfo{author}{Kim, S.}
\newblock \bibinfo{title}{Coherence resonance in a hodgkin-huxley neuron}.
\newblock \emph{\bibinfo{journal}{Phys. Rev. E}} \textbf{\bibinfo{volume}{57}},
  \bibinfo{pages}{3292} (\bibinfo{year}{1998}).

\bibitem{hille2001ion}
\bibinfo{author}{Hille, B.} \emph{et~al.}
\newblock \emph{\bibinfo{title}{Ion channels of excitable membranes}}, vol.
  \bibinfo{volume}{507} (\bibinfo{publisher}{Sinauer Sunderland, MA},
  \bibinfo{year}{2001}).

\bibitem{plant1976mathematical}
\bibinfo{author}{Plant, R.} \& \bibinfo{author}{Kim, M.}
\newblock \bibinfo{title}{Mathematical description of a bursting pacemaker
  neuron by a modification of the hodgkin-huxley equations}.
\newblock \emph{\bibinfo{journal}{Biophys. J.}} \textbf{\bibinfo{volume}{16}},
  \bibinfo{pages}{227--244} (\bibinfo{year}{1976}).

\bibitem{andersen2009towards}
\bibinfo{author}{Andersen, S.~S.}, \bibinfo{author}{Jackson, A.~D.} \&
  \bibinfo{author}{Heimburg, T.}
\newblock \bibinfo{title}{Towards a thermodynamic theory of nerve pulse
  propagation}.
\newblock \emph{\bibinfo{journal}{Prog. Neurobiol.}}
  \textbf{\bibinfo{volume}{88}}, \bibinfo{pages}{104--113}
  (\bibinfo{year}{2009}).

\bibitem{pakdaman2010fluid}
\bibinfo{author}{Pakdaman, K.}, \bibinfo{author}{Thieullen, M.} \&
  \bibinfo{author}{Wainrib, G.}
\newblock \bibinfo{title}{Fluid limit theorems for stochastic hybrid systems
  with application to neuron models}.
\newblock \emph{\bibinfo{journal}{Adv. Appl. Probab.}}
  \textbf{\bibinfo{volume}{42}}, \bibinfo{pages}{761--794}
  (\bibinfo{year}{2010}).

\bibitem{nagumo1962active}
\bibinfo{author}{Nagumo, J.}, \bibinfo{author}{Arimoto, S.} \&
  \bibinfo{author}{Yoshizawa, S.}
\newblock \bibinfo{title}{An active pulse transmission line simulating nerve
  axon}.
\newblock \emph{\bibinfo{journal}{Proc. IRE}} \textbf{\bibinfo{volume}{50}},
  \bibinfo{pages}{2061--2070} (\bibinfo{year}{1962}).

\bibitem{niebur1993theory}
\bibinfo{author}{Niebur, E.} \& \bibinfo{author}{Erd\"{o}s, P.}
\newblock \bibinfo{title}{Theory of the locomotion of nematodes: control of the
  somatic motor neurons by interneurons}.
\newblock \emph{\bibinfo{journal}{Math. Biosci.}}
  \textbf{\bibinfo{volume}{118}}, \bibinfo{pages}{51--82}
  (\bibinfo{year}{1993}).

\bibitem{bryden2004simulation}
\bibinfo{author}{Bryden, J.} \& \bibinfo{author}{Cohen, N.}
\newblock \bibinfo{title}{A simulation model of the locomotion controllers for
  the nematode caenorhabditis elegans}.
\newblock In \emph{\bibinfo{booktitle}{From Animals to Animats 8: Proceedings
  of the Eighth International Conference on the Simulation of Adaptive
  Behavior}}, \bibinfo{pages}{183--192} (\bibinfo{organization}{MIT Press},
  \bibinfo{year}{2004}).

\bibitem{arena2010insect}
\bibinfo{author}{Arena, P.}, \bibinfo{author}{Patan{\'e}, L.} \&
  \bibinfo{author}{Termini, P.~S.}
\newblock \bibinfo{title}{An insect brain computational model inspired by
  drosophila melanogaster: simulation results}.
\newblock In \emph{\bibinfo{booktitle}{Neural Networks (IJCNN), The 2010
  International Joint Conference on}}, \bibinfo{pages}{1--8}
  (\bibinfo{organization}{IEEE}, \bibinfo{year}{2010}).

\bibitem{markram2006blue}
\bibinfo{author}{Markram, H.}
\newblock \bibinfo{title}{The blue brain project}.
\newblock \emph{\bibinfo{journal}{Nat. Rev. Neurosci.}}
  \textbf{\bibinfo{volume}{7}}, \bibinfo{pages}{153} (\bibinfo{year}{2006}).

\bibitem{kishimoto1979existence}
\bibinfo{author}{Kishimoto, K.} \& \bibinfo{author}{Amari, S.-i.}
\newblock \bibinfo{title}{Existence and stability of local excitations in
  homogeneous neural fields}.
\newblock \emph{\bibinfo{journal}{J. Math. Biol.}}
  \textbf{\bibinfo{volume}{7}}, \bibinfo{pages}{303--318}
  (\bibinfo{year}{1979}).

\bibitem{pinto2001spatially}
\bibinfo{author}{Pinto, D.~J.} \& \bibinfo{author}{Ermentrout, G.~B.}
\newblock \bibinfo{title}{Spatially structured activity in synaptically coupled
  neuronal networks: I. traveling fronts and pulses}.
\newblock \emph{\bibinfo{journal}{SIAM J Appl Math}}
  \textbf{\bibinfo{volume}{62}}, \bibinfo{pages}{206--225}
  (\bibinfo{year}{2001}).

\bibitem{david2003neural}
\bibinfo{author}{David, O.} \& \bibinfo{author}{Friston, K.~J.}
\newblock \bibinfo{title}{A neural mass model for meg/eeg:: coupling and
  neuronal dynamics}.
\newblock \emph{\bibinfo{journal}{NeuroImage}} \textbf{\bibinfo{volume}{20}},
  \bibinfo{pages}{1743--1755} (\bibinfo{year}{2003}).

\bibitem{david2004evaluation}
\bibinfo{author}{David, O.}, \bibinfo{author}{Cosmelli, D.} \&
  \bibinfo{author}{Friston, K.~J.}
\newblock \bibinfo{title}{Evaluation of different measures of functional
  connectivity using a neural mass model}.
\newblock \emph{\bibinfo{journal}{Neuroimage}} \textbf{\bibinfo{volume}{21}},
  \bibinfo{pages}{659--673} (\bibinfo{year}{2004}).

\bibitem{kuramoto1975lecture}
\bibinfo{author}{Kuramoto, Y.} \& \bibinfo{author}{Araki, H.}
\newblock \bibinfo{title}{Lecture notes in physics, international symposium on
  mathematical problems in theoretical physics} (\bibinfo{year}{1975}).

\bibitem{ward2003synchronous}
\bibinfo{author}{Ward, L.~M.}
\newblock \bibinfo{title}{Synchronous neural oscillations and cognitive
  processes}.
\newblock \emph{\bibinfo{journal}{Trends Cogn. Sci.}}
  \textbf{\bibinfo{volume}{7}}, \bibinfo{pages}{553--559}
  (\bibinfo{year}{2003}).

\bibitem{fries2005mechanism}
\bibinfo{author}{Fries, P.}
\newblock \bibinfo{title}{A mechanism for cognitive dynamics: neuronal
  communication through neuronal coherence}.
\newblock \emph{\bibinfo{journal}{Trends Cogn. Sci.}}
  \textbf{\bibinfo{volume}{9}}, \bibinfo{pages}{474--480}
  (\bibinfo{year}{2005}).

\bibitem{palmigiano2017flexible}
\bibinfo{author}{Palmigiano, A.}, \bibinfo{author}{Geisel, T.},
  \bibinfo{author}{Wolf, F.} \& \bibinfo{author}{Battaglia, D.}
\newblock \bibinfo{title}{Flexible information routing by transient synchrony}.
\newblock \emph{\bibinfo{journal}{Nat Neurosci}} \textbf{\bibinfo{volume}{20}},
  \bibinfo{pages}{1014--1022} (\bibinfo{year}{2017}).

\bibitem{schnitzler2005normal}
\bibinfo{author}{Schnitzler, A.} \& \bibinfo{author}{Gross, J.}
\newblock \bibinfo{title}{Normal and pathological oscillatory communication in
  the brain}.
\newblock \emph{\bibinfo{journal}{Nat. Rev. Neurosci.}}
  \textbf{\bibinfo{volume}{6}}, \bibinfo{pages}{285} (\bibinfo{year}{2005}).

\bibitem{cabral2011role}
\bibinfo{author}{Cabral, J.}, \bibinfo{author}{Hugues, E.},
  \bibinfo{author}{Sporns, O.} \& \bibinfo{author}{Deco, G.}
\newblock \bibinfo{title}{Role of local network oscillations in resting-state
  functional connectivity}.
\newblock \emph{\bibinfo{journal}{Neuroimage}} \textbf{\bibinfo{volume}{57}},
  \bibinfo{pages}{130--139} (\bibinfo{year}{2011}).

\bibitem{petersson1999statistical1}
\bibinfo{author}{Petersson, K.~M.}, \bibinfo{author}{Nichols, T.~E.},
  \bibinfo{author}{Poline, J.-B.} \& \bibinfo{author}{Holmes, A.~P.}
\newblock \bibinfo{title}{Statistical limitations in functional neuroimaging.
  i. non-inferential methods and statistical models}.
\newblock \emph{\bibinfo{journal}{Philos. Trans. R. Soc. Lond., B, Biol. Sci.}}
  \textbf{\bibinfo{volume}{354}}, \bibinfo{pages}{1239--1260}
  (\bibinfo{year}{1999}).

\bibitem{petersson1999statistical2}
\bibinfo{author}{Petersson, K.~M.}, \bibinfo{author}{Nichols, T.~E.},
  \bibinfo{author}{Poline, J.-B.} \& \bibinfo{author}{Holmes, A.~P.}
\newblock \bibinfo{title}{Statistical limitations in functional neuroimaging
  ii. signal detection and statistical inference}.
\newblock \emph{\bibinfo{journal}{Philos. Trans. R. Soc. Lond., B, Biol. Sci.}}
  \textbf{\bibinfo{volume}{354}}, \bibinfo{pages}{1261--1281}
  (\bibinfo{year}{1999}).

\bibitem{bancaud1973}
\bibinfo{author}{Bancaud, J.} \& \bibinfo{author}{Talairach, J.}
\newblock \bibinfo{title}{Methodology of stereo eeg exploration and surgical
  intervention in epilepsy}.
\newblock \emph{\bibinfo{journal}{Rev. Otoneuroophtalmol.}}
  \textbf{\bibinfo{volume}{45}}, \bibinfo{pages}{315--–328}
  (\bibinfo{year}{1973}).

\bibitem{chauvel1966stereo}
\bibinfo{author}{Chauvel, P.}, \bibinfo{author}{Vignal, J.},
  \bibinfo{author}{Biraben, A.}, \bibinfo{author}{Badier, J.} \&
  \bibinfo{author}{Scarabin, J.}
\newblock \emph{\bibinfo{title}{Stereoelectroencephalography}},
  \bibinfo{pages}{80--108} (\bibinfo{publisher}{Springer Verlag},
  \bibinfo{year}{1996}).

\bibitem{todaro2018mapping}
\bibinfo{author}{Todaro, C.}, \bibinfo{author}{Marzetti, L.},
  \bibinfo{author}{Valdes~Sosa, P.~A.}, \bibinfo{author}{Valdes-Hernandez,
  P.~A.} \& \bibinfo{author}{Pizzella, V.}
\newblock \bibinfo{title}{Mapping brain activity with electrocorticography:
  Resolution properties and robustness of inverse solutions}.
\newblock \emph{\bibinfo{journal}{Brain Topogr}} \textbf{\bibinfo{volume}{Epub
  Ahead of Print}} (\bibinfo{year}{2018}).

\bibitem{menon1999spatial}
\bibinfo{author}{Menon, R.~S.} \& \bibinfo{author}{Kim, S.-G.}
\newblock \bibinfo{title}{Spatial and temporal limits in cognitive neuroimaging
  with fmri}.
\newblock \emph{\bibinfo{journal}{Trends Cogn. Sci.}}
  \textbf{\bibinfo{volume}{3}}, \bibinfo{pages}{207--216}
  (\bibinfo{year}{1999}).

\bibitem{aguirre2014functional}
\bibinfo{author}{Aguirre, G.~K.}
\newblock \bibinfo{title}{Functional neuroimaging: technical, logical, and
  social perspectives}.
\newblock \emph{\bibinfo{journal}{Hastings Cent. Rep.}}
  \textbf{\bibinfo{volume}{44}}, \bibinfo{pages}{S8--S18}
  (\bibinfo{year}{2014}).

\bibitem{ciric2017benchmarking}
\bibinfo{author}{Ciric, R.} \emph{et~al.}
\newblock \bibinfo{title}{Benchmarking of participant-level confound regression
  strategies for the control of motion artifact in studies of functional
  connectivity}.
\newblock \emph{\bibinfo{journal}{Neuroimage}} \textbf{\bibinfo{volume}{154}},
  \bibinfo{pages}{174--187} (\bibinfo{year}{2017}).

\bibitem{avants2011reproducible}
\bibinfo{author}{Avants, B.~B.} \emph{et~al.}
\newblock \bibinfo{title}{A reproducible evaluation of {ANTs} similarity metric
  performance in brain image registration}.
\newblock \emph{\bibinfo{journal}{Neuroimage}} \textbf{\bibinfo{volume}{54}},
  \bibinfo{pages}{2033--2044} (\bibinfo{year}{2011}).

\bibitem{amari2003synchronous}
\bibinfo{author}{Amari, S.-i.}, \bibinfo{author}{Nakahara, H.},
  \bibinfo{author}{Wu, S.} \& \bibinfo{author}{Sakai, Y.}
\newblock \bibinfo{title}{Synchronous firing and higher-order interactions in
  neuron pool}.
\newblock \emph{\bibinfo{journal}{Neural Comput.}}
  \textbf{\bibinfo{volume}{15}}, \bibinfo{pages}{127--142}
  (\bibinfo{year}{2003}).

\bibitem{sizemore2017cliques}
\bibinfo{author}{Sizemore, A.~E.} \emph{et~al.}
\newblock \bibinfo{title}{Cliques and cavities in the human connectome}.
\newblock \emph{\bibinfo{journal}{J Comput Neurosci}}
  \textbf{\bibinfo{volume}{Epub Ahead of Print}} (\bibinfo{year}{2017}).

\bibitem{giusti2016twos}
\bibinfo{author}{Giusti, C.}, \bibinfo{author}{Ghrist, R.} \&
  \bibinfo{author}{Bassett, D.~S.}
\newblock \bibinfo{title}{Two's company, three (or more) is a simplex:
  {A}lgebraic-topological tools for understanding higher-order structure in
  neural data}.
\newblock \emph{\bibinfo{journal}{J Comput Neurosci}}
  \textbf{\bibinfo{volume}{41}}, \bibinfo{pages}{1--14} (\bibinfo{year}{2016}).

\bibitem{giusti2015clique}
\bibinfo{author}{Giusti, C.}, \bibinfo{author}{Pastalkova, E.},
  \bibinfo{author}{Curto, C.} \& \bibinfo{author}{Itskov, V.}
\newblock \bibinfo{title}{Clique topology reveals intrinsic geometric structure
  in neural correlations}.
\newblock \emph{\bibinfo{journal}{Proc Natl Acad Sci U S A}}
  \textbf{\bibinfo{volume}{112}}, \bibinfo{pages}{13455--13460}
  (\bibinfo{year}{2015}).

\bibitem{reimann2017cliques}
\bibinfo{author}{Reimann, M.~W.} \emph{et~al.}
\newblock \bibinfo{title}{Cliques of neurons bound into cavities provide a
  missing link between structure and function}.
\newblock \emph{\bibinfo{journal}{Front Comput Neurosci}}
  \textbf{\bibinfo{volume}{11}}, \bibinfo{pages}{48} (\bibinfo{year}{2017}).

\bibitem{battaglia2012dynamic}
\bibinfo{author}{Battaglia, D.}, \bibinfo{author}{Witt, A.},
  \bibinfo{author}{Wolf, F.} \& \bibinfo{author}{Geisel, T.}
\newblock \bibinfo{title}{Dynamic effective connectivity of inter-areal brain
  circuits}.
\newblock \emph{\bibinfo{journal}{PLoS Comput Biol}}
  \textbf{\bibinfo{volume}{8}}, \bibinfo{pages}{e1002438}
  (\bibinfo{year}{2012}).

\bibitem{zylberg2017robust}
\bibinfo{author}{Zylberberg, J.}, \bibinfo{author}{Pouget, A.},
  \bibinfo{author}{Latham, P.~E.} \& \bibinfo{author}{Shea-Brown, E.}
\newblock \bibinfo{title}{Robust information propagation through noisy neural
  circuits}.
\newblock \emph{\bibinfo{journal}{PLoS Comput Biol}}
  \textbf{\bibinfo{volume}{13}}, \bibinfo{pages}{e1005497}
  (\bibinfo{year}{2017}).

\bibitem{kirst2016dynamic}
\bibinfo{author}{Kirst, C.}, \bibinfo{author}{Timme, M.} \&
  \bibinfo{author}{Battaglia, D.}
\newblock \bibinfo{title}{Dynamic information routing in complex networks}.
\newblock \emph{\bibinfo{journal}{Nat Commun}} \textbf{\bibinfo{volume}{7}},
  \bibinfo{pages}{11061} (\bibinfo{year}{2016}).

\bibitem{mcintyre2004uncovering}
\bibinfo{author}{McIntyre, C.~C.}, \bibinfo{author}{Savasta, M.},
  \bibinfo{author}{Kerkerian-Le~Goff, L.} \& \bibinfo{author}{Vitek, J.~L.}
\newblock \bibinfo{title}{Uncovering the mechanism (s) of action of deep brain
  stimulation: activation, inhibition, or both}.
\newblock \emph{\bibinfo{journal}{Clin. Neurophysiol.}}
  \textbf{\bibinfo{volume}{115}}, \bibinfo{pages}{1239--1248}
  (\bibinfo{year}{2004}).

\bibitem{lozano2013probing}
\bibinfo{author}{Lozano, A.~M.} \& \bibinfo{author}{Lipsman, N.}
\newblock \bibinfo{title}{Probing and regulating dysfunctional circuits using
  deep brain stimulation}.
\newblock \emph{\bibinfo{journal}{Neuron}} \textbf{\bibinfo{volume}{77}},
  \bibinfo{pages}{406--424} (\bibinfo{year}{2013}).

\bibitem{liu2016control}
\bibinfo{author}{Liu, Y.-Y.} \& \bibinfo{author}{Barab{\'a}si, A.-L.}
\newblock \bibinfo{title}{Control principles of complex systems}.
\newblock \emph{\bibinfo{journal}{Rev. Mod. Phys.}}
  \textbf{\bibinfo{volume}{88}}, \bibinfo{pages}{035006}
  (\bibinfo{year}{2016}).

\bibitem{schiff2012neural}
\bibinfo{author}{Schiff, S.~J.}
\newblock \emph{\bibinfo{title}{Neural control engineering: the emerging
  intersection between control theory and neuroscience}}
  (\bibinfo{publisher}{MIT Press}, \bibinfo{year}{2012}).

\bibitem{kim2018role}
\bibinfo{author}{Kim, J.~Z.} \emph{et~al.}
\newblock \bibinfo{title}{Role of graph architecture in controlling dynamical
  networks with applications to neural systems}.
\newblock \emph{\bibinfo{journal}{Nature Physics}}
  \textbf{\bibinfo{volume}{Epub Ahead of Print}} (\bibinfo{year}{2018}).

\bibitem{gu2015controllability}
\bibinfo{author}{Gu, S.} \emph{et~al.}
\newblock \bibinfo{title}{Controllability of structural brain networks}.
\newblock \emph{\bibinfo{journal}{Nat Commun}} \textbf{\bibinfo{volume}{6}},
  \bibinfo{pages}{8414} (\bibinfo{year}{2015}).

\bibitem{jeganathan2018fronto}
\bibinfo{author}{Jeganathan, J.} \emph{et~al.}
\newblock \bibinfo{title}{Fronto-limbic dysconnectivity leads to impaired brain
  network controllability in young people with bipolar disorder and those at
  high genetic risk}.
\newblock \emph{\bibinfo{journal}{Neuroimage Clin}}
  \textbf{\bibinfo{volume}{19}}, \bibinfo{pages}{71--81}
  (\bibinfo{year}{2018}).

\bibitem{muldoon2016stimulation}
\bibinfo{author}{Muldoon, S.~F.} \emph{et~al.}
\newblock \bibinfo{title}{Stimulation-based control of dynamic brain networks}.
\newblock \emph{\bibinfo{journal}{PLoS Comput Biol}}
  \textbf{\bibinfo{volume}{12}}, \bibinfo{pages}{e1005076}
  (\bibinfo{year}{2016}).

\bibitem{taylor2015optimal}
\bibinfo{author}{Taylor, P.~N.} \emph{et~al.}
\newblock \bibinfo{title}{Optimal control based seizure abatement using patient
  derived connectivity}.
\newblock \emph{\bibinfo{journal}{Front Neurosci}}
  \textbf{\bibinfo{volume}{9}}, \bibinfo{pages}{202} (\bibinfo{year}{2015}).

\bibitem{medaglia2018network}
\bibinfo{author}{Medaglia, J.~D.} \emph{et~al.}
\newblock \bibinfo{title}{Network controllability in the inferior frontal gyrus
  relates to controlled language variability and susceptibility to {TMS}}.
\newblock \emph{\bibinfo{journal}{J Neurosci}} \textbf{\bibinfo{volume}{38}},
  \bibinfo{pages}{6399--6410} (\bibinfo{year}{2018}).

\bibitem{holt2016phasic}
\bibinfo{author}{Holt, A.~B.}, \bibinfo{author}{Wilson, D.},
  \bibinfo{author}{Shinn, M.}, \bibinfo{author}{Moehlis, J.} \&
  \bibinfo{author}{Netoff, T.~I.}
\newblock \bibinfo{title}{Phasic burst stimulation: {A} closed-loop approach to
  tuning deep brain stimulation parameters for {P}arkinson's disease}.
\newblock \emph{\bibinfo{journal}{PLoS Comput Biol}}
  \textbf{\bibinfo{volume}{12}}, \bibinfo{pages}{e1005011}
  (\bibinfo{year}{2016}).

\bibitem{holmes1918disturbances}
\bibinfo{author}{Holmes, G.}
\newblock \bibinfo{title}{Disturbances of vision by cerebral lesions}.
\newblock \emph{\bibinfo{journal}{The British journal of ophthalmology}}
  \textbf{\bibinfo{volume}{2}}, \bibinfo{pages}{353} (\bibinfo{year}{1918}).

\bibitem{owen1990planning}
\bibinfo{author}{Owen, A.~M.}, \bibinfo{author}{Downes, J.~J.},
  \bibinfo{author}{Sahakian, B.~J.}, \bibinfo{author}{Polkey, C.~E.} \&
  \bibinfo{author}{Robbins, T.~W.}
\newblock \bibinfo{title}{Planning and spatial working memory following frontal
  lobe lesions in man}.
\newblock \emph{\bibinfo{journal}{Neuropsychologia}}
  \textbf{\bibinfo{volume}{28}}, \bibinfo{pages}{1021--1034}
  (\bibinfo{year}{1990}).

\bibitem{walsh2000transcranial}
\bibinfo{author}{Walsh, V.} \& \bibinfo{author}{Cowey, A.}
\newblock \bibinfo{title}{Transcranial magnetic stimulation and cognitive
  neuroscience}.
\newblock \emph{\bibinfo{journal}{Nat. Rev. Neurosci.}}
  \textbf{\bibinfo{volume}{1}}, \bibinfo{pages}{73} (\bibinfo{year}{2000}).

\bibitem{amassian1993measurement}
\bibinfo{author}{Amassian, V.~E.} \emph{et~al.}
\newblock \bibinfo{title}{Measurement of information processing delays in human
  visual cortex with repetitive magnetic coil stimulation}.
\newblock \emph{\bibinfo{journal}{Brain Res.}} \textbf{\bibinfo{volume}{605}},
  \bibinfo{pages}{317--321} (\bibinfo{year}{1993}).

\bibitem{pascual1994modulation}
\bibinfo{author}{Pascual-Leone, A.}, \bibinfo{author}{Grafman, J.} \&
  \bibinfo{author}{Hallett, M.}
\newblock \bibinfo{title}{Modulation of cortical motor output maps during
  development of implicit and explicit knowledge}.
\newblock \emph{\bibinfo{journal}{Science}} \textbf{\bibinfo{volume}{263}},
  \bibinfo{pages}{1287--1289} (\bibinfo{year}{1994}).

\bibitem{pascual1991induction}
\bibinfo{author}{Pascual-Leone, A.}, \bibinfo{author}{Gates, J.~R.} \&
  \bibinfo{author}{Dhuna, A.}
\newblock \bibinfo{title}{Induction of speech arrest and counting errors with
  rapid-rate transcranial magnetic stimulation}.
\newblock \emph{\bibinfo{journal}{Neurology}} \textbf{\bibinfo{volume}{41}},
  \bibinfo{pages}{697--702} (\bibinfo{year}{1991}).

\bibitem{walsh1998task}
\bibinfo{author}{Walsh, V.}, \bibinfo{author}{Ellison, A.},
  \bibinfo{author}{Battelli, L.} \& \bibinfo{author}{Cowey, A.}
\newblock \bibinfo{title}{Task--specific impairments and enhancements induced
  by magnetic stimulation of human visual area v5}.
\newblock \emph{\bibinfo{journal}{Proc. R. Soc. Lond., B, Biol. Sci.}}
  \textbf{\bibinfo{volume}{265}}, \bibinfo{pages}{537--543}
  (\bibinfo{year}{1998}).

\bibitem{kringelbach2007translational}
\bibinfo{author}{Kringelbach, M.~L.}, \bibinfo{author}{Jenkinson, N.},
  \bibinfo{author}{Owen, S.~L.} \& \bibinfo{author}{Aziz, T.~Z.}
\newblock \bibinfo{title}{Translational principles of deep brain stimulation}.
\newblock \emph{\bibinfo{journal}{Nat. Rev. Neurosci.}}
  \textbf{\bibinfo{volume}{8}}, \bibinfo{pages}{623} (\bibinfo{year}{2007}).

\bibitem{george1999transcranial}
\bibinfo{author}{George, M.~S.}, \bibinfo{author}{Lisanby, S.~H.} \&
  \bibinfo{author}{Sackeim, H.~A.}
\newblock \bibinfo{title}{Transcranial magnetic stimulation: applications in
  neuropsychiatry}.
\newblock \emph{\bibinfo{journal}{Archives of General Psychiatry}}
  \textbf{\bibinfo{volume}{56}}, \bibinfo{pages}{300--311}
  (\bibinfo{year}{1999}).

\bibitem{perlmutter2006deep}
\bibinfo{author}{Perlmutter, J.~S.} \& \bibinfo{author}{Mink, J.~W.}
\newblock \bibinfo{title}{Deep brain stimulation}.
\newblock \emph{\bibinfo{journal}{Annu. Rev. Neurosci.}}
  \textbf{\bibinfo{volume}{29}}, \bibinfo{pages}{229--257}
  (\bibinfo{year}{2006}).

\bibitem{tass1998detection}
\bibinfo{author}{Tass, P.} \emph{et~al.}
\newblock \bibinfo{title}{Detection of n: m phase locking from noisy data:
  Application to magnetoencephalography}.
\newblock \emph{\bibinfo{journal}{Phys. Rev. Lett.}}
  \textbf{\bibinfo{volume}{81}}, \bibinfo{pages}{3291} (\bibinfo{year}{1998}).

\bibitem{santaniello2015therapeutic}
\bibinfo{author}{Santaniello, S.} \emph{et~al.}
\newblock \bibinfo{title}{Therapeutic mechanisms of high-frequency stimulation
  in parkinson’s disease and neural restoration via loop-based
  reinforcement}.
\newblock \emph{\bibinfo{journal}{Proceedings of the National Academy of
  Sciences}} \textbf{\bibinfo{volume}{112}}, \bibinfo{pages}{E586--E595}
  (\bibinfo{year}{2015}).

\bibitem{zeki1993vision}
\bibinfo{author}{Zeki, S.}
\newblock \emph{\bibinfo{title}{A vision of the brain}}
  (\bibinfo{publisher}{Blackwell Scientific Publ.}, \bibinfo{year}{1993}).

\bibitem{chiken2014disrupting}
\bibinfo{author}{Chiken, S.} \& \bibinfo{author}{Nambu, A.}
\newblock \bibinfo{title}{Disrupting neuronal transmission: mechanism of dbs?}
\newblock \emph{\bibinfo{journal}{Front. Syst. Neurosci.}}
  \textbf{\bibinfo{volume}{8}}, \bibinfo{pages}{33} (\bibinfo{year}{2014}).

\bibitem{berenyi2012closed}
\bibinfo{author}{Ber{\'e}nyi, A.}, \bibinfo{author}{Belluscio, M.},
  \bibinfo{author}{Mao, D.} \& \bibinfo{author}{Buzs{\'a}ki, G.}
\newblock \bibinfo{title}{Closed-loop control of epilepsy by transcranial
  electrical stimulation}.
\newblock \emph{\bibinfo{journal}{Science}} \textbf{\bibinfo{volume}{337}},
  \bibinfo{pages}{735--737} (\bibinfo{year}{2012}).

\bibitem{kedzior2016cognitive}
\bibinfo{author}{Kedzior, K.~K.}, \bibinfo{author}{Gierke, L.},
  \bibinfo{author}{Gellersen, H.~M.} \& \bibinfo{author}{Berlim, M.~T.}
\newblock \bibinfo{title}{Cognitive functioning and deep transcranial magnetic
  stimulation (dtms) in major psychiatric disorders: a systematic review}.
\newblock \emph{\bibinfo{journal}{J. Psychiatr. Res.}}
  \textbf{\bibinfo{volume}{75}}, \bibinfo{pages}{107--115}
  (\bibinfo{year}{2016}).

\bibitem{ching2013real}
\bibinfo{author}{Ching, S.} \emph{et~al.}
\newblock \bibinfo{title}{Real-time closed-loop control in a rodent model of
  medically induced coma using burst suppression}.
\newblock \emph{\bibinfo{journal}{Anesthesiology}}
  \textbf{\bibinfo{volume}{119}}, \bibinfo{pages}{848--860}
  (\bibinfo{year}{2013}).

\bibitem{holt2014origins}
\bibinfo{author}{Holt, A.~B.} \& \bibinfo{author}{Netoff, T.~I.}
\newblock \bibinfo{title}{Origins and suppression of oscillations in a
  computational model of parkinson’s disease}.
\newblock \emph{\bibinfo{journal}{J. Comput. Neurosci.}}
  \textbf{\bibinfo{volume}{37}}, \bibinfo{pages}{505--521}
  (\bibinfo{year}{2014}).

\bibitem{heck2014two}
\bibinfo{author}{Heck, C.~N.} \emph{et~al.}
\newblock \bibinfo{title}{Two-year seizure reduction in adults with medically
  intractable partial onset epilepsy treated with responsive neurostimulation:
  final results of the {RNS} {S}ystem {P}ivotal trial}.
\newblock \emph{\bibinfo{journal}{Epilepsia}} \textbf{\bibinfo{volume}{55}},
  \bibinfo{pages}{432--441} (\bibinfo{year}{2014}).

\bibitem{crinion2007spatial}
\bibinfo{author}{Crinion, J.} \emph{et~al.}
\newblock \bibinfo{title}{Spatial normalization of lesioned brains: performance
  evaluation and impact on fmri analyses}.
\newblock \emph{\bibinfo{journal}{Neuroimage}} \textbf{\bibinfo{volume}{37}},
  \bibinfo{pages}{866--875} (\bibinfo{year}{2007}).

\bibitem{santaniello2011closed}
\bibinfo{author}{Santaniello, S.}, \bibinfo{author}{Fiengo, G.},
  \bibinfo{author}{Glielmo, L.} \& \bibinfo{author}{Grill, W.~M.}
\newblock \bibinfo{title}{Closed-loop control of deep brain stimulation: a
  simulation study}.
\newblock \emph{\bibinfo{journal}{IEEE Trans. Neural. Syst. Rehabil. Eng.}}
  \textbf{\bibinfo{volume}{19}}, \bibinfo{pages}{15--24}
  (\bibinfo{year}{2011}).

\bibitem{iudice2015structural}
\bibinfo{author}{Iudice, F.~L.}, \bibinfo{author}{Garofalo, F.} \&
  \bibinfo{author}{Sorrentino, F.}
\newblock \bibinfo{title}{Structural permeability of complex networks to
  control signals}.
\newblock \emph{\bibinfo{journal}{Nat. Commun.}} \textbf{\bibinfo{volume}{6}},
  \bibinfo{pages}{8349} (\bibinfo{year}{2015}).

\bibitem{posner2004attention}
\bibinfo{author}{Posner, M.~I.}, \bibinfo{author}{Snyder, C.~R.} \&
  \bibinfo{author}{Solso, R.}
\newblock \bibinfo{title}{Attention and cognitive control}.
\newblock \emph{\bibinfo{journal}{Cogn. Psychol.}}
  \textbf{\bibinfo{volume}{205}} (\bibinfo{year}{2004}).

\bibitem{fuster1971neuron}
\bibinfo{author}{Fuster, J.~M.} \& \bibinfo{author}{Alexander, G.~E.}
\newblock \bibinfo{title}{Neuron activity related to short-term memory}.
\newblock \emph{\bibinfo{journal}{Science}} \textbf{\bibinfo{volume}{173}},
  \bibinfo{pages}{652--654} (\bibinfo{year}{1971}).

\bibitem{goldman1970localization}
\bibinfo{author}{Goldman, P.~S.} \& \bibinfo{author}{Rosvold, H.~E.}
\newblock \bibinfo{title}{Localization of function within the dorsolateral
  prefrontal cortex of the rhesus monkey}.
\newblock \emph{\bibinfo{journal}{Exp. Neurol.}} \textbf{\bibinfo{volume}{27}},
  \bibinfo{pages}{291--304} (\bibinfo{year}{1970}).

\bibitem{bechara1994insensitivity}
\bibinfo{author}{Bechara, A.}, \bibinfo{author}{Damasio, A.~R.},
  \bibinfo{author}{Damasio, H.} \& \bibinfo{author}{Anderson, S.~W.}
\newblock \bibinfo{title}{Insensitivity to future consequences following damage
  to human prefrontal cortex}.
\newblock \emph{\bibinfo{journal}{Cognition}} \textbf{\bibinfo{volume}{50}},
  \bibinfo{pages}{7--15} (\bibinfo{year}{1994}).

\bibitem{dias1996dissociation}
\bibinfo{author}{Dias, R.}, \bibinfo{author}{Robbins, T.} \&
  \bibinfo{author}{Roberts, A.}
\newblock \bibinfo{title}{Dissociation in prefrontal cortex of affective and
  attentional shifts}.
\newblock \emph{\bibinfo{journal}{Nature}} \textbf{\bibinfo{volume}{380}},
  \bibinfo{pages}{69} (\bibinfo{year}{1996}).

\bibitem{gu2017optimal}
\bibinfo{author}{Gu, S.} \emph{et~al.}
\newblock \bibinfo{title}{Optimal trajectories of brain state transitions}.
\newblock \emph{\bibinfo{journal}{Neuroimage}} \textbf{\bibinfo{volume}{148}},
  \bibinfo{pages}{305--317} (\bibinfo{year}{2017}).

\bibitem{betzel2016optimally}
\bibinfo{author}{Betzel, R.~F.}, \bibinfo{author}{Gu, S.},
  \bibinfo{author}{Medaglia, J.~D.}, \bibinfo{author}{Pasqualetti, F.} \&
  \bibinfo{author}{Bassett, D.~S.}
\newblock \bibinfo{title}{Optimally controlling the human connectome: the role
  of network topology}.
\newblock \emph{\bibinfo{journal}{Sci. Rep.}} \textbf{\bibinfo{volume}{6}},
  \bibinfo{pages}{30770} (\bibinfo{year}{2016}).

\bibitem{pasqualetti2014controllability}
\bibinfo{author}{Pasqualetti, F.}, \bibinfo{author}{Zampieri, S.} \&
  \bibinfo{author}{Bullo, F.}
\newblock \bibinfo{title}{Controllability metrics, limitations and algorithms
  for complex networks}.
\newblock \emph{\bibinfo{journal}{IEEE Trans. Control Network Syst.}}
  \textbf{\bibinfo{volume}{1}}, \bibinfo{pages}{40--52} (\bibinfo{year}{2014}).

\bibitem{tang2017developmental}
\bibinfo{author}{Tang, E.} \emph{et~al.}
\newblock \bibinfo{title}{Developmental increases in white matter network
  controllability support a growing diversity of brain dynamics}.
\newblock \emph{\bibinfo{journal}{Nat. Commun.}} \textbf{\bibinfo{volume}{8}},
  \bibinfo{pages}{1252} (\bibinfo{year}{2017}).

\bibitem{cornblath2018sex}
\bibinfo{author}{Cornblath, E.~J.} \emph{et~al.}
\newblock \bibinfo{title}{Sex differences in network controllability as a
  predictor of executive function in youth}.
\newblock \emph{\bibinfo{journal}{arXiv}} \textbf{\bibinfo{volume}{1801}},
  \bibinfo{pages}{04623}.

\bibitem{tang2018control}
\bibinfo{author}{Tang, E.} \& \bibinfo{author}{Bassett, D.~S.}
\newblock \bibinfo{title}{Control of dynamics in brain networks}.
\newblock \emph{\bibinfo{journal}{Reviews of Modern Physics}}
  \textbf{\bibinfo{volume}{In Press}} (\bibinfo{year}{2018}).

\bibitem{adamantidis2007neural}
\bibinfo{author}{Adamantidis, A.~R.}, \bibinfo{author}{Zhang, F.},
  \bibinfo{author}{Aravanis, A.~M.}, \bibinfo{author}{Deisseroth, K.} \&
  \bibinfo{author}{De~Lecea, L.}
\newblock \bibinfo{title}{Neural substrates of awakening probed with
  optogenetic control of hypocretin neurons}.
\newblock \emph{\bibinfo{journal}{Nature}} \textbf{\bibinfo{volume}{450}},
  \bibinfo{pages}{420} (\bibinfo{year}{2007}).

\bibitem{deisseroth2011optogenetics}
\bibinfo{author}{Deisseroth, K.}
\newblock \bibinfo{title}{Optogenetics}.
\newblock \emph{\bibinfo{journal}{Nat. Methods}} \textbf{\bibinfo{volume}{8}},
  \bibinfo{pages}{26} (\bibinfo{year}{2011}).

\bibitem{gunaydin2010ultrafast}
\bibinfo{author}{Gunaydin, L.~A.} \emph{et~al.}
\newblock \bibinfo{title}{Ultrafast optogenetic control}.
\newblock \emph{\bibinfo{journal}{Nature neuroscience}}
  \textbf{\bibinfo{volume}{13}}, \bibinfo{pages}{387} (\bibinfo{year}{2010}).

\bibitem{grosenick2015closed}
\bibinfo{author}{Grosenick, L.}, \bibinfo{author}{Marshel, J.~H.} \&
  \bibinfo{author}{Deisseroth, K.}
\newblock \bibinfo{title}{Closed-loop and activity-guided optogenetic control}.
\newblock \emph{\bibinfo{journal}{Neuron}} \textbf{\bibinfo{volume}{86}},
  \bibinfo{pages}{106--139} (\bibinfo{year}{2015}).

\bibitem{prakash2012two}
\bibinfo{author}{Prakash, R.} \emph{et~al.}
\newblock \bibinfo{title}{Two-photon optogenetic toolbox for fast inhibition,
  excitation and bistable modulation}.
\newblock \emph{\bibinfo{journal}{Nat. Methods}} \textbf{\bibinfo{volume}{9}},
  \bibinfo{pages}{1171} (\bibinfo{year}{2012}).

\bibitem{rickgauer2014simultaneous}
\bibinfo{author}{Rickgauer, J.~P.}, \bibinfo{author}{Deisseroth, K.} \&
  \bibinfo{author}{Tank, D.~W.}
\newblock \bibinfo{title}{Simultaneous cellular-resolution optical perturbation
  and imaging of place cell firing fields}.
\newblock \emph{\bibinfo{journal}{Nat. Neurosci.}}
  \textbf{\bibinfo{volume}{17}}, \bibinfo{pages}{1816} (\bibinfo{year}{2014}).

\bibitem{becker2018large}
\bibinfo{author}{Becker, C.~O.}, \bibinfo{author}{Bassett, D.} \&
  \bibinfo{author}{Preciado, V.~M.}
\newblock \bibinfo{title}{Large-scale dynamic modeling of task-{fMRI} signals
  via subspace system identification}.
\newblock \emph{\bibinfo{journal}{J Neural Eng}} \textbf{\bibinfo{volume}{Epub
  Ahead of print}} (\bibinfo{year}{2018}).

\bibitem{coron2007control}
\bibinfo{author}{Coron, J.-M.}
\newblock \emph{\bibinfo{title}{Control and nonlinearity}}.
\newblock \bibinfo{number}{136} (\bibinfo{publisher}{American Mathematical
  Soc.}, \bibinfo{year}{2007}).

\bibitem{klickstein2017locally}
\bibinfo{author}{Klickstein, I.}, \bibinfo{author}{Shirin, A.} \&
  \bibinfo{author}{Sorrentino, F.}
\newblock \bibinfo{title}{Locally optimal control of complex networks}.
\newblock \emph{\bibinfo{journal}{Phys. Rev. Let.}}
  \textbf{\bibinfo{volume}{119}}, \bibinfo{pages}{268301}
  (\bibinfo{year}{2017}).

\bibitem{haynes1970nonlinear}
\bibinfo{author}{Haynes, G.} \& \bibinfo{author}{Hermes, H.}
\newblock \bibinfo{title}{Nonlinear controllability via lie theory}.
\newblock \emph{\bibinfo{journal}{SIAM J. Control}}
  \textbf{\bibinfo{volume}{8}}, \bibinfo{pages}{450--460}
  (\bibinfo{year}{1970}).

\bibitem{sussmann1972controllability}
\bibinfo{author}{Sussmann, H.~J.} \& \bibinfo{author}{Jurdjevic, V.}
\newblock \bibinfo{title}{Controllability of nonlinear systems}.
\newblock \emph{\bibinfo{journal}{Differ. Equ.}} \textbf{\bibinfo{volume}{12}},
  \bibinfo{pages}{95--116} (\bibinfo{year}{1972}).

\bibitem{hermann1977nonlinear}
\bibinfo{author}{Hermann, R.} \& \bibinfo{author}{Krener, A.}
\newblock \bibinfo{title}{Nonlinear controllability and observability}.
\newblock \emph{\bibinfo{journal}{IEEE Trans. Automat. Contr.}}
  \textbf{\bibinfo{volume}{22}}, \bibinfo{pages}{728--740}
  (\bibinfo{year}{1977}).

\bibitem{cornelius2013realistic}
\bibinfo{author}{Cornelius, S.~P.}, \bibinfo{author}{Kath, W.~L.} \&
  \bibinfo{author}{Motter, A.~E.}
\newblock \bibinfo{title}{Realistic control of network dynamics}.
\newblock \emph{\bibinfo{journal}{Nat. Commun.}} \textbf{\bibinfo{volume}{4}},
  \bibinfo{pages}{1942} (\bibinfo{year}{2013}).

\bibitem{whalen2015observability}
\bibinfo{author}{Whalen, A.~J.}, \bibinfo{author}{Brennan, S.~N.},
  \bibinfo{author}{Sauer, T.~D.} \& \bibinfo{author}{Schiff, S.~J.}
\newblock \bibinfo{title}{Observability and controllability of nonlinear
  networks: The role of symmetry}.
\newblock \emph{\bibinfo{journal}{Phys. Rev. X}} \textbf{\bibinfo{volume}{5}},
  \bibinfo{pages}{011005} (\bibinfo{year}{2015}).

\bibitem{isidori2013nonlinear}
\bibinfo{author}{Isidori, A.}
\newblock \emph{\bibinfo{title}{Nonlinear control systems}}
  (\bibinfo{publisher}{Springer Science \& Business Media},
  \bibinfo{year}{2013}).

\bibitem{chopra2009exponential}
\bibinfo{author}{Chopra, N.} \& \bibinfo{author}{Spong, M.~W.}
\newblock \bibinfo{title}{On exponential synchronization of kuramoto
  oscillators}.
\newblock \emph{\bibinfo{journal}{IEEE Trans. Automat. Contr.}}
  \textbf{\bibinfo{volume}{54}}, \bibinfo{pages}{353--357}
  (\bibinfo{year}{2009}).

\bibitem{lynn2017statistical}
\bibinfo{author}{Lynn, C.~W.} \& \bibinfo{author}{Lee, D.~D.}
\newblock \bibinfo{title}{Statistical mechanics of influence maximization with
  thermal noise}.
\newblock \emph{\bibinfo{journal}{EPL}} \textbf{\bibinfo{volume}{117}},
  \bibinfo{pages}{66001} (\bibinfo{year}{2017}).

\bibitem{lynn2018maximizing}
\bibinfo{author}{Lynn, C.~W.} \& \bibinfo{author}{Lee, D.~D.}
\newblock \bibinfo{title}{Maximizing activity in ising networks via the tap
  approximation}.
\newblock In \emph{\bibinfo{booktitle}{Association for the Advancement of
  Artificial Intelligence}}, \bibinfo{pages}{679--686} (\bibinfo{year}{2018}).

\bibitem{amunts2015architectonic}
\bibinfo{author}{Amunts, K.} \& \bibinfo{author}{Zilles, K.}
\newblock \bibinfo{title}{Architectonic mapping of the human brain beyond
  {B}rodmann}.
\newblock \emph{\bibinfo{journal}{Neuron}} \textbf{\bibinfo{volume}{88}},
  \bibinfo{pages}{1086--1107} (\bibinfo{year}{2015}).

\bibitem{cohen2011measuring}
\bibinfo{author}{Cohen, M.~R.} \& \bibinfo{author}{Kohn, A.}
\newblock \bibinfo{title}{Measuring and interpreting neuronal correlations}.
\newblock \emph{\bibinfo{journal}{Nat Neurosci}} \textbf{\bibinfo{volume}{14}},
  \bibinfo{pages}{811--819} (\bibinfo{year}{2011}).

\bibitem{heuvel2011rich}
\bibinfo{author}{van~den Heuvel, M.~P.} \& \bibinfo{author}{Sporns, O.}
\newblock \bibinfo{title}{Rich-club organization of the human connectome}.
\newblock \emph{\bibinfo{journal}{J Neurosci}} \textbf{\bibinfo{volume}{31}},
  \bibinfo{pages}{15775--15786} (\bibinfo{year}{2011}).

\bibitem{stiso2018spatial}
\bibinfo{author}{Stiso, J.} \& \bibinfo{author}{Bassett, D.~S.}
\newblock \bibinfo{title}{Spatial embedding imposes constraints on the network
  architectures of neural systems}.
\newblock \emph{\bibinfo{journal}{arXiv}} \textbf{\bibinfo{volume}{1807}},
  \bibinfo{pages}{04691} (\bibinfo{year}{2018}).

\bibitem{heuvel2016comparative}
\bibinfo{author}{van~den Heuvel, M.~P.}, \bibinfo{author}{Bullmore, E.~T.} \&
  \bibinfo{author}{Sporns, O.}
\newblock \bibinfo{title}{Comparative connectomics}.
\newblock \emph{\bibinfo{journal}{Trends Cogn Sci}}
  \textbf{\bibinfo{volume}{20}}, \bibinfo{pages}{345--361}
  (\bibinfo{year}{2016}).

\bibitem{persichetti2015value}
\bibinfo{author}{Persichetti, A.~S.}, \bibinfo{author}{Aguirre, G.~K.} \&
  \bibinfo{author}{Thompson-Schill, S.~L.}
\newblock \bibinfo{title}{Value is in the eye of the beholder: early visual
  cortex codes monetary value of objects during a diverted attention task}.
\newblock \emph{\bibinfo{journal}{J Cogn Neurosci}}
  \textbf{\bibinfo{volume}{27}}, \bibinfo{pages}{893--901}
  (\bibinfo{year}{2015}).

\bibitem{dore2018brain}
\bibinfo{author}{Dore, B.~P.} \emph{et~al.}
\newblock \bibinfo{title}{Brain activity tracks population information sharing
  by capturing consensus judgments of value}.
\newblock \emph{\bibinfo{journal}{Cereb Cortex}} \bibinfo{pages}{Aug 28}
  (\bibinfo{year}{2018}).

\bibitem{constantinescu2016organizing}
\bibinfo{author}{Constantinescu, A.~O.}, \bibinfo{author}{O'Reilly, J.~X.} \&
  \bibinfo{author}{Behrens, T. E.~J.}
\newblock \bibinfo{title}{Organizing conceptual knowledge in humans with a
  gridlike code}.
\newblock \emph{\bibinfo{journal}{Science}} \textbf{\bibinfo{volume}{352}},
  \bibinfo{pages}{1464--1468} (\bibinfo{year}{2016}).

\bibitem{papadopoulos2018network}
\bibinfo{author}{Papadopoulos, L.}, \bibinfo{author}{Porter, M.~A.},
  \bibinfo{author}{Daniels, K.~E.} \& \bibinfo{author}{Bassett, D.~S.}
\newblock \bibinfo{title}{Network analysis of particles and grains}.
\newblock \emph{\bibinfo{journal}{Journal of Complex Networks}}
  \textbf{\bibinfo{volume}{6}}, \bibinfo{pages}{485--–565}
  (\bibinfo{year}{2018}).

\bibitem{bianconi2015complex}
\bibinfo{author}{Bianconi, G.}, \bibinfo{author}{Rahmede, C.} \&
  \bibinfo{author}{Wu, Z.}
\newblock \bibinfo{title}{Complex quantum network geometries: {E}volution and
  phase transitions}.
\newblock \emph{\bibinfo{journal}{Phys Rev E Stat Nonlin Soft Matter Phys}}
  \textbf{\bibinfo{volume}{92}}, \bibinfo{pages}{022815}
  (\bibinfo{year}{2015}).

\bibitem{kivela2014multilayer}
\bibinfo{author}{Kivel, M.} \emph{et~al.}
\newblock \bibinfo{title}{Multilayer networks}.
\newblock \emph{\bibinfo{journal}{J. Complex Netw.}}
  \textbf{\bibinfo{volume}{2}}, \bibinfo{pages}{203--271}
  (\bibinfo{year}{2014}).

\bibitem{domenico2016physics}
\bibinfo{author}{De~Domenico, M.}, \bibinfo{author}{Granell, C.},
  \bibinfo{author}{Porter, M.~A.} \& \bibinfo{author}{Arenas, A.}
\newblock \bibinfo{title}{The physics of spreading processes in multilayer
  networks} \textbf{\bibinfo{volume}{12}}, \bibinfo{pages}{901--906}
  (\bibinfo{year}{2016}).

\bibitem{newman2016structure}
\bibinfo{author}{Newman, M. E.~J.} \& \bibinfo{author}{Clauset, A.}
\newblock \bibinfo{title}{Structure and inference in annotated networks}.
\newblock \emph{\bibinfo{journal}{Nature Communications}}
  \textbf{\bibinfo{volume}{7}}, \bibinfo{pages}{11863} (\bibinfo{year}{2016}).

\bibitem{bassett2014cross}
\bibinfo{author}{Bassett, D.~S.}, \bibinfo{author}{Wymbs, N.~F.},
  \bibinfo{author}{Porter, M.~A.}, \bibinfo{author}{Mucha, P.~J.} \&
  \bibinfo{author}{Grafton, S.~T.}
\newblock \bibinfo{title}{Cross-linked structure of network evolution}.
\newblock \emph{\bibinfo{journal}{Chaos}} \textbf{\bibinfo{volume}{24}},
  \bibinfo{pages}{013112} (\bibinfo{year}{2014}).

\bibitem{porter2009communities}
\bibinfo{author}{Porter, M.~A.}, \bibinfo{author}{Onnela, J.-P.} \&
  \bibinfo{author}{Mucha, P.~J.}
\newblock \bibinfo{title}{Communities in networks}.
\newblock \emph{\bibinfo{journal}{Notices of the AMS}}
  \textbf{\bibinfo{volume}{56}}, \bibinfo{pages}{1082--1097}
  (\bibinfo{year}{2009}).

\bibitem{fortunato2010community}
\bibinfo{author}{Fortunato, S.}
\newblock \bibinfo{title}{Community detection in graphs}.
\newblock \emph{\bibinfo{journal}{Physics reports}}
  \textbf{\bibinfo{volume}{486}}, \bibinfo{pages}{75--174}
  (\bibinfo{year}{2010}).

\bibitem{fortunato2016community}
\bibinfo{author}{Fortunato, S.} \& \bibinfo{author}{Hric, D.}
\newblock \bibinfo{title}{Community detection in networks: {A} user guide}.
\newblock \emph{\bibinfo{journal}{Physics Reports}}
  \textbf{\bibinfo{volume}{659}}, \bibinfo{pages}{1--44}
  (\bibinfo{year}{2016}).

\bibitem{gosak2018network}
\bibinfo{author}{Gosak, M.} \emph{et~al.}
\newblock \bibinfo{title}{Network science of biological systems at different
  scales: {A} review}.
\newblock \emph{\bibinfo{journal}{Phys Life Rev}}
  \textbf{\bibinfo{volume}{24}}, \bibinfo{pages}{118--135}
  (\bibinfo{year}{2018}).

\bibitem{richiardi2015correlated}
\bibinfo{author}{Richiardi, J.} \emph{et~al.}
\newblock \bibinfo{title}{Correlated gene expression supports synchronous
  activity in brain networks}.
\newblock \emph{\bibinfo{journal}{Science}} \textbf{\bibinfo{volume}{348}},
  \bibinfo{pages}{1241--1244} (\bibinfo{year}{2015}).

\bibitem{romero2018structural}
\bibinfo{author}{Romero-Garcia, R.} \emph{et~al.}
\newblock \bibinfo{title}{Structural covariance networks are coupled to
  expression of genes enriched in supragranular layers of the human cortex}.
\newblock \emph{\bibinfo{journal}{Neuroimage}} \textbf{\bibinfo{volume}{171}},
  \bibinfo{pages}{256--267} (\bibinfo{year}{2018}).

\bibitem{whitaker2016adolescence}
\bibinfo{author}{Whitaker, K.~J.} \emph{et~al.}
\newblock \bibinfo{title}{Adolescence is associated with genomically patterned
  consolidation of the hubs of the human brain connectome}.
\newblock \emph{\bibinfo{journal}{Proc Natl Acad Sci U S A}}
  \textbf{\bibinfo{volume}{113}}, \bibinfo{pages}{9105--9110}
  (\bibinfo{year}{2016}).

\bibitem{hardingham2018lineage}
\bibinfo{author}{Hardingham, G.~E.}, \bibinfo{author}{Pruunsild, P.},
  \bibinfo{author}{Greenberg, M.~E.} \& \bibinfo{author}{Bading, H.}
\newblock \bibinfo{title}{Lineage divergence of activity-driven transcription
  and evolution of cognitive ability}.
\newblock \emph{\bibinfo{journal}{Nat Rev Neurosci}}
  \textbf{\bibinfo{volume}{19}}, \bibinfo{pages}{9--15} (\bibinfo{year}{2018}).

\bibitem{luke2007network}
\bibinfo{author}{Luke, D.~A.} \& \bibinfo{author}{Harris, J.~K.}
\newblock \bibinfo{title}{Network analysis in public health: history, methods,
  and applications}.
\newblock \emph{\bibinfo{journal}{Annu Rev Public Health}}
  \textbf{\bibinfo{volume}{28}}, \bibinfo{pages}{69--93}
  (\bibinfo{year}{2007}).

\bibitem{braun2018from}
\bibinfo{author}{Braun, U.} \emph{et~al.}
\newblock \bibinfo{title}{From maps to multi-dimensional network mechanisms of
  mental disorders}.
\newblock \emph{\bibinfo{journal}{Neuron}} \textbf{\bibinfo{volume}{97}},
  \bibinfo{pages}{14--31} (\bibinfo{year}{2018}).

\bibitem{schmalzle2017brain}
\bibinfo{author}{Schmalzle, R.} \emph{et~al.}
\newblock \bibinfo{title}{Brain connectivity dynamics during social interaction
  reflect social network structure}.
\newblock \emph{\bibinfo{journal}{Proc Natl Acad Sci U S A}}
  \textbf{\bibinfo{volume}{114}}, \bibinfo{pages}{5153--5158}
  (\bibinfo{year}{2017}).

\bibitem{parkinson2018similar}
\bibinfo{author}{Parkinson, C.}, \bibinfo{author}{Kleinbaum, A.~M.} \&
  \bibinfo{author}{Wheatley, T.}
\newblock \bibinfo{title}{Similar neural responses predict friendship}.
\newblock \emph{\bibinfo{journal}{Nat Commun}} \textbf{\bibinfo{volume}{9}},
  \bibinfo{pages}{332} (\bibinfo{year}{2018}).

\bibitem{parkinson2014common}
\bibinfo{author}{Parkinson, C.}, \bibinfo{author}{Liu, S.} \&
  \bibinfo{author}{Wheatley, T.}
\newblock \bibinfo{title}{A common cortical metric for spatial, temporal, and
  social distance}.
\newblock \emph{\bibinfo{journal}{J Neurosci}} \textbf{\bibinfo{volume}{34}},
  \bibinfo{pages}{1979--1987} (\bibinfo{year}{2014}).

\bibitem{falk2017brain}
\bibinfo{author}{Falk, E.~B.} \& \bibinfo{author}{Bassett, D.~S.}
\newblock \bibinfo{title}{Brain and social networks: {F}undamental building
  blocks of human experience}.
\newblock \emph{\bibinfo{journal}{Trends Cogn Sci}}
  \textbf{\bibinfo{volume}{21}}, \bibinfo{pages}{674--690}
  (\bibinfo{year}{2017}).

\bibitem{rieke1999spikes}
\bibinfo{author}{Rieke, F.}, \bibinfo{author}{Warland, D.},
  \bibinfo{author}{de~Ruyter~van Steveninck, R.} \& \bibinfo{author}{Bialek,
  W.}
\newblock \emph{\bibinfo{title}{Spikes: exploring the neural code}}
  (\bibinfo{publisher}{MIT Press}, \bibinfo{year}{1997}).

\bibitem{cover2012elements}
\bibinfo{author}{Cover, T.~M.} \& \bibinfo{author}{Thomas, J.~A.}
\newblock \emph{\bibinfo{title}{Elements of information theory}}
  (\bibinfo{publisher}{John Wiley \& Sons}, \bibinfo{year}{2012}).

\bibitem{shannon1948mathematical}
\bibinfo{author}{Shannon, C.~E.}
\newblock \bibinfo{title}{A mathematical theory of communication}.
\newblock \emph{\bibinfo{journal}{Bell Syst. Tech. J.}}
  \textbf{\bibinfo{volume}{27}} (\bibinfo{year}{1948}).

\bibitem{mackay1952limiting}
\bibinfo{author}{MacKay, D.~M.} \& \bibinfo{author}{McCulloch, W.~S.}
\newblock \bibinfo{title}{The limiting information capacity of a neuronal
  link}.
\newblock \emph{\bibinfo{journal}{Bull. Math. Biophys.}}
  \textbf{\bibinfo{volume}{14}}, \bibinfo{pages}{127--135}
  (\bibinfo{year}{1952}).

\bibitem{attneave1954some}
\bibinfo{author}{Attneave, F.}
\newblock \bibinfo{title}{Some informational aspects of visual perception.}
\newblock \emph{\bibinfo{journal}{Psychol. Rev.}}
  \textbf{\bibinfo{volume}{61}}, \bibinfo{pages}{183} (\bibinfo{year}{1954}).

\bibitem{barlow1961possible}
\bibinfo{author}{Barlow, H.~B.}
\newblock \bibinfo{title}{Possible principles underlying the transformations of
  sensory messages}  (\bibinfo{year}{1961}).

\bibitem{van1988real}
\bibinfo{author}{van Steveninck, R. d.~R.} \& \bibinfo{author}{Bialek, W.}
\newblock \bibinfo{title}{Real-time performance of a movement-sensitive neuron
  in the blowfly visual system: coding and information transfer in short spike
  sequences}.
\newblock \emph{\bibinfo{journal}{Proc. R. Soc. Lond. B}}
  \textbf{\bibinfo{volume}{234}}, \bibinfo{pages}{379--414}
  (\bibinfo{year}{1988}).

\bibitem{strong1998entropy}
\bibinfo{author}{Strong, S.~P.}, \bibinfo{author}{Koberle, R.},
  \bibinfo{author}{van Steveninck, R. R. d.~R.} \& \bibinfo{author}{Bialek, W.}
\newblock \bibinfo{title}{Entropy and information in neural spike trains}.
\newblock \emph{\bibinfo{journal}{Phys. Rev. Lett.}}
  \textbf{\bibinfo{volume}{80}}, \bibinfo{pages}{197} (\bibinfo{year}{1998}).

\bibitem{paninski2003estimation}
\bibinfo{author}{Paninski, L.}
\newblock \bibinfo{title}{Estimation of entropy and mutual information}.
\newblock \emph{\bibinfo{journal}{Neural Comput.}}
  \textbf{\bibinfo{volume}{15}}, \bibinfo{pages}{1191--1253}
  (\bibinfo{year}{2003}).

\bibitem{nemenman2004entropy}
\bibinfo{author}{Nemenman, I.}, \bibinfo{author}{Bialek, W.} \&
  \bibinfo{author}{van Steveninck, R. d.~R.}
\newblock \bibinfo{title}{Entropy and information in neural spike trains:
  Progress on the sampling problem}.
\newblock \emph{\bibinfo{journal}{Phys. Rev. E}} \textbf{\bibinfo{volume}{69}},
  \bibinfo{pages}{056111} (\bibinfo{year}{2004}).

\bibitem{schreiber2000measuring}
\bibinfo{author}{Schreiber, T.}
\newblock \bibinfo{title}{Measuring information transfer}.
\newblock \emph{\bibinfo{journal}{Phys. Rev. Lett.}}
  \textbf{\bibinfo{volume}{85}}, \bibinfo{pages}{461} (\bibinfo{year}{2000}).

\bibitem{vicente2011transfer}
\bibinfo{author}{Vicente, R.}, \bibinfo{author}{Wibral, M.},
  \bibinfo{author}{Lindner, M.} \& \bibinfo{author}{Pipa, G.}
\newblock \bibinfo{title}{Transfer entropy---a model-free measure of effective
  connectivity for the neurosciences}.
\newblock \emph{\bibinfo{journal}{J. Comput. Neurosci.}}
  \textbf{\bibinfo{volume}{30}}, \bibinfo{pages}{45--67}
  (\bibinfo{year}{2011}).

\bibitem{jaynes1957information}
\bibinfo{author}{Jaynes, E.~T.}
\newblock \bibinfo{title}{Information theory and statistical mechanics}.
\newblock \emph{\bibinfo{journal}{Phys. Rev.}} \textbf{\bibinfo{volume}{106}},
  \bibinfo{pages}{620} (\bibinfo{year}{1957}).

\bibitem{kailath1980linear}
\bibinfo{author}{Kailath, T.}
\newblock \emph{\bibinfo{title}{Linear Systems}}
  (\bibinfo{publisher}{Prentice-Hall, Inc.}, \bibinfo{year}{1980}).

\bibitem{liu2011controllability}
\bibinfo{author}{Liu, Y.-Y.}, \bibinfo{author}{Slotine, J.-J.} \&
  \bibinfo{author}{Barab{\'a}si, A.-L.}
\newblock \bibinfo{title}{Controllability of complex networks}.
\newblock \emph{\bibinfo{journal}{Nature}} \textbf{\bibinfo{volume}{473}},
  \bibinfo{pages}{167--173} (\bibinfo{year}{2011}).

\bibitem{klickstein2017energy}
\bibinfo{author}{Klickstein, I.}, \bibinfo{author}{Shirin, A.} \&
  \bibinfo{author}{Sorrentino, F.}
\newblock \bibinfo{title}{Energy scaling of targeted optimal control of complex
  networks}.
\newblock \emph{\bibinfo{journal}{Nat. Commun.}} \textbf{\bibinfo{volume}{8}},
  \bibinfo{pages}{15145} (\bibinfo{year}{2017}).

\bibitem{yan2017network}
\bibinfo{author}{Yan, G.} \emph{et~al.}
\newblock \bibinfo{title}{Network control principles predict neuron function in
  the {C}aenorhabditis elegans connectome}.
\newblock \emph{\bibinfo{journal}{Nature}} \textbf{\bibinfo{volume}{550}},
  \bibinfo{pages}{519--523} (\bibinfo{year}{2017}).

\end{thebibliography}

%% Here is the endmatter stuff: Supplementary Info, etc.
%% Use \item's to separate, default label is "Acknowledgements"

\begin{addendum}

\item[Acknowledgements] We are grateful to Lia Papadopoulos, Jason Z. Kim, and Vivek Buch for helpful comments on an earlier version of this manuscript. We also thank Ann E. Sizemore for artistic inspiration. D.S.B. and C.W.L. acknowledge support from the John D. and Catherine T. MacArthur Foundation, the Alfred P. Sloan Foundation, the ISI Foundation, the Paul Allen Foundation, the Army Research Laboratory (W911NF-10-2-0022), the Army Research Office (Bassett-W911NF-14-1-0679, Grafton-W911NF-16-1-0474, DCIST- W911NF-17-2-0181), the Office of Naval Research, the National Institute of Mental Health (2-R01-DC-009209-11, R01-MH112847, R01-MH107235, R21-M MH-106799), the National Institute of Child Health and Human Development (1R01HD086888-01), National Institute of Neurological Disorders and Stroke (R01 NS099348), and the National Science Foundation (BCS-1441502, BCS-1430087, NSF PHY-1554488 and BCS-1631550). We also thank Vecteezy for supplying vector art.
 
\item[Competing interests] The authors declare no competing interests.
 
\item[Corresponding author] Correspondence and requests for materials should be addressed to D.S.B. \\ (dsb@seas.upenn.edu).
 
\end{addendum}

\end{document}